\documentclass[12pt]{article}
\usepackage[top=1in, bottom=1in, left=1in, right=1in]{geometry}
\usepackage{setspace}
\usepackage{booktabs}
\usepackage{rotating}
\usepackage{graphicx}
\usepackage[section]{placeins} %Placeins.sty keeps floats `in their place', preventing them from floating past a "\FloatBarrier" command into another section.  To use it, declare "\usepackage{placeins}" and insert "\FloatBarrier" at places that floats should not move past, perhaps at every "\section".  
\usepackage[small, bf]{caption}
\usepackage{subfig}
\usepackage{pdfpages}\usepackage{pdflscape}
\usepackage{textcomp}
\usepackage{longtable}
\usepackage{nicefrac}
\usepackage{adjustbox}	% to adjust the size of objects to fit into a page
\usepackage[hyphens]{url}
\usepackage{bbm}

\usepackage{natbib}
\usepackage[sectionbib]{bibunits}

\bibpunct{(}{)}{;}{a}{,}{,}

\usepackage{amsmath, amsfonts, amssymb, amsthm}
\usepackage[pdftex]{hyperref}
\hypersetup{
    colorlinks=true,
    linkcolor=red,
    citecolor=blue,
    filecolor=magenta,      
    urlcolor=cyan,
}

\usepackage{mathtools}

\newtheorem{proposition}{Proposition}[section]
\newtheorem{lemma}{Lemma}[section]
\newtheorem{corollary}{Corollary}[section]

\theoremstyle{definition}

\newtheorem{assumption}{Assumption}[section]

\newtheorem*{example*}{Example}
\newtheorem{remark}{Remark}

\usepackage{titlesec}
\titlespacing*{\section}{0pt}{1.5ex plus 1ex minus .2ex}{0.8ex plus .2ex}
\titlespacing*{\subsection}{0pt}{1.2ex plus 1ex minus .2ex}{0.8ex plus .2ex}
\titlespacing*{\subsubsection}{0pt}{1.2ex plus 1ex minus .2ex}{0.8ex plus .2ex}

\newcommand{\p}{\mathbb{P}}
\newcommand{\e}{\mathbb{E}}
\newcommand{\E}{\mathbb{E}}
\newcommand{\En}{\mathbb{E}_n}

\newcommand{\reals}{\mathbb{R}}

\newcommand{\cH}{\mathcal{H}}

\newcommand{\cU}{\mathcal{U}}
\newcommand{\cX}{\mathcal{X}}

\newcommand{\cZ}{\mathcal{Z}}

\newcommand{\indep}{\perp \!\!\!\perp}

\newcommand{\etahat}{\hat \eta}
\newcommand{\muhat}{\hat{\mu}}
\newcommand{\pihat}{\hat{\pi}}

\newcommand{\deltamin}{\underline{\delta}}
\newcommand{\deltamax}{\overline{\delta}}

\newcommand{\outcomereg}{\mu_{1}}
\newcommand{\outcomereghat}{\hat{\mu}_1}

\newcommand{\muup}{\overline{\mu}^{*}}
\newcommand{\mulow}{\underline{\mu}^{*}}

\newcommand{\sumoverfolds}{\sum_{k=1}^{K}}

\newcommand{\sumoveri}{\sum_{i=1}^{n}}

\usepackage{color-edits}
\addauthor[Amanda]{am}{blue}
\addauthor[Alex]{ac}{purple}
\addauthor[Ashesh]{ar}{cyan}
\addauthor[Edward]{ed}{red}

\newcommand{\perf}{\mbox{perf}}
\newcommand{\disp}{\mbox{disp}}
\usepackage{accents}

\usepackage{multirow}
\usepackage{floatrow}

\usepackage{wrapfig}

\usepackage[linesnumbered,ruled,vlined]{algorithm2e}
\usepackage{algpseudocode}
\SetKwInput{KwInput}{Input}                % Set the Input
\SetKwInput{KwOutput}{Output}              % set the Output

\usepackage{subfiles} 

\usepackage[subtle]{savetrees}

\title{Robust Design and Evaluation of \\ Predictive Algorithms under Unobserved Confounding\thanks{This paper was previously titled: ``Counterfactual Risk Assessments under Unmeasured Confounding.'' First version: September 2022. We thank Rohan Alur, Alexandra Chouldechova, Peter Bergman, Peter Hull, Matt Masten, Sendhil Mullainathan, Jonathan Roth, and numerous seminar participants for helpful comments.
We especially thank our partners at Commonwealth Bank of Australia.
All errors are our own.}}
\date{\today}
\author{
    Ashesh Rambachan\thanks{MIT, Department of Economics: \texttt{asheshr@mit.edu}} 
    \and
    Amanda Coston\thanks{UC Berkeley, Department of Statistics: \texttt{acoston@berkeley.edu} }
    \and
    Edward H. Kennedy\thanks{ Carnegie Mellon University, Department of Statistics and Data Science: \texttt{edward@stat.cmu.edu} }
}
\begin{document}

{\singlespacing
\maketitle
\thispagestyle{empty} 
\setcounter{page}{0}

%%%%%%%%%%%%
% Abstract %
%%%%%%%%%%%%
\begin{abstract}
Predictive algorithms inform consequential decisions in settings with selective labels: outcomes are observed only for units selected by past decision makers.
This creates an identification problem under unobserved confounding --- when selected and unselected units differ in unobserved ways that affect outcomes. 
We propose a framework for robust design and evaluation of predictive algorithms that bounds how much outcomes may differ between selected and unselected units with the same observed characteristics.
These bounds formalize common empirical strategies including proxy outcomes and instrumental variables. 
Our estimators work across bounding strategies and performance measures such as conditional likelihoods, mean square error, and true/false positive rates. 
Using administrative data from a large Australian financial institution, we show that varying confounding assumptions substantially affects credit risk predictions and fairness evaluations across income groups.

\end{abstract}
}

%%%%%%%%%%%%%%%%
% Introduction %
%%%%%%%%%%%%%%%%
\newpage \clearpage
\section{Introduction}

Predictive algorithms inform high-stakes decisions in pretrial release, consumer lending, health care, and many other domains. 
Designing and evaluating such tools is often complicated by the ``selective labels'' problem: outcomes are observed only for units selected by past decision makers. 
For example, pretrial risk assessments predict the likelihood defendants would commit pretrial misconduct if they were released before trial; but we only observe whether a defendant committed misconduct if a judge decided to release them \citep[e.g.,][]{KLLLM(18), Rambachan(21), AngelovaEtAl(22)}. 
Consumer credit scores predict the likelihood of default if an applicant were granted a loan; but we only observe whether an applicant defaulted if the financial institution approved them and they accepted the loan terms \citep[e.g.,][]{BlattnerNelson(21), CostonEtAl(21)-Rashomon}.\footnote{The selective labels problem arises when analyzing predictive algorithms in many other empirical settings as well, such as medical diagnosis \citep[e.g.,][]{MullainathanObermeyer(21)}, child protective services \citep[e.g.,][]{ChouldechovaEtAl(18)}, hiring \citep[e.g.,][]{LiRaymondBergman(20)}, education \citep[][]{BergmanEtAl(21)}, and tax audits \citep[e.g.,][]{BlackEtAl(22)-TaxAudits, BattagliniEtAl(22), ElzaynEtAl(23)}.}

The selective labels problem creates an identification challenge under unobserved confounding--when selected and unselected units differ in ways not captured by observed covariates.
Judges may release defendants based on courtroom interactions not in administrative data; loan applicants may accept offers based on competing credit options unavailable to researchers.
Ignoring unobserved confounding leads to inaccurate predictions and misleading evaluations of predictive algorithms.\footnote{While there exists methods for designing and evaluating predictive algorithms in computer science from selectively observed data \citep[e.g.,][]{Coston(20)-Evaluation, CostonEtAl(21)-Rashomon, MishlerEtAl(21), mishler2021fade, guerdan2023counterfactual}, they tackle this challenge assuming that the historical selection decision was unconfounded conditional on observed covariates.}
Existing solutions are limited. 
Applied researchers rely on informal heuristics---imputing missing outcomes using proxy variables \citep[][]{ChouldechovaEtAl(18), BlattnerNelson(21), MullainathanObermeyer(21)} or applying ad-hoc adjustments to observed outcomes among selected units.
Recent work uses instrumental variables but typically focuses on specific estimands \citep[][]{lakkaraju2017selective, ArnoldDobbieHull(20)-discrimination_alg} or predictive algorithms that are linear in covariates \citep[][]{ArnoldEtAl(24)}.

We develop a framework for robust design and evaluation of predictive algorithms under unobserved confounding. 
We bound how much outcomes may differ between selected and unselected units with the same observed characteristics. 
These bounds formalize common empirical strategies --- including proxy outcomes and instrumental variables --- and partially identify key predictive estimands such as conditional likelihoods, mean square error, calibration, and true/false positive rates. 
Our estimation approach works across multiple bounding strategies and performance measures, flexibly incorporating machine learning methods.

We consider data $(X_i, D_i, Y_i)$ for $i = 1, \hdots, n$ drawn i.i.d. from a joint distribution $\p(\cdot)$, where $X_i$ is a vector of covariates and $D_i \in \{0, 1\}$ is a selection decision by past decision makers (e.g., judges or loan applicants). 
The observed outcome satisfies $Y_i = D_i \times Y_i^*$ for some true outcome $Y_i^* \in \{0, 1\}$; for example, $Y_i^*$ is whether a defendant would commit pretrial misconduct if released before trial or whether an applicant would default if granted a loan.

Researchers have two objectives. 
First, design \textit{predictive algorithms} $s(X_i)$ that predict the probability $Y_i^* = 1$ given the covariates $X_i$ and construct data-driven decision rules.
Second, evaluate existing predictive algorithms through performance measures, such as \textit{overall performance} given by $\perf(s; \beta) = \e[\beta_0(X_i; s) + \beta_1(X_i; s) Y_i^*]$,
and \textit{class-specific performance} given by $\perf_{+}(s; \beta) = \e[\beta_0(X_i; s) \mid Y_i^* = 1]$ and $\perf_{-}(s; \beta) = \e[\beta_0(X_i; s) \mid Y_i^* = 0]$, where $\beta_0(X_i; s), \beta_1(X_i; s)$ are researcher-specified functions. 
Alternative choices recover popular diagnostics such as mean square error, true positive rate, and false positive rate.
Comparing these measures across sensitive attributes evaluates fairness properties \citep[e.g.,][]{MitchellEtAl(19), barocas-hardt-narayanan}.

Rather than assuming unconfounded selection, we bound the confounding function $\delta(x) := \p(Y_i^* = 1 \mid D_i = 0, X_i = x) - \p(Y_i^* = 1 \mid D_i = 1, X_i = x)$ using researcher-specified bounding functions that may depend on covariates and identified nuisance parameters. 
The confounding function measures how much unselected units differ from selected units with the same characteristics; bounding it formalizes assumptions about unobserved confounding severity without requiring a full selection model. 
Under these bounds, the conditional likelihood and predictive performance measures are partially identified.

Our framework encompasses popular strategies for addressing selective labels as special cases.
When researchers observe proxy outcomes correlated with $Y_i^*$ --- such as default on other credit products \citep{BlattnerNelson(21)} or long-term health outcomes \citep{MullainathanObermeyer(21), ChanGentzkowYu(21)} --- this implies specific bounds. 
When researchers have instrumental variables like randomly assigned decision makers \citep{lakkaraju2017selective, KLLLM(18)}, this implies alternative bounds. 
Our framework accommodates other plausible identifying assumptions.

We apply this framework to two objectives. 
First, we develop estimators for the bounds on the conditional likelihood $P(Y_i^* = 1 \mid X_i = x)$ (e.g., the pretrial misconduct rate or default rate given covariates), enabling design of predictive algorithms robust to unobserved confounding.
Building on advances in causal machine learning \citep[][]{NieWager(20), kennedy2020optimal},  we propose pseudo-outcome regression estimators with sample splitting: estimate nuisance parameters on one fold, construct influence-function-based pseudo-outcomes, then regress them on covariates using machine learning on a second fold.
A key innovation is an oracle inequality extending \citet{kennedy2020optimal}'s pointwise analysis to integrated mean square error: our feasible estimators approximate an infeasible oracle up to a smoothed, doubly-robust product of first-stage errors.
This applies to any regression method satisfying a mild stability condition, including random forests, kernel methods, and series estimators.
We leverage this to characterize integrated mean square error convergence rates and establish regret bounds for plug-in decision rules, connecting estimation guarantees to decision-theoretic performance.

Second, we develop estimators for bounds on the performance of a given predictive algorithm $s(\cdot)$ under unobserved confounding.
For overall performance measures $\perf(s; \beta)$ --- which nest mean square error, calibration, and precision --- we construct debiased estimators using crossfitting and debiased machine learning \citep[e.g.,][]{RobinsEtAl(08), ZhengvanderLaan(11), ChernozhukovEtAl(18)-DML}.
Under mild regularity conditions, these estimators are $\sqrt{n}$-consistent and jointly asymptotically normal.
For class-specific performance $\perf_{+}(s; \beta)$ and $\perf_{-}(s; \beta)$ --- such as true and false positive rates --- we show sharp bounds correspond to optimal values of linear fractional programs with nuisances in both the objective and constraints.
We estimate using two folds: estimate nuisances and bounding functions on the first fold, then solve the sample optimization program on the second fold.
We establish partial double robustness: estimation error decomposes into sampling error, a doubly-robust product for objective nuisances, and constraint estimation errors entering through root mean square error.
Proving this exploits the optimization structure of linear fractional programs.\footnote{Since linear fractional programs arise in several related sensitivity analysis frameworks in causal inference \citep[e.g.,][]{AronowLee(2013), MiratrixWagerZubizaretta(18), KallusMaoZhou(18)-IntervalEstimation, ZhaoSmallBhattacharya(19), KallusZhou(21)}, our results may be broadly applicable beyond the selective labels setting.}
Our estimation strategies accommodate multiple bounding strategies, flexibly incorporating domain-specific knowledge while maintaining statistical guarantees.

As an empirical illustration, we apply our framework to administrative data on personal loan applications from a large Australian bank. 
First, we construct confounding-robust credit risk scores under varying confounding assumptions. 
The resulting algorithms produce substantially different risk rankings that could meaningfully affect credit access.
Second, inspired by recent work in consumer finance \citep[][]{BlattnerNelson(21), FusterEtAl(22)predictably, DiMaggioEtAl(22)-Fintech}, we evaluate a benchmark credit score's performance across income groups.
An analysis ignoring unobserved confounding suggests the score is more accurate (in mean square error) for higher-income applicants. 
However, we show this disparity disappears at plausible levels of unobserved confounding, illustrating how researchers can assess the robustness of fairness conclusions.

Our framework prioritizes breadth and applicability. Rather than bespoke methods for each combination of bounding strategy and performance measure, we provide a unified approach for a variety of assumptions and performance measures. 
We deliver practical estimation procedures that enable researchers to robustly design and evaluate predictive algorithms under unobserved confounding.

\vspace{-1em}
\paragraph{Related Work:}
Our analysis builds on foundational work on partial identification with selective labels \citep[e.g.,][]{Manski(89), Manski(90), Manski(94)-SampleSelectionBounds}. 
We generalize \citet[][]{Manski(03)}'s ``approximate mean independence'' assumption by incorporating covariates and nuisance parameters into the bounding functions, enabling flexible formalization of empirical heuristics using proxy outcomes and other strategies. See also \cite{diaz2013sensitivity, luedtke2015statistics, diaz2018sensitivity} for similar bounding assumptions without covariates.
We characterize identified sets for novel estimands arising in predictive algorithm evaluation.

Instrumental variable strategies are widely used to address unobserved confounding in this setting. \citet{lakkaraju2017selective} and \citet{KLLLM(18)} evaluate specific performance measures by returning a single point in the identified set.
\cite{ArnoldDobbieHull(20)-discrimination_alg, ArnoldEtAl(24)} extrapolate across instrument values using identification-at-infinity arguments. We instead derive covariate-dependent bounds following \cite{Manski(94)-SampleSelectionBounds}, accommodating IVs as a special case while also encompassing other identifying strategies.\footnote{In related work, \cite{LevisEtAl(23)-CovariateAssistedIV} analyze covariate-assisted Balke-Pearl bounds on the average treatment effect.
\cite{semenova2023generalized} generalizes the \cite[][]{Lee(09)} bounds on the average treatment effect on the always takers, which requires a binary instrument that monotonically affects selection, to incorporate covariate information. Monotonicity is particularly unlikely to hold when the instrument arises from the random assignment of decision makers. See, for example, \citet[][]{FrandsenLefgrenLeslie(19), ArnoldDobbieHull(20)-discrimination_bail, ChanGentzkowYu(21)}.}

Our work relates to sensitivity analysis in causal inference, where researchers bound treatment effects under relaxations of unconfoundedness. 
Influential frameworks include the marginal sensitivity model \citep[e.g.,][]{Tan(06), KallusMaoZhou(18)-IntervalEstimation, ZhaoSmallBhattacharya(19), DornGuoKallus(21), KallusZhou(21), DornGuo(21)}, Rosenbaum's sensitivity model \citep[e.g.,][]{Rosenbaum(02), yadlowsky2018bounds}, and partial $c$-dependence \citep[e.g.,][]{MastenPoirier(18), MastenPoirier(20)}. 
These models bound how unobservables affect selection propensities—assumptions that may be difficult to formulate when decisions involve human judgment. We instead bound outcome differences, which may be more intuitive in applied settings. Nonetheless, these models imply specific bounding functions in our framework, as we discuss in Remark \ref{remark:causal}. 
\citet{chernozhukov2022long} develop omitted variable bounds for linear functionals; this covers our overall performance measures but not class-specific performance.

%%%%%%%%%%%%%%%%%%%%%%%%%%%%
% Identification Framework %
%%%%%%%%%%%%%%%%%%%%%%%%%%%%
\section{Setup and Identification Framework}\label{section: model set-up}

We consider data $(X_i, D_i, Y_i)$ for $i = 1, \hdots, n$ drawn i.i.d. from a joint distribution $\p(\cdot)$, where $X_i \in \mathcal{X} \subseteq \mathbb{R}^{d}$ are covariates and $D_i \in \{0, 1\}$ is the decision. 
There is a ``selective labels'' problem: the observed outcome satisfies $Y_i = D_i \, Y_i^*$ for some true outcome $Y_i^* \in \{0, 1\}$ \citep[e.g.,][]{KLLLM(18)}.
We assume $0 < \p(Y_i^* = 1) < 1$ so that we face a non-trivial prediction problem and that the decision satisfies strict overlap (i.e., there exists $\epsilon > 0$ such that $\p(D_i = 1 \mid X_i = x) \geq \epsilon$ with probability one). 

\vspace{-1em}
\paragraph{Example: lending decisions} We observe data on loan applications, where $X_i$ contains information reported on the application such as income and prior credit history, $D_i$ is whether the application was granted a loan, and $Y_i = D_i \, Y_i^*$ is whether the applicant defaulted if granted the loan.
We study credit scores that predict whether a new applicant would default $Y_i^* = 1$ \citep[e.g.,][]{BlattnerNelson(21), DiMaggioEtAl(22)-Fintech, FusterEtAl(22)predictably}. $\blacktriangle$

\vspace{-1em}
\paragraph{Example: pretrial release} We observe data on defendants, where $X_i$ contains information about the defendant such as their current charge and prior conviction history, $D_i$ is whether the defendant was released before trial, and $Y_i = D_i \, Y_i^*$ is whether the defendant committed pretrial misconduct if released. 
We study pretrial risk tools that predict whether a new defendant would commit pretrial misconduct $Y_i^* = 1$ \citep[e.g.,][]{KLLLM(18), ArnoldDobbieHull(20)-discrimination_alg, AngelovaEtAl(22)}. $\blacktriangle$

\vspace{-1em}
\paragraph{Example: medical testing} We observe data on patients, where $X_i$ contains patient information such as vital signs at admission and prior medical history, $D_i$ is whether a doctor decided to conduct a diagnostic test such as a cardiac stress test or lab test, and $Y_i = D_i \, Y_i^*$ is whether the patient was diagnosed with a heart attack or bacterial infection.
We study medical risk tools that predict whether a new patient suffered a heart attack or bacterial infection $Y_i^* = 1$ \citep[e.g.,][]{MullainathanObermeyer(21), HuangEtAl(22)}. $\blacktriangle$

\medskip

Let $\pi_d(x) := \p(D_i = d \mid X_i = x)$ for $d \in \{0, 1\}$ be the propensity score and $\mu_{1}(x) := \p(Y_i^* = 1 \mid D_i = 1, X_i = x)$.
Averages of a random variable $V_i$ are $\mathbb{E}_n[V_i] := n^{-1} \sum_{i=1}^{n} V_i$.
We let $\| \cdot \|$ denote the appropriate $L_2$-norm by context, where $\|f \| = \left( \int f(v)^2 d\p(v) \right)^{1/2}$ for a measurable function $f(\cdot)$ and $\|v\| = \left( \sum_{j=1}^{k} v_j^2 \right)^{1/2}$ for a vector $v \in \mathbb{R}^k$.

\subsection{Predictive Algorithms and Performance Measures}\label{section: estimands} 

A \textit{predictive algorithm} $s(\cdot) \colon \cX \rightarrow [0, 1]$ predicts the probability $Y_i^* = 1$ given covariates $X_i$, therefore estimating the \textit{conditional likelihood} $\mu^*(x) := \p(Y_i^* = 1 \mid X_i = x)$. 
Given a predictive algorithm, researchers evaluate its performance in two ways.

First, given a predictive algorithm, researchers construct decision rules $d(\cdot) \colon \cX \rightarrow \{0, 1\}$ by applying a threshold rule; that is, defining $d(X_i) = 1\{s(X_i) \geq \tau\}$ for some threshold $\tau \in [0, 1]$. 
Alternative decision rules are evaluated based on their \textit{expected payoff}
\begin{equation}\label{eqn: definiton of expected welfare}
U(d) := \E[ (-u_{1,1}(X_i) Y_i^* + u_{1,0}(X_i) (1 - Y_i^*)) d(X_i)  + (-u_{0,0}(X_i) (1-Y_i^*) + u_{0,1}(X_i) Y_i^*) (1 - d(X_i)],
\end{equation}
where each $u_{d,y^*}(\cdot) \geq 0$ for $d,y^* \in \{0,1\}^2$ specifies the known payoff associated with a combination of decision $D_i$ and outcome $Y_i^*$ at features $X_i$.
Payoffs are normalized so that $\sum_{d,y^* \in \{0,1\}^2} u_{d,y^*}(x) = 1$ for all $x \in \cX$.
In consumer lending, the profitability of approving customers that would not default $u_{1,0}(\cdot)$ may vary based on features like the requested loan size.
In pretrial release, the societal benefits of releasing defendants that would not commit misconduct $u_{1,0}(\cdot)$ may vary based on the defendant's age, charge severity, or prior history of pretrial misconduct \citep[e.g., ][]{Baughman(17)}.

Second, given a predictive algorithm $s(\cdot)$, researchers evaluate its performance over the full population or specific groups.
The target parameters are its \textit{overall performance}
\begin{equation}\label{eqn: overall performance}
\perf(s; \beta) = \e[\beta_0(X_i; s) + \beta_1(X_i; s) Y_i^*]
\end{equation}
and its \textit{positive class} and \textit{negative class performance}
\begin{equation}\label{eqn: class performance}
\perf_{+}(s; \beta) = \e[\beta_0(X_i; s) \mid Y_i^* = 1] \mbox{ and } \perf_{-}(s; \beta) = \e[\beta_0(X_i; s) \mid Y_i^* = 0],
\end{equation}
where $\beta_0(X_i; s), \beta_1(X_i; s)$ are researcher-specified functions that may depend on covariates and the predictive algorithm. 
Alternative choices of $\beta_0(X_i; s), \beta_1(X_i ; s)$ recover popular evaluations used in empirical research. 
We discuss several leading examples.

\vspace{-1em}
\paragraph{Example: mean square error and calibration} \textit{Mean square error} is $\E[\left( s(X_i) - Y_i^* \right)^2] = \perf(s; \beta)$ for $\beta_0(X_i; s) = s^2(X_i)$, $\beta_1(X_i; s) = 1 - 2s(X_i)$.
\textit{Calibration} at prediction bin $[r_1, r_2] \subseteq [0,1]$ is $\E[Y_i^* \mid r_1 \leq s(X_i) \leq r_2] = \perf(s; \beta)$ for $\beta_0(X_i; s) := 0$, $\beta_1(X_i; s) := \frac{1\{r_1 \leq s(X_i) \leq r_2\}}{\p\left(r_1 \leq s(X_i) \leq r_2 \right)}$ assuming $\p(r_1 \leq s(X_i) \leq r_2) > 0$. $\blacktriangle$

\vspace{-1em}
\paragraph{Example: generalized true positive and false positive rates}
Generalized \textit{true positive rate} is $\E[s(X_i) \mid Y_i^* = 1] = \perf_{+}(s; \beta)$ for $\beta_0(X_i; s) = s(X_i)$, and generalized \textit{false positive rate} is $\E[s(X_i) \mid Y_i^* = 0] = \perf_{-}(s; \beta)$ for $\beta_0(X_i; s) = s(X_i)$. $\blacktriangle$

\vspace{-1em}
\paragraph{Example: ROC curve} 
The \textit{true positive rate} or \textit{recall} at threshold $\tau$ is $\p\left( s(X_i) \geq \tau \mid Y_i^* = 1\right) = \perf_{+}(s; \beta)$ for $\beta_{0}(X_i; s) = 1\{s(X_i) \geq \tau\}$.
The \textit{false positive rate} at threshold $\tau$ is $\p\left(s(X_i) \geq \tau \mid Y_i^* = 0\right) = \perf_{-}(s; \beta)$.
The \textit{Receiver Operating Characteristic (ROC) curve} $\{( \p(s(X_i) \geq \tau \mid Y_i^* = 0 ), \p( s(X_i) \geq \tau \mid Y_i^* = 1 ) ) \colon \tau \in [0, 1] \}$ summarizes the predictive algorithm's ability to separate the positive class $Y_i^* = 1$ from the negative class $Y_i^* = 0$ as $\tau$ varies. $\blacktriangle$

\vspace{-1em}
\paragraph{Example: precision-recall curve} 
The \textit{precision} at threshold $\tau$ is $\p\left( Y_i^* = 1 \mid s(X_i) \geq \tau \right) := \perf(s; \beta)$ for $\beta_0(X_i; s) = 0$ and $\beta_1(X_i; s) = \frac{1\{ s(X_i) \geq \tau \}}{\p\left( s(X_i) \geq \tau \right)}$ assuming $\p(s(X_i) \geq \tau) > 0$.
The \textit{precision-recall curve} $\{ \left( \p\left( s(X_i) \geq \tau \mid Y_i^* = 1\right), \p\left( Y_i^* = 1 \mid s(X_i) \geq \tau \right) \right) \colon \tau \in [0, 1] \}$ summarizes the predictive algorithm's ability to accurately classify $Y_i^* = 1$ as $\tau$ varies. $\blacktriangle$

\medskip

\noindent Returning to the earlier examples, \cite{BlattnerNelson(21)}, \cite{FusterEtAl(22)predictably}, and \cite{DiMaggioEtAl(22)-Fintech} compare mean square error, ROC curves, and precision-recall curves of alternative predictive algorithms for default likelihood in consumer finance.
\cite{KLLLM(18)} and \cite{MullainathanObermeyer(21)} analyze predictive algorithms in pretrial release and medical diagnosis by reporting calibration at various prediction bins and ROC curves.\footnote{In analyzing pretrial risk assessments, \cite{lakkaraju2017selective} and \cite{AngelovaEtAl(22)} consider the ``failure rate'' or ``counterfactual misconduct rate'' $\p(Y_i^* = 1, s(X_i) \leq \tau)$ at threshold $\tau \in [0,1]$. This is recovered by the overall performance $\perf(s; \beta)$ for $\beta_0(X_i; s) = 0, \beta_1(X_i;s) = 1\{s(X_i) \leq \tau\}$.}
 
\subsection{Identification}\label{section: identification of estimands}

Selective labels create an identification problem: since $Y_i^*$ is only observed when $D_i = 1$, the conditional likelihood $\p(Y_i^* = 1 \mid D_i = 0, X_i)$ is not point identified from the data alone. 
Consequently, neither are the conditional likelihood, expected payoffs for any decision rule, nor predictive performance measures defined in Section \ref{section: estimands}.

A natural approach would be to assume unconfounded selection --- that is, $Y_i^* \indep D_i \mid X_i$ \citep[e.g.,][]{Coston(20)-Evaluation, mishler2021fade}. 
Under unconfoundedness, the conditional likelihood among unselected units equals that among selected units with the same covariates: $P(Y_i^* = 1 \mid D_i = 0, X_i) = P(Y_i^* = 1 \mid D_i = 1, X_i)$.
However, this assumption is often implausible in practice. 
Selected and unselected units may differ systematically in ways not captured by observed covariates $X_i$.
In consumer lending, applicants' decisions to accept loan terms may depend on competing offers from other lenders or their private assessment of their ability to repay --- information unavailable to the researcher.
In pretrial release, judges may rely on courtroom interactions or details from case files that are not recorded in administrative data.
Consequently, unobserved confounding --- unobservables correlated with both the outcome $Y_i^*$ and the decision $D_i$ --- is a central challenge.

Rather than imposing unconfoundedness, we partially identify the conditional likelihood, expected payoffs, and predictive performance measures by placing restrictions on how much outcomes can differ between unselected and selected units with the same covariates. 
These restrictions formalize researcher beliefs about the severity of unobserved confounding in a transparent way and nest popular empirical strategies as special cases.
The key to our approach is bounding the \textit{confounding function} $\delta(x) := \p(Y_i^* = 1 \mid D_i = 0, X_i = x) - \p(Y_i^* = 1 \mid D_i = 1, X_i = x)$, which summarizes the difference in outcomes between unselected and selected units conditional on covariates. 
When $\delta(x) = 0$ for all $x \in \mathcal{X}$, the selection decision is unconfounded. 
When $\delta(x) \neq 0$, unobserved confounding is present: units with the same observed covariates have systematically different outcomes depending on whether they were selected.

\begin{assumption}\label{asm: bounding assumption}
For researcher-specified bounding functions $\underline{\delta}(\cdot; \eta), \overline{\delta}(\cdot; \eta)$ with (possibly infinite-dimensional) nuisance parameters $\eta$, the confounding function satisfies 
\begin{equation}\label{eqn: bounding assumption}    
\underline{\delta}(x; \eta) \leq \delta(x) \leq \overline{\delta}(x; \eta) \mbox{ for all } x \in \mathcal{X}. 
\end{equation}
Let $\Delta$ denote the set of all confounding functions satisfying Equation \eqref{eqn: bounding assumption}.
\end{assumption}

\noindent Assumption \ref{asm: bounding assumption} restricts how much outcomes can differ between unselected and selected units with the same observed covariates. 
In consumer lending, this bounds the difference in default rates between unfunded and funded applicants conditional on application characteristics. 
In pretrial release, this bounds the difference in misconduct rates between detained and released defendants conditional on charge and criminal history.

Assumption \ref{asm: bounding assumption} allows researchers to encode beliefs about unobserved confounding without committing to a specific selection model. 
It nests two important special cases: unconfoundedness with $\underline{\delta}(x; \eta) = \overline{\delta}(x; \eta) = 0$ for all $x \in \cX$, and the assumption-free, worst-case bounds with $\underline{\delta}(x; \eta) = -\p(Y_i^* = 1 \mid D_i = 1, X_i = x)$, $\overline{\delta}(x; \eta) = 1 - \p(Y_i^* = 1 \mid D_i= 1, X_i = x)$ \citep[e.g.,][]{Manski(89), Manski(94)-SampleSelectionBounds}.
Thus Assumption \ref{asm: bounding assumption} interpolates between strong identifying assumptions and possibly uninformative bounds. 

More broadly, Assumption \ref{asm: bounding assumption} follows the structure of widely-used partial identification frameworks in econometrics, most notably \citet[][]{Manski(03)}'s "approximate mean independence" assumption.
It extends this by conditioning on covariates and allowing bounds to depend on identified nuisance parameters. 
In this sense, Assumption \ref{asm: bounding assumption} formalizes restrictions in terms that align with domain knowledge: researchers in these settings naturally reason about outcome differences rather than, say, how unobservables affect selection propensities. 
As shown in Section \ref{section: choice of bounding function}, the bounding functions can be chosen through direct calibration based on domain expertise or derived indirectly from additional data features such as proxy outcomes or instrumental variables.
Each strategy is instantiated by specifying both the functional form of the bounding functions and the collection of nuisance parameters $\eta$ (possibly infinite-dimensional) to be estimated.

Under Assumption \ref{asm: bounding assumption}, the conditional likelihood and evaluation estimands are partially identified. 
We assume throughout that the chosen bounding functions yield valid conditional probabilities. Specifically, for all $\delta \in \Delta$ and $x \in \mathcal{X}$, $\mu_1(x) + \pi_0 \delta(x; \eta) \in [0, 1]$ and $\delta(x; \eta) \in [-1, 1]$. 
This is a mild restriction in practice; as we discuss below, for common bounding functions, these constraints can be enforced through the researcher's specification of sensitivity parameters or are satisfied automatically, as we will discuss next. 
If these constraints are violated and the researcher proceeds with our identification framework, the resulting bounds will be conservative (non-sharp) but may still be informative.

Since $\mu^*(x) = \mu_1(x) + \pi_0(x) \delta(x)$, 
the \textit{identified set} is the set of all values consistent with the bounds on $\delta(x)$,
\begin{equation*}
    \mathcal{H}(\mu^{*}(x)) = \left\{ \mu_{1}(x) + \delta(x) \pi_0(x) \mbox{ such that } \delta(x) \in \Delta \right\}.
\end{equation*}
The identified sets for the performance measures, $\mathcal{H}(\perf(s; \beta))$ and $\mathcal{H}(\perf_{+}(s; \beta))$, and for the expected payoff of a decision rule $\mathcal{H}(U(d))$ are defined analogously.
As shorthand, let $\beta_{0,i} := \beta_0(X_i; s), \beta_{1,i} = \beta_{1}(X_i; s)$ and $\underline{\delta}_i := \underline{\delta}(X_i; \eta)$, $\overline{\delta}_i := \overline{\delta}_i(X_i; \eta)$.

\begin{lemma}\label{lem: bounds under MOSM}
For all $x \in \cX$, $\mathcal{H}(\mu^*(x)) = \left[\mulow(x), \muup(x) \right],$
where $\muup(x) = \mu_{1}(x) + \pi_0(x) \overline{\delta}(x; \eta) $, $\mulow(x) = \mu_1(x) + \pi_0(x) \underline{\delta}(x; \eta)$. 
Furthermore,
\begin{align*}
& \mathcal{H}(\perf(s; \beta)) = \left[ \underline{\perf}(s; \beta), \overline{\perf}(s; \beta) \right] \mbox{ and }\mathcal{H}(\perf_{+}(s; \beta)) = \left[ \underline{\perf}_{+}(s; \beta), \overline{\perf}_{+}(s; \beta) \right],
\end{align*}
where $\overline{\perf}(s; \beta) = \mathbb{E}[\beta_{0,i} + \beta_{1,i} \mu_{1}(X_i) + \beta_{1,i} \pi_0(X_i) \left( 1\{\beta_{1,i} > 0\} \overline{\delta}_{i} + 1\{\beta_{1,i} \leq 0\} \underline{\delta}_{i}  \right) ]$, $\underline{\perf}(s; \beta) = \mathbb{E}[\beta_{0,i} + \beta_{1,i} \mu_{1}(X_i) + \beta_{1,i} \pi_0(X_i) \left( 1\{\beta_{1,i} \leq 0\} \overline{\delta}_{i} + 1\{\beta_{1,i} > 0\} \underline{\delta}_{i}  \right) ]$, and
\begin{align*}
& \overline{\perf}_{+}(s; \beta) = \sup_{\delta \in \Delta} \mathbb{E}[\mu_{1}(X_i) + \pi_0(X_i) \delta(X_i)]^{-1} \mathbb{E}[ \beta_{0,i} \mu_{1}(X_i) +  \beta_{0,i} \pi_0(X_i) \delta(X_i) ], \\
& \underline{\perf}_{+}(s; \beta) = \inf_{\delta \in \Delta} \mathbb{E}[\mu_{1}(X_i) + \pi_0(X_i) \delta(X_i)]^{-1} \mathbb{E}[ \beta_{0,i} \mu_{1}(X_i) +  \beta_{0,i} \pi_0(X_i) \delta(X_i) ].
\end{align*}
% Negative class performance has an analogous characterization.
\end{lemma}

\begin{lemma}\label{lem: sharp bounds on ECU under MOSM}
For $d(\cdot) \colon \cX \rightarrow \{0, 1\}$ and $u_{d, y^*,i} = u_{d,y^*}(X_i)$, $\cH(U(d)) = [\underline{U}(d), \overline{U}(d)]$ for $\underline{U}(d), \overline{U}(d)$ defined in the proof.
\end{lemma}

The remainder of the paper applies this partial identification framework to the two objectives for predictive algorithms commonly seen in applied work. 
Section \ref{section: DR Learner, main text} addresses algorithm design: we estimate bounds on the conditional likelihood, thereby constructing predictive algorithms robust to unobserved confounding. 
Section \ref{section: robust evaluation, main text} addresses algorithm evaluation: we estimate bounds on performance measures for a given predictive algorithm.
Before proceeding to these estimation results, we next discuss how researchers might choose the bounding functions in Assumption \ref{asm: bounding assumption}.

\begin{remark}[Connection to Algorithmic Fairness]\label{remark: fairness}
Differences in predictive performance across sensitive attributes (e.g., race, gender) formalize violations of popular fairness criteria studied in the algorithmic fairness literature \citep[see][for reviews]{MitchellEtAl(19), barocas-hardt-narayanan}. 
Specifically, differences in positive and negative class performance across groups characterize violations of ``equality of opportunity'' \citep{hardt2016equality}, ``equalized odds'' \citep{chouldechova2017fair, KleinbergMullainathanRaghavan(17)}, and ``group-level discrimination'' \citep{ArnoldDobbieHull(20)-discrimination_alg, ArnoldDobbieHull(20)-discrimination_bail}. Differences in overall performance across groups characterize violations of ``bounded group loss'' \citep{AgarwalEtAl(19)-FairRegression}. In Online Appendix \ref{section: identification and estimation of predictive disparities}, we partially identify these performance disparities under Assumption \ref{asm: bounding assumption}, enabling robust fairness audits.
\end{remark}

\subsection{Choice of Bounding Functions}\label{section: choice of bounding function}

How should the bounding functions $\underline{\delta}(\cdot; \eta)$ and $\overline{\delta}(\cdot; \eta)$ be chosen?
The choice encodes assumptions about how much unselected units might differ from selected units with the same observed characteristics. In consumer lending: How much more likely are unfunded applicants to default than funded applicants with the same credit profile? In pretrial release: How much more likely are detained defendants to commit misconduct than released defendants with the same criminal history? 
We next show how alternative bounding functions formalize three common strategies for addressing selective labels in applied research.

\subsubsection{Observed Outcome Bounds}\label{section: observed outcome bounds}

A natural starting point is to assume that the unobserved outcome rate among unselected units is similar to the observed outcome rate among selected units with the same characteristics. 
For instance, in pretrial release, detained defendants with a given criminal history might be 1.5 times as likely to commit misconduct as released defendants with the same history. 
In consumer lending, unfunded applicants might be twice as likely to default as funded applicants with the same credit profile. 

\textit{Observed outcome bounds} formalize this intuition by specifying, for researcher-chosen constants $\underline{\Gamma}, \overline{\Gamma} > 0$ and all $x \in \mathcal{X}$,
\begin{equation}
    \underline{\delta}(x; \eta) := (\underline{\Gamma} - 1) \mu_1(x) \mbox{ and } \overline{\delta}(x; \eta) := (\overline{\Gamma} - 1) \mu_1(x).
\end{equation}
Under this choice, the unobserved conditional probability among unselected units satisfies $\underline{\Gamma} \mu_1(x) \leq P(Y_i^* = 1 \mid D_i = 0, X_i = x) \leq \overline{\Gamma} \mu_1(x)$ --- that is, it is bounded by constant multiples of the observed conditional probability $\mu_1(x)$. 
Setting $\underline{\Gamma} = \overline{\Gamma} = 1$ recovers unconfoundedness.
For observed outcome bounds, the probability constraints are satisfied for all choices of $\underline{\Gamma}, \overline{\Gamma}$ in a neighborhood of one under strict overlap on the decision and observed outcome probabilities being bounded away from zero and one.

This approach formalizes strategies commonly used in applied work. \citet[][]{KLLLM(18)} evaluate a constructed pretrial risk tool under varying assumptions on how much misconduct rates among detained defendants differ from released defendants (their Table 5). 
In consumer finance, researchers impute default rates for rejected applicants by applying multiplicative adjustments to observed default rates among accepted applicants \citep[e.g.,][]{HandHenley(93), ZengZhao(14)}. Observed outcome bounds flexibly encode such adjustments while allowing the imputed values to vary richly with covariates.

\begin{remark}[Connection to Sensitivity Analysis Models in Causal Inference]\label{remark:causal}
Several influential sensitivity analysis frameworks --- including the marginal sensitivity model \citep{Tan(06), ZhaoSmallBhattacharya(19), DornGuo(21)}, Rosenbaum's $\Gamma$-sensitivity model \citep{Rosenbaum(02), Yadlowsky(21)evaluating}, and partial $c$-dependence \citep{MastenPoirier(18), MastenPoirier(20)} --- place bounds on how unobservables affect selection propensities. 
While conceptually distinct from bounding outcome differences, these models imply observed outcome bounds with specific choices of $\underline{\Gamma}, \overline{\Gamma}$ (see Online Appendix \ref{section: connections to existing sensitivity analysis models}). 
The connection between outcome-based bounds and selection-based sensitivity models arises through Bayes' rule; though the observed outcome bounds we derive may not exhaust all implications of the original sensitivity model. 
\end{remark}

\subsubsection{Proxy Outcome Bounds}\label{section: proxy variable bounds}

Sometimes researchers observe a \textit{proxy outcome} $\widetilde{Y}_i \in \{0, 1\}$ for both selected and unselected units that is statistically related to the true outcome $Y_i^*$. 
This proxy can help discipline assumptions about unobserved confounding. 
For example, in consumer lending, whether an applicant defaulted on other credit products (e.g., credit cards) serves as a proxy for whether they would default on the loan of interest \citep[][]{BlattnerNelson(21)}. 
In medical testing, long-term health outcomes or total medical spending serves as a proxy for whether a patient suffered an acute condition \citep[][]{ObermeyerEtAl(19), ChanGentzkowYu(21), MullainathanObermeyer(21)}. 

The key assumption is that the proxy outcome is equally informative about the true outcome for both selected and unselected units. 
Formally, suppose the proxy outcome satisfies, for all $x \in \mathcal{X}$,
\begin{equation}\label{eqn: proxy assumption}
P(Y^*_i = \tilde{Y}_i \mid D_i = 0, X_i = x) = P(Y^*_i = \tilde{Y}_i \mid D_i = 1, X_i = x).
\end{equation} 
Conditional on the covariates, the probability that the true and proxy outcome agree is the same regardless of selection status. 
This assumption, combined with the observed distribution of $\widetilde{Y}_i$ among unselected units, pins down specific bounding functions on $\delta(x)$. 

Under mild restrictions satisfied by the proxy outcomes used in \cite{BlattnerNelson(21)} and \cite{MullainathanObermeyer(21)}, the bounding functions simplify to 
\begin{align*}
& \underline{\delta}(x; \eta) = 1 - \gamma_1(x) - \widetilde{\mu}_0(x) - \mu_1(x), \\
& \overline{\delta}(x; \eta) = 1 - \gamma_1(x) + \widetilde{\mu}_0(x) - \mu_1(x),
\end{align*}
where $\gamma_1(x) := \p(Y_i^* = \tilde{Y}_i \mid D_i = 1, X_i = x)$ is the probability the true and proxy outcome agree among selected units and $\widetilde{\mu}_0(x) := \p(\tilde{Y}_i = 1 \mid D_i = 0, X_i = x)$ is the observed proxy outcome rate among unselected units. 
These bounding functions satisfy the probability constraints under the maintained proxy assumptions under strict overlap on the decision and observed outcome probabilities being bounded away from zero and one. 
Online Appendix \ref{section: proxy outcome bounds} provides the general characterization without these simplifying restrictions. 
To our knowledge, proxy outcome assumptions of this form have not been explored in the causal inference sensitivity analysis literature.

\subsubsection{Instrumental Variable Bounds}\label{section: instrumental variable bounds}

Sometimes researchers observe an \textit{instrumental variable}
$Z_i \in \cZ$ with finite support that generates quasi-random variation in selection decisions but does not directly affect outcomes (i.e., $Y_i^*$ independent of $Z_i$ given $X_i$). 
For example, in pretrial release, judges randomly assigned to cases vary in their propensity to release defendants \citep[][]{KLLLM(18), ArnoldDobbieHull(20)-discrimination_bail, Rambachan(21), AngelovaEtAl(22)}. 
The day-of-week affects whether certain tests are ordered \citep[e.g.,][]{MullainathanObermeyer(21)}, and state-level variation in banking regulations affects loan approval rates \citep[e.g.,][]{BlattnerNelson(21)}. 

An instrument $Z_i$ implies specific bounding functions through a classic result of \citet[][]{Manski(94)-SampleSelectionBounds}. 
For any fixed instrument value $z \in \mathcal{Z}$, the confounding function satisfies, for all $x \in \mathcal{X}$, 
$$
    \underline{\delta}_{z}(x; \eta) \leq \delta(x) \leq \overline{\delta}_{z}(x; \eta),
$$
where $\underline{\delta}_{z}(x) = \left( \e[Y_i D_i \mid X_i = x, Z_i = z] - \mu_1(x) \right)/\pi_0(x)$ and $\overline{\delta}_z(x) := ( \pi_0(x, z) + \e[Y_i D_i \mid X_i = x, Z_i = z] - \mu_1(x) )/\pi_0(x)$. 
Our identification and estimation results apply to these bounds for any fixed $z \in \cZ$. 

Researchers can obtain sharper bounds by aggregating across all instrument values: $\max_{z \in \cZ} \ \underline{\delta}_{z}(x; \eta) \leq \delta(x) \leq \min_{z \in \cZ} \ \overline{\delta}_{z}(x; \eta)$.
These intersection bounds \citep[][]{Chernozhukov-IntersectionBounds} are tighter but involve maximum and minimum operators, creating non-smoothness that complicates estimation.
Following \citet{LevisEtAl(23)-CovariateAssistedIV}, we address this challenge using smooth approximations to the intersection bounds.
Online Appendix \ref{section: log-sum-exp bounds for IV} discusses both approaches: bounds based on a fixed instrument value and smooth approximations to intersection bounds.\footnote{\citet{LevisEtAl(23)-CovariateAssistedIV} and \citet{semenova2023adaptive} consider an alternative strategy for dealing with pointwise intersection bounds that invokes a ``margin condition'' on the separation of the maximal value of the intersection bounds. 
Under such a margin condition, existing results would apply to bounds on overall performance.
Extending this approach to estimators for the conditional likelihood bounds or positive class performance bounds is not immediate.
We leave the full analysis of margin conditions for future work.}

%%%%%%%%%%%%%%%%%
% Robust Design %
%%%%%%%%%%%%%%%%%
\section{Robust Design of Predictive Algorithms}\label{section: DR Learner, main text}

In this section, we propose estimators for the bounds on the conditional likelihood $\mu^*(\cdot)$ under Assumption \ref{asm: bounding assumption}.
We build on recent work studying the estimation of heterogeneous treatment effects in causal inference \citep[e.g.,][]{KunzelEtAl(19), NieWager(20), Kennedy(22)-towardsoptimal}, and our estimators use cross-fitting and doubly-robust bias corrections.
Since the bounds on the conditional likelihood are an infinite-dimensional parameter, we establish finite-sample bounds on our estimator's integrated mean square error (MSE) convergence using a novel oracle inequality for pseudo-outcome regression procedures.
These convergence guarantees enable us to construct plug-in decision rules with provable regret bounds.

To simplify exposition, we develop our estimator assuming the researcher specifies observed outcome bounds (Section \ref{section: observed outcome bounds}) and illustrate it with two folds. 
The analysis for multiple folds and cross-fitting is straightforward. 
We extend to proxy outcome bounds in Online Appendix \ref{section: proxy outcome bounds} and instrumental variable bounds in Online Appendix \ref{section: log-sum-exp bounds for IV}.

\vspace{-1em}
\paragraph{Estimation Procedure for Observed Outcome Bounds:}
Under Assumption \ref{asm: bounding assumption}, the unobserved conditional probability among unselected units can be decomposed as $P(Y^*_i = 1 | D_i = 0, X_i = x) = \mu_1(x) + \delta(x)$, where $\mu_1(x)$ is the observed rate among selected units and $\delta(x)$ measures confounding. 
Our estimation strategy constructs pseudo-outcomes based on influence functions that combine observed outcomes among selected units with extrapolation to unselected units.
The pseudo-outcomes correct for both observable selection (via inverse propensity weighting) and unobservable confounding (via the sensitivity parameters $\underline{\Gamma}, \overline{\Gamma}$).

We estimate the bounds on the conditional likelihood under observed outcome bounds with the following steps.
Split the data into two folds and estimate the nuisance functions $\etahat = \left( \muhat_1(\cdot), \pihat_1(\cdot) \right)$ on the first fold.
On the second fold, construct the pseudo-outcomes $\phi_{\mu,i}(\etahat) + \left( \overline{\Gamma} - 1 \right) \phi_{\pi \mu, i}(\etahat)$ for the upper bound and $\phi_{\mu,i}(\etahat) + \left( \underline{\Gamma} - 1 \right) \phi_{\pi \mu, i}(\etahat)$ for the lower bound, where $\phi_{\mu,i}(\eta) := \mu_1(X_i) + \frac{D_i}{\pi_1(X_i)} (Y_i - \mu_1(X_i))$ and $\phi_{\pi \mu, i}(\eta) := \left((1-D_i) - \pi_0(X_i)\right) \mu_1(X_i) + \frac{D_i}{\pi_1(X_i)} (Y_i - \mu_1(X_i)) \pi_0(X_i) + \pi_0(X_i) \mu_1(X_i)$ are the influence functions for $\e\left[ \mu_1(X_i)\right]$ and $\e\left[ \pi_0(X_i) \mu_1(X_i) \right]$ respectively.
These can be derived using standard arguments \citep[e.g.,][]{Hines_2022}.
We regress the estimated pseudo-outcomes on the covariates $X_i$ using a researcher-specified nonparametric regression procedure in the second fold. 
This yields our estimators $\widehat{\underline{\mu}}(\cdot)$, $\widehat{\overline{\mu}}(\cdot)$.

\subsection{Bound on Integrated Mean Square Error Convergence}\label{section: main text, L2 inequality}

We analyze the integrated MSE convergence rate of our proposed estimators by comparing them against an infeasible oracle nonparametric regression.
This oracle has access to the true nuisance functions $\eta$ and estimates the conditional likelihood bounds by regressing the true pseudo-outcomes $\phi_{\mu,i}(\eta) + \left( \overline{\Gamma} - 1 \right) \phi_{\pi \mu, i}(\eta)$ for the upper bound and $\phi_{\mu,i}(\eta) + \left( \underline{\Gamma} - 1 \right) \phi_{\pi \mu, i}(\eta)$ for the lower bound on $X_i$ in the second fold using the same nonparametric regression procedure specified by the researcher. 
We denote these oracle estimators by $\widehat{\underline{\mu}}_{oracle}(\cdot), \widehat{\overline{\mu}}_{oracle}(\cdot)$.

This oracle comparison is useful because it decomposes the integrated MSE of our proposed estimators into two sources.
The first is approximation error from using a finite-sample nonparametric regression procedure to estimate the infinite-dimensional conditional likelihood bounds --- this is captured by the oracle's integrated MSE. 
The second is estimation error from first-stage nuisance function estimation. 
Our next result shows that the latter only affects our proposed estimator through a smoothed, doubly-robust remainder term.

\begin{proposition}\label{prop: oracle result for learning, outcome regression bounds}
Let $\widehat{\E}_n[\cdot \mid X_i = x]$ denote the second-stage pseudo-outcome regression estimator.
Suppose $\widehat{\E}_n[\cdot \mid X_i = x]$ satisfies the $L_2(\p)$-stability condition (Assumption \ref{asm: L2 stability condition}), and $\p(\epsilon \leq \hat{\pi}_{1}(X_i) \leq 1 - \epsilon) = 1$ for some $\epsilon > 0$. 
Define $\tilde{R}(x) = \widehat{\E}_n[ (\pi_1(X_i) - \pihat_1(X_i)) (\outcomereg(X_i) - \outcomereghat(X_i)) \mid X_i = x ]$, and $R_{oracle}^2 = \E[ \| \widehat{\overline{\mu}}_{oracle}(\cdot) - \overline{\mu}^*(\cdot) \|^2]$. 
Then, 
\begin{align*}
& \| \widehat{\overline{\mu}}(\cdot) - \overline{\mu}^*(\cdot) \| \leq \|\widehat{\overline{\mu}}_{oracle}(\cdot) - \overline{\mu}^*(\cdot) \| + \epsilon^{-1} \sqrt{2} ( \overline{\Gamma} - 1 ) \| \tilde{R}(\cdot) \| + o_\p(R_{oracle})
\end{align*}
The analogous result holds for $\widehat{\underline{\mu}}(x)$.
\end{proposition}

\noindent Proposition \ref{prop: oracle result for learning, outcome regression bounds} delivers two insights.
First, our analysis is modular: researchers can pair choices of nuisance function estimators in the first step with any choice of nonparametric regression method (satisfying Assumption \ref{asm: L2 stability condition}) in the second step, and the integrated MSE is characterized by the oracle rate plus a doubly-robust remainder term. 
Second, Proposition \ref{prop: oracle result for learning, outcome regression bounds} 
shows that the cost of estimating the nuisance functions in the first step only enters through the product of nuisance function errors.

To prove this result, we establish an oracle inequality on the $L_2(\p)$-error of pseudo-outcome regression procedures (Lemma \ref{lemma: L2 oracle inequality}), extending \cite{Kennedy(22)-towardsoptimal}'s pointwise analysis. 
This result may be of independent interest. 
The $L_2(\p)$-stability condition (Assumption \ref{asm: L2 stability condition}) on $\widehat{\E}_n[\cdot \mid X_i = x]$ is quite mild in practice. 
It is satisfied by a variety of generic linear smoothers such as linear regression, series regression, nearest neighbor matching, random forest models, and several others (Proposition \ref{prop: linear smoothers satisfy L2 stability}).
Once again, in this sense, Proposition \ref{prop: oracle result for learning, outcome regression bounds} is agnostic to both the researcher's choice of ${\widehat{\E}_n[\cdot \mid X_i = x]}$ and nuisance function estimators.

Furthermore, Proposition \ref{prop: oracle result for learning, outcome regression bounds} can be applied in settings where nuisance functions satisfy additional smoothness or sparsity conditions, and for particular choices of the second-stage regression estimator and nuisance function estimators.
Known results on mean-squared error convergence rates of nonparametric regression procedures can then be used to characterize $\|\widehat{\overline{\mu}}_{oracle}(\cdot) - \overline{\mu}^*(\cdot) \|$ and $\| \tilde{R}(\cdot) \|$, yielding specific rates of convergence in terms of primitives such as underlying smoothness assumptions and sample size.
This follows in the spirit of recent work on optimal estimation of heterogeneous treatment effects \citep[][]{Kennedy(22)-towardsoptimal, KennedyEtAl(23)-Minimax}.

\subsection{Regret Bound for Plug-In Decision Rules}

Researchers often apply a threshold rule to an estimated predictive algorithm to make decisions. 
For example, \cite{KLLLM(18), Rambachan(21), AngelovaEtAl(22)} compare outcomes under alternative decisions rules that threshold a predictive algorithm in pretrial release.
We show that plug-in decision rules based on our estimators of the conditional likelihood bounds achieve low regret, ensuring that the statistical uncertainty from estimation does not lead to substantially worse decisions than if the true conditional probabilities were known.

Since the expected payoff $U(d)$ of any decision rule $d(\cdot) \colon \cX \rightarrow \{0, 1\}$ is partially identified (Lemma \ref{lem: sharp bounds on ECU under MOSM}), we evaluate decision rules by comparing their worst-case expected welfare \citep[e.g.,][]{Manski(07)}.
An immediate consequence of Lemma \ref{lem: sharp bounds on ECU under MOSM} is that the optimal max-min decision rule $d^*(\cdot) \in \arg \max_{d(\cdot) \colon \cX \rightarrow [0,1]} \underline{U}(d)$ is a threshold rule:
$$
d^*(X_i) = 1\{ \widetilde{\mu}^*(X_i) \leq u_{1,0,i} + u_{0,0,i}\},
$$
where $\widetilde{\mu}^*(x) = (u_{1,1,i} + u_{1,0,i}) \overline{\mu}^*(x) + (u_{0,0,i} + u_{0,1,i}) \underline{\mu}^*(x)$ is a welfare-weighted average of the conditional probability bounds. 
We construct a feasible plug-in decision rule $\widehat{d}(X_i)$ by substituting in our nonparametric estimators $\widehat{\overline{\mu}}(\cdot)$, $\widehat{\underline{\mu}}(\cdot)$. 

The regret of the plug-in decision rule is $R(\hat{d}) = \underline{U}(d^*) - \underline{U}(\widehat{d})$, which measures the welfare loss from using estimated rather than true conditional probabilities.
Our next result bounds squared regret by the integrated MSE of the oracle estimators plus the smoothed, doubly-robust remainder term.

\begin{proposition}\label{prop: regret bound for feasible plug-in decision rule}
Under the same conditions as Proposition \ref{prop: oracle result for learning, outcome regression bounds},
$$
R(\widehat{d})^2 \leq 2 \| \widehat{\overline{\mu}}_{oracle}(\cdot) - \overline{\mu}^*(\cdot) \| + 2 \| \widehat{\underline{\mu}}_{oracle}(\cdot) - \underline{\mu}^*(\cdot) \| + 4 \epsilon^{-1} \sqrt{2} \| \tilde{R}(x) \| + o_\p( R_{oracle} ).
$$
\end{proposition}

\noindent The proof proceeds in two steps. First, we show that the regret of the plug-in decision rule is bounded by the integrated MSE of our estimators for the conditional probability bounds --- this follows because the optimal decision rule is a threshold rule, and threshold rules are sensitive to estimation errors measured in $L_2(\p)$ norm. 
Second, we apply Proposition \ref{prop: oracle result for learning, outcome regression bounds} to decompose this integrated MSE into the oracle rate plus the doubly-robust remainder term. 
Consequently, worst-case regret converges to zero whenever the oracle's integrated MSE converges to zero. 
Plug-in decision rules based on our nonparametric estimators are therefore robust: they achieve nearly optimal welfare despite estimation uncertainty.

%%%%%%%%%%%%%%%%%%%%%
% Robust Evaluation %
%%%%%%%%%%%%%%%%%%%%%
\section{Robust Evaluation of Predictive Algorithms}\label{section: robust evaluation, main text}

In this section, we construct estimators for the bounds on the performance of a given predictive algorithm $s(\cdot)$ under Assumption \ref{asm: bounding assumption}.
This setting arises empirically in two scenarios: (i) when researchers evaluate an externally provided algorithm (e.g., a proprietary credit score or a commercial risk assessment tool), and (ii) when researchers conduct hold-out evaluation of an algorithm trained on separate data. 
In both cases, the algorithm $s(\cdot)$ is taken as fixed, and the goal is to assess its performance under varying assumptions about unobserved confounding.
Our estimators are consistent for the bounds on the identified set as sample size grows large, and we characterize their convergence rates in terms of errors in the first-step estimation of nuisance parameters.

\subsection{Bounds on Overall Performance}\label{section: estimating overall performance bounds}

We first construct estimators for the bounds on overall performance $\perf(s; \beta)$. 
To simplify exposition, we develop estimators for observed outcome bounds (Section \ref{section: observed outcome bounds}).
We extend to proxy outcome bounds and instrumental variable bounds in Online Appendices \ref{section: proxy outcome bounds}-\ref{section: log-sum-exp bounds for IV} respectively.
Under observed outcome bounds, Lemma \ref{lem: bounds under MOSM} implies that the upper bound can be written as 
$$
    \overline{\perf}(s; \beta) := \e\left[ \beta_{0,i} + \beta_{1,i} \mu_1(X_i) + \beta_{1,i} \left( 1\{ \beta_{1,i} > 0 \} \left( \overline{\Gamma} - 1 \right) + 1\{ \beta_{1,i} \leq 0 \} (\underline{\Gamma} - 1) \right) \pi_0(X_i) \mu_1(X_i) \right].
$$
The lower bound $\underline{\perf}(s; \beta)$ has an analogous expression. 
Both are linear functionals of known functions of the data and identified nuisance parameters $\eta = (\mu_1, \pi_0)$.
This linearity enables us to construct debiased estimators using standard arguments based on influence functions and cross-fitting \citep[e.g.,][]{ChernozhukovEtAl(18)-DML, ChernozhukovEtAl(22)-LocallRobust, Kennedy(22)-IFreview}.

\vspace{-1em}
\paragraph{Estimation Procedure for Observed Outcome Bounds:} For completeness, we sketch the construction of our estimators based on $K$-fold cross-fitting. 
We randomly split the data into $K$ disjoint folds.
For each fold $k$, we estimate the nuisance functions $\widehat{\eta}_{-k}$ using all observations not in the $k$-th fold and construct 
$$
\overline{\perf}_{i}(\etahat_{-k}) := \beta_{0,i} + \beta_{1,i} \phi_{\mu,i}(\etahat_{-k}) + \beta_{1,i} \left( 1\{ \beta_{1,i} > 0 \} \left( \overline{\Gamma} - 1 \right) + 1\{ \beta_{1,i} \leq 0 \} (\underline{\Gamma} - 1) \right) \phi_{\pi \mu, i}(\etahat_{-k})
$$
for each observation $i$ in the $k$-th fold, where $\phi_{\mu,i}(\eta)$ and $\phi_{\pi \mu, i}(\eta)$ are defined as before.
We then average over all observations $\widehat{\overline{\perf}}(s; \beta) := \e_n\left[ \overline{\perf}_i(\etahat_{-K_i}) \right],$
where $K_i$ is the fold containing observation $i$. 
Our estimator for the lower bound $\widehat{\underline{\perf}}(s;\beta)$ is defined analogously.

Our next result characterizes the convergence rate and asymptotic distribution of our estimators for the bounds on overall performance under observed outcome bounds.

\begin{proposition}\label{prop: overall performance estimator, nonparametric outcome bounds}
Define $R_{1,n}^{k} := \|\hat{\mu}_{1,-k}(\cdot) - \mu_{1}(\cdot)\| \|\hat{\pi}_{1,-k}(\cdot) - \pi_1(\cdot)\|$ for each fold $k$.
Assume (i) there exists $M < \infty$ such that $\| \beta_1(\cdot) \| \leq M$; (ii) $\p(\pi_1(X_i) \geq \delta) = 1$ for some $\delta > 0$, (iii) there exists $\epsilon > 0$ such that $\p(\hat{\pi}_{1,-k}(X_i) \geq \epsilon) = 1$ for each fold $k$, and (iv) $\|\hat{\mu}_{1, -k}(\cdot) - \mu_{1}(\cdot)\| = o_P(1)$ and $\|\hat{\pi}_{1, -k}(\cdot) - \pi_1(\cdot)\| = o_P(1)$ for each fold $k$.
Then, 
$$\left| \widehat{\overline{\perf}}(s; \beta) - \overline{\perf}(s; \beta) \right| = O_\p\left(1 / \sqrt{n} + \sumoverfolds R_{1,n}^{k}\right),$$ 
and the analogous result holds for $\widehat{\underline{\perf}}(s; \beta)$.
If further $R_{1,n}^{k} = o_\p(1/\sqrt{n})$ for each fold $k$, then 
\begin{equation*}
\sqrt{n} \left( \begin{pmatrix} \widehat{\overline{\perf}}(s; \beta) \\ \widehat{\underline{\perf}}(s; \beta)\end{pmatrix} - \begin{pmatrix} \overline{\perf}(s; \beta) \\ \underline{\perf}(s; \beta) \end{pmatrix} \right) \xrightarrow{d} N\left( 0, \Sigma \right)
\end{equation*}
for $\Sigma = Cov\left( (\overline{\perf}_i(\eta), \underline{\perf}_i(\eta))^\prime \right)$.
\end{proposition}

\noindent The estimators converge at rate $O_{\p}(1/\sqrt{n})$ plus a term that captures the nuisance estimation error. 
Nuisance errors enter multiplicatively --- through the product of propensity score and outcome regression errors --- rather than additively. 
The rate condition required for the estimators to be jointly asymptotically normal is satisfied under the standard condition that all nuisance function estimators converge at a rate faster than $O_\p(n^{-1/4})$.
This is the familiar condition from debiased machine learning, satisfied by a wide range of nonparametric regression and modern machine learning methods.

Online Appendix \ref{section: variance estimation for overall performance bounds} constructs a consistent estimator of the asymptotic covariance matrix in Proposition \ref{prop: overall performance estimator, nonparametric outcome bounds}.
This implies that confidence intervals for the identified set $\cH(\perf(s; \beta))$ can be constructed using standard methods \citep[e.g.,][]{ImbensManski(04), Stoye(09)--PICI}.

\subsection{Bounds on Positive Class Performance}\label{section: estimating pos class bounds}

We next estimate bounds on positive class performance $\perf_{+}(s; \beta)$. 
Unlike overall performance, this is a ratio of expectations, and the bounds are given by the optimal values of linear fractional programs as shown in Lemma \ref{lem: bounds under MOSM}.
This creates additional challenges in estimation.
In this section, we will develop estimators for these bounds and establish their convergence properties.

\subsubsection{Estimation via Linear Fractional Programming:}
We describe the construction of our estimators using sample-splitting across two folds; extending to cross-fitting with multiple folds is straightforward. 
We randomly split the data into two folds (we assume $n$ is divisible by $2$ and each fold contains $n/2$ observations for simplicity).
In the first fold, we estimate the nuisance functions $\etahat := (\widehat{\mu}_1(\cdot), \widehat{\pi}_1(\cdot))$ and the bounding functions $(\widehat{\underline{\delta}}(\cdot), \widehat{\overline{\delta}}(\cdot))$ using one of the strategies in Section \ref{subsec: estimating bounds}.
On the second fold, we construct $\phi_{\mu,i}(\etahat)$ and solve the sample analogue
\begin{equation}\label{equation: sample pos class LFP}
    \widehat{\overline{\perf}}_{+}(s; \beta) = \max_{\tilde{\delta} \in \widehat{\Delta}_n} \frac{\mathbb{E}_n\left[ \beta_{0,i} \phi_{\mu, i}(\etahat) + \beta_{0,i} \tilde{\delta}_i \right]}{\mathbb{E}_n\left[ \phi_{\mu, i}(\etahat) + \tilde{\delta}_i \right]},
\end{equation}
where $\widehat{\Delta}_n := \left\{ \tilde{\delta} \colon (1 - D_i) \widehat{\underline{\delta}}(X_i) \leq \tilde{\delta}_i \leq (1 - D_i) \widehat{\overline{\delta}}(X_i) \right\}$ and $\mathbb{E}_{n}[\cdot]$ denotes the sample average over the second fold.
The optimization problem in Equation \eqref{equation: sample pos class LFP} can be solved efficiently using standard techniques. 
It can be written as a linear program by applying the Charnes-Cooper transformation \citep[][]{CharnesCooper(62)}.
See Online Appendix \ref{section: LP reduction for positive class bounds} for details. 
The construction of this estimator involves two key ideas. 
First, we replace $\mu_1(\cdot)$ in the objective with its estimated influence function $\phi_{\mu,i}(\etahat)$. This orthogonalizes against first-stage estimation errors in nuisance functions that only enter the objective.
Second, we plug in estimated bounding functions into the constraint set. 

We provide an error decomposition for the positive class performance bounds. 
This result characterizes how errors in estimating the nuisance parameters $(\mu_1(\cdot), \pi_1(\cdot))$ and the bounding functions $(\underline{\delta}(\cdot), \overline{\delta}(\cdot))$ propagate to the final estimator. 

\begin{proposition}\label{proposition: partial double robustness for pos class estimator}
Define $R_{1,n} = \|\hat{\mu}_{1}(\cdot) - \mu_{1}(\cdot)\| \|\hat{\pi}_{1}(\cdot) - \pi_{1}(\cdot)\|$.
Assume (i) there exists $M < \infty$ such that $\| \beta_0(\cdot) \| \leq M$;
(ii) there exists $\delta > 0$ such that $\p(\pi_1(X_i) \geq \delta) = 1$; 
(iii) there exists $\epsilon > 0$ such that $\p(\hat{\pi}_{1}(X_i) \geq \epsilon) = 1$; 
(iv) $\|\hat{\mu}_{1}(\cdot) - \mu_{1}(\cdot)\| = o_P(1)$ and $\|\hat{\pi}_{1}(\cdot) - \pi_{1}(\cdot)\| = o_P(1)$; 
and (v) $\widehat{\Delta}_n$ is non-empty with probability one.
Then, 
\begin{align*}
\left\| \widehat{\overline{\perf}}_{+}(s; \beta) - \overline{\perf}_{+}(s; \beta) \right\| = O_\p\left( 1/\sqrt{n} + R_{1,n} + \sqrt{\mathbb{E}_{n}[(\widehat{\underline{\delta}}(X_i) - \underline{\delta}(X_i))^2]} + \sqrt{\mathbb{E}_{n}[(\widehat{\overline{\delta}}(X_i) - \overline{\delta}(X_i))^2]} \right).
\end{align*}
\end{proposition}

Proposition \ref{proposition: partial double robustness for pos class estimator} reveals the structure of how estimation errors affect our estimator.
The first term $O_{\p}(1/\sqrt{n})$ reflects sampling error that would arise even if the nuisance functions $(\mu_1(\cdot), \pi_1(\cdot))$ and the bounding functions $(\underline{\delta}(\cdot), \overline{\delta}(\cdot))$ were known. 
The second term $O_{\p}(R_{1,n})$ reflects errors from estimating the nuisances $(\mu_1(\cdot), \pi_1(\cdot))$ that appear in the numerator and denominator of the linear fractional program. 
These enter through a doubly robust product, and this behavior is inherited from the influence function structure underlying the numerator and denominator.
The final term captures errors from estimating the bounding functions $(\underline{\delta}(\cdot), \overline{\delta}(\cdot))$ that define the constraint set $\Delta$. 
This error enters through the root mean square error. 
This error can be controlled using standard nonparametric regression techniques, and researchers can trade off bias and variance in familiar ways.

For this reason, Proposition \ref{proposition: partial double robustness for pos class estimator} establishes a ``partial'' double robustness: 
given estimated constraints, our estimator is doubly robust to errors in the objective nuisances $(\mu_1(\cdot), \pi_1(\cdot))$.
Constraint estimation errors enter additively through RMSE.
Achieving full double robustness --- where constraint estimation errors would also enter with product structure --- remains an open problem, which we discuss in Remark \ref{remark: path to full double robustness}.
Remark \ref{remark: double robustness for nonsharp bounds} describes a simple alternative approach that obtains fully doubly-robust estimators for non-sharp bounds on positive class performance.

Proposition \ref{proposition: partial double robustness for pos class estimator} and our estimator are practically valuable.
Researchers can use state-of-the-art machine learning methods for estimating the objective nuisances $(\mu_1(\cdot), \pi_1(\cdot))$, while separately controlling the constraint estimation error through their choice of estimator for the bounding functions.
The error decomposition applies to \textit{any} estimator of the bounding functions with favorable $L_2(\p)$ properties, and we discuss alternative strategies and their tradeoffs in Section \ref{subsec: estimating bounds}.
This modularity is a key practical advantage of our framework.

\begin{remark}[Connection to existing work]\label{remark: connection to LFP literature}
Our analysis extends existing work on linear fractional programs, including \citet{AronowLee(2013)}, \citet{MiratrixWagerZubizaretta(18)}, \citet{KallusZhou(18)-confoundingrobust}, and \citet{ZhaoSmallBhattacharya(19)}.
These papers analyze linear fractional programs that arise when bounding population means or treatment effects under alternative sensitivity analysis models.
Our analysis advances in two key respects:
(i) we establish partial double robustness for objective nuisances -- a property not achieved in prior work; and
(ii) we provide a unified framework that accommodates multiple bounding strategies (observed outcome bounds, proxy outcome bounds, instrumental variable bounds) through a modular approach to constraint estimation.
\end{remark}

\begin{remark}[Paths toward full double robustness]\label{remark: path to full double robustness}
The proof of Proposition \ref{proposition: partial double robustness for pos class estimator} reveals that the optimizer in the linear fractional program can be characterized as having a threshold structure in $\beta_{0,i}$. 
% Formally, the feasible set $\Delta$ effectively reduces to monotone (non-decreasing) selectors in $\beta_{0,i}$, so the search is over monotone classification rules rather than arbitrary functions. 
The twist is that the cutoff determining the threshold is itself not fixed but rather a functional of the estimated constraints. 
Consequently, estimation error in the constraints affects not only which functions are feasible but also where the optimal threshold lies.
While related, recent work on conditional linear programs \citep[][]{benmichael2025partialidentificationconditionallinear} and aggregated intersection bounds \citep[][]{semenova2023adaptive} are not directly applicable.
While the linear fractional program can be reduced to a linear program through the Charnes-Cooper transformation, the transformation uses a global normalization that couples all observations together, whereas existing work exploits conditional separability across $X$-cells. We leave full resolution of this question for future work and view our partial double robustness result as an important first step.
\end{remark}

\begin{remark}[Estimation of non-sharp bounds]\label{remark: double robustness for nonsharp bounds}
Our estimator targets sharp bounds on positive class performance, but an alternative approach that restores double robustness at the cost of sharpness is possible. 
The idea is straightforward: separately estimate the numerator and the denominator using overall performance estimators (Section \ref{section: estimating overall performance bounds}) and then combine via the delta method. 
Concretely, define 
$$
\widetilde{\overline{\perf}}_{+}(s; \beta) = \frac{\sup_{\delta \in \Delta} \mathbb{E}[ \beta_{0,i} \mu_1(X_i) + \beta_{0,i}  \pi_0(X_i) \delta(X_i) ] }{\inf_{\delta \in \Delta} \mathbb{E}[\mu_1(X_i) + \pi_0(X_i) \delta(X_i)]}.
$$
Clearly, $\widetilde{\overline{\perf}}_{+}(s; \beta) \geq \overline{\perf}_{+}(s; \beta)$. The numerator and the denominator can be separately estimated using the methods in Section \ref{section: estimating overall performance bounds}. 
This delivers: (i) full double robustness with respect to all nuisances, and (ii) standard inference with straightforward variance estimation. 
The cost is that we target non-sharp bounds.
Whether this trade-off is attractive depends on the application.
\end{remark}

\begin{remark}[Interpretation of non-empty constraint sets]\label{remark: interpertation of non-empty constraint sets}
Proposition \ref{proposition: partial double robustness for pos class estimator} assumes the estimated constraint set $\widehat{\Delta}_n$ is non-empty.
If finite-sample variability leads to $\widehat{\underline{\delta}}(x) > \widehat{\overline{\delta}}(x)$ for some $x \in \mathcal{X}$, researchers can either:
(i) clip the bounds to enforce feasibility, which might introduce bias captured by the root mean square terms in Proposition \ref{proposition: partial double robustness for pos class estimator}; or (ii) restrict attention to the feasible covariate region $\tilde{X}_n = \{x : \widehat{\underline{\delta}}(x) \leq \widehat{\overline{\delta}}(x)\}$, thus redefining the estimand.
The latter approach is analogous to trimming observations with extreme estimated propensity scores in causal inference, though this may complicate analysis \citep[e.g.,][]{HotzEtAl(09), Sasaki_Ura_2022}. 
\end{remark}

\subsubsection{Estimation of the Bounding Functions}\label{subsec: estimating bounds}

The error decomposition in Proposition \ref{proposition: partial double robustness for pos class estimator} applies to any estimator of the bounding functions with favorable mean square error properties. This modularity gives researchers flexibility. 
We discuss two natural strategies and their relative merits, focusing on observed outcome bounds for concreteness.

The simplest approach directly substitutes nuisance estimates into the bounding function formulas. 
For observed outcome bounds, this means $\widehat{\underline{\delta}}(x) = (\underline{\Gamma} - 1) \widehat{\mu}_1(x)$ and $\overline{\delta}(x) = (\overline{\Gamma} - 1) \widehat{\mu}_1(x)$.
The mean square error properties of the estimated bounding functions are then inherited from those of $\widehat{\mu}_1(x)$. 
This approach requires no additional estimation beyond the first-stage nuisances.
An alternative approach is to construct influence-function-based pseudo-outcomes and regress them on covariates using the full sample (i.e., our pseudo-outcome regression procedure in Section \ref{section: DR Learner, main text}). 
We develop this approach below as it may offer bias-variance advantages in practice.
Ultimately, both strategies are valid under our framework. 

To understand why pseudo-outcome regression may be advantageous, consider that $\widehat{\mu}_1(x)$ is typically constructed using only observations with $D_i = 1$. 
Consequently, standard regression procedures target the mean square error conditional on $D_i = 1$
However, the error bound in Proposition \ref{proposition: partial double robustness for pos class estimator} depends on the \textit{unconditional} mean square error.
These two objectives could lead to different bias-variance tradeoffs when the selected population ($D_i=1$) is small relative to the full population or there is covariate shift between the selected and unselected populations. 
In such settings, an estimator that directly targets the unconditional MSE may achieve better performance. Pseudo-outcome regression accomplishes this by constructing unbiased pseudo-outcomes using observations from both the selected and unselected populations, then regressing these pseudo-outcomes on the full sample.

\vspace{-1em} 
\paragraph{Estimation Procedure for Observed Outcome Bounds:} 
To illustrate how Proposition \ref{proposition: partial double robustness for pos class estimator} may be applied, we estimate bounding functions using our pseudo-outcome regression procedure. This illustrates how our partial double robustness result (Proposition \ref{proposition: partial double robustness for pos class estimator}) can be combined with analyses of nonparametric regression estimators --- in particular, leveraging our oracle inequality for pseudo-outcome regression (Appendix \ref{section: oracle inequality}).
 
We now randomly split the data into three folds (we assume $n$ is divisible by $3$ and each fold contains $n/3$ observations for simplicity).
We construct nuisance function estimates $\etahat$ using the first fold. 
On the second fold, we construct the pseudo-outcomes $\phi_{\mu,i}(\etahat)$ and regress the estimated pseudo-outcomes on the covariates $X_i$ using a researcher-specified nonparametric regression procedure, yielding $\widehat{\delta}(\cdot)$.
On the third fold, we solve the sample analogue in Equation \eqref{equation: sample pos class LFP}.
Our analysis in Section \ref{section: DR Learner, main text} immediately allows us to analyze the errors from the estimated bounding functions.

\begin{corollary}\label{cor: DR learner rate for pos class}
Under the same conditions as Proposition \ref{proposition: partial double robustness for pos class estimator}, if further $\widehat{\E}_n[\cdot \mid X_i = x]$ satisfies the $L_2(\p)$-stability condition (Assumption \ref{asm: L2 stability condition}), then
\begin{align*}
& \sqrt{\mathbb{E}_{n}[(\widehat{\delta}(X_i) - \mu_1(X_i))^2]} = O_{\p}\left( 1/\sqrt{n} + \epsilon^{-1} \| \tilde{R}(\cdot) \| + R_{oracle} \right),
\end{align*}
where $\tilde{R}(x) = \widehat{\E}_n[ (\pi_1(X_i) - \pihat_1(X_i)) (\outcomereg(X_i) - \outcomereghat(X_i)) \mid X_i = x ]$, and now $R_{oracle}^2 = \E[ \| \widehat{\delta}_{oracle}(\cdot) - \mu_1(\cdot) \|^2]$ for the oracle pseudo-outcome regression procedure.
\end{corollary}

% To add as a footnote: When constraint functions are estimated parametrically (δ(x; θ) with finite-dimensional θ), the Zhao et al. (2019) percentile bootstrap approach for directionally differentiable parameters could potentially provide inference for the bounds. We leave this extension to future work.

%%%%%%%%%%%%%%%%%%%%%%%%%
% Empirical Application %
%%%%%%%%%%%%%%%%%%%%%%%%%
\section{Empirical Application to Credit Risk Scores}\label{section: empirical application, main text}

We validate the finite-sample performance of our estimators through Monte Carlo simulations in Appendix \ref{section: monte carlo simulations}. 
We now apply our framework to credit risk prediction, where financial institutions use scores $s(\cdot)$ to predict default risk. 
We observe data on past loan applications but only observe defaults $Y_i = D_i Y_i^*$ for funded applications ($D_i = 1$). 
It is implausible to assume unconfounded funding decisions. Applicants who accept loan offers may differ systematically from those who decline in ways that affect their default risk.\footnote{Lending institutions face two distinct inference problems. First, "policy rejects" with $\p(D_i = 1 \mid X_i) = 0$ (i.e., overlap violations) due to underwriting rules require extrapolation-based methods that are beyond our scope.  Second, even among fundable applicants with $\p(D_i = 1 \mid X_i) > 0$, those accepting versus declining offers may differ in unobserved ways --- the selective labels problem we address. Extending our framework to handle policy rejects is an important direction for future research.}

We analyze 372,346 personal loan applications submitted to Commonwealth Bank of Australia from July 2017 to July 2019. These loans are repaid in monthly installments and used to purchase vehicles, refinance debt, or cover major expenses \citep[][]{CostonEtAl(21)-Rashomon}. Loan amounts ranged up to AU\$50,000 (median AU\$10,000) with a median offered interest rate of 14.9\% per annum. We observe rich application-level covariates including reported income, occupation, and credit history at CommBank. We only observe whether an applicant defaulted within 5 months $Y_i^* \in \{0, 1\}$ if the application was funded $D_i = 1$. 
Approximately one-third of applications were funded, with a 2.0\% default rate among funded loans.

We consider two exercises. 
First, we construct credit risk scores robust to varying confounding assumptions. The resulting algorithms produce substantially different risk rankings that could affect credit access. 
Second, following recent work in consumer finance \citep[][]{BlattnerNelson(21), FusterEtAl(22)predictably, DiMaggioEtAl(22)-Fintech}, we evaluate how a benchmark score's accuracy varies across income groups. 
While an analysis ignoring confounding suggests better performance for higher-income applicants, this disparity disappears under plausible confounding levels.

\subsection{Bounding Individual Default Risk}\label{section: DR-Learner, CBA application}

We construct robust credit risk scores using our estimator from Section \ref{section: DR Learner, main text}, exploring how varying confounding assumptions affect individual default risk predictions.
We estimate the upper bound on the conditional default probability with observed outcome bounds, $\overline{\mu}(\cdot)$. 
We split applications into two folds: the first estimates $\pi_1(\cdot)$, $\mu_1(\cdot)$ using random forests; the second regresses estimated pseudo-outcomes on application-level covariates $X_i$ using cross-validated Lasso regression.
We set $\underline{\Gamma} = 1$ and vary $\overline{\Gamma} \in \{1, 2, 3\}$.

We compare our robust estimator (with $\underline{\Gamma} = 1, \overline{\Gamma} = 2$) against a benchmark score trained only on funded applications. 
Figure \ref{figure: DR-Learner, CBA application}'s left panel shows the joint distribution of their predictions. 
For each benchmark decile, the heatmap displays the percentage of applications falling into each decile of our robust predictions. Applications exhibit substantial reranking: among those in the benchmark's 5th decile, 18.7\% shift to extreme deciles (1-3 or 8-10) under the robust score. 
These differences meaningfully affect credit access. For instance, if only the least risky third of applications receive funding, 10.1\% would have their decision reversed when comparing benchmark versus robust scores.

%We compare our robust estimator's predictions of default risk against a benchmark credit score that predicts default risk among funded applications. 
%The left panel of Figure \ref{figure: DR-Learner, CBA application} summarizes the joint distribution of the benchmark credit score's predictions and our estimator's predictions, where our estimator is constructed assuming $\underline{\Gamma} = 1, \overline{\Gamma} = 2$.
%Among applications in each decile of the benchmark credit score, the left panel plots the percentage of applications at each decile of our estimator's predictions.
%Among applications in the 3rd decile and 8th decile of the benchmark risk score's predicted risk distribution, for example, 13.8\% of applications are in the top half of our estimator's predicted risk distribution and 4.7\% of applications are in the bottom half of our estimator's predicted risk distribution respectively. 
%Among applications in the 5th decile of the benchmark risk score, 18.7\% of applications are flipped to either the bottom (deciles 1-3) or top (deciles 8-10) of our estimator's predicted risk distribution. 
%Such differences meaningfully affect who would receive access to credit. 
%If, for example, only the bottom third of least risky applications are funded, then 10.1\% of all applications would have their decision flipped between the benchmark risk score and the robust risk score.

Figure \ref{figure: DR-Learner, CBA application}'s right panel examines how estimated coefficients vary with $\overline{\Gamma}$ (variable definitions in Online Appendix Table \ref{table: DR-Learner, CBA detailed variable description}). 
Confounding assumptions affect different covariates heterogeneously. 
Some coefficients—total net income and credit bureau score—remain zero across all $\overline{\Gamma}$ values. 
Others, like the number of recent credit card applications, receive zero weight in the benchmark but non-zero weight for all $\overline{\Gamma}$>1. 
Still others, such as occupation type and maximum recent delinquency, exhibit large magnitude changes as $\overline{\Gamma}$ varies, highlighting how confounding assumptions shape which applicant characteristics the algorithm prioritizes.

%The right panel of Figure \ref{figure: DR-Learner, CBA application} investigates these differences in predicted default risk by comparing how the estimated coefficients on a subset of characteristics vary as $\overline{\Gamma}$ varies.
%Online Appendix Table \ref{table: DR-Learner, CBA detailed variable description} provides a more detailed description of the variable names in the right panel.
%Altering assumptions on unobserved confounding leads to effectively no difference in the estimated coefficients for some notable characteristics. 
%The estimated coefficient on the application's total net income and credit bureau score do not vary as $\overline{\Gamma}$ varies and are always equal to zero.
%In contrast, for some characteristics such as the number of credit card applications submitted by all applicants on the application, the benchmark risk score places no weight them, whereas all estimators with $\overline{\Gamma} > 1$ incorporate them into the model and assign a non-zero weight.
%For other characteristics like the applicant's occupation type or their maximum delinquency over the last 12 months, varying our assumptions on selection lead to large changes in the magnitudes of the estimated coefficients. 

\subsection{Bounding the Predictive Performance of a Credit Risk Score}\label{section: robust audit, CBA application}

We now evaluate an existing credit risk score's performance. 
We split applications into training and evaluation data. 
On the training data, we construct a benchmark score predicting 5-month default risk among funded applications using random forests. 
On the evaluation data, we analyze mean square error and ROC curves, following recent mortgage lending studies \citep[][]{BlattnerNelson(21), FusterEtAl(22)predictably}. 
Using observed outcome bounds with $\underline{\Gamma} = 1$, we assess how the benchmark's estimated performance varies as we increase $\overline{\Gamma}$, allowing unfunded applicants to have progressively higher default rates than funded applicants with similar characteristics.

%We next robustly evaluate the predictive performance of an existing credit risk score. 
%We split our sample of personal loan applications into two subsets, which we will refer to as the training data and evaluation data.
%In the training data, we construct a benchmark credit risk score that predicts 5 month default risk among funded applications $\p(Y_i^* = 1 \mid D_i = 1, X_i = x)$ using a random forest.
%In the evaluation data, we analyze this benchmark credit risk score's mean square error and ROC curve, following recent empirical analyses of credit risk scores in mortgage lending in \cite{BlattnerNelson(21), FusterEtAl(22)predictably}.
%We specify observed outcome bounds, restricting the unobserved default rate among unfunded applications to be not too different than the observed 5 month default rate among funded applicants conditional on covariates. 
% We evaluate the benchmark credit risk score as we vary the choice of $\overline{\Gamma}$, fixing $\underline{\Gamma} = 1$ throughout. 

Figure \ref{figure: full population evaluation, CBA application}(a) shows how estimated MSE bounds vary with $\overline{\Gamma}$. When $\overline{\Gamma} = 1$ (the unconfounded case), both bounds equal the benchmark's MSE on funded applications (dashed red line; $0.143$). As $\overline{\Gamma}$ increases, allowing unfunded applicants higher default rates, bounds widen but remain informative. 
At $\overline{\Gamma} = 2$ --- permitting unfunded applicants to default at twice the rate of funded applicants with similar characteristics --- the estimated bounds are $[0.138, 0.182]$.
Figure \ref{figure: full population evaluation, CBA application}(b) examines how estimated ROC curves vary with $\overline{\Gamma}$. 
We summarize ROC curves using Area Under the Curve (AUC), computed via the trapezoidal rule. 
Under unconfoundedness ($\overline{\Gamma} = 1$), the benchmark's AUC is $0.637$. 
At $\overline{\Gamma} = 1.5$, AUC bounds are $[0.575, 0.705]$. 
As with MSE, bounds widen with $\overline{\Gamma}$ but remain informative at empirically plausible confounding levels.

We finally investigate the benchmark credit risk score's predictive disparities across income groups. 
We define $G_i \in \{0, 1\}$ to be whether an application is below or above the median personal income of all submitted applications.
We examine how the benchmark credit risk score's mean square error and ROC curve vary across these income groups, thereby studying whether it is a ``noisier'' signal of 5-month default risk for lower income applications, as in \cite{BlattnerNelson(21), FusterEtAl(22)predictably}.

Figure \ref{figure: evaluation by income group, CBA application}(a) compares the mean square error of the benchmark credit risk across income groups. 
Interestingly, for $\overline{\Gamma} = 1$, the mean square error of the benchmark credit risk score is significantly larger on applications below the median income (0.169) than those above it (0.136). 
% In other words, the benchmark credit risk score's mean square error is 24.2\% larger on funded applications below the median income than those above it.
This difference persists as we vary assumptions on unobserved confounding.
The upper bound on the mean square error for applications above the median income is only larger than the lower bound on the mean square error for applications below the median income for values $\overline{\Gamma} \geq 1.69$. 
Ruling out this mean square error disparity across income groups would require that unfunded applications are at least 1.69 times as likely to default within 5 months as funded applications conditional on covariates.\footnote{Rather than examining disparity bounds across a range of $\overline{\Gamma}$ values, researchers could compute the ``identification breakdown point'' $\overline{\Gamma}^*$: the minimal value of $\overline{\Gamma}$ (holding $\underline{\Gamma}=1$) such that $0 \in H(\text{disp}(s;\beta))$. 
A natural estimator is $\widehat{\overline{\Gamma}}^*$, the minimal $\overline{\Gamma}$ such that zero is contained in the estimated $(1-\alpha)$ confidence set $\widehat{H}(\text{disp}(s;\beta))$. Since our confidence sets have valid coverage by the results in Section \ref{section: estimating overall performance bounds} and Appendix \ref{section: identification and estimation of predictive disparities}, $P(\widehat{\overline{\Gamma}}^* \geq \overline{\Gamma}^*) \geq 1-\alpha$.} 
Figure \ref{figure: evaluation by income group, CBA application}(b) also compares the ROC curves of the benchmark credit risk score across income groups. 
Intriguingly, there exists effectively no disparity across income groups based on the ROC curves.
When $\overline{\Gamma} = 1$, the implied AUC for applications above the median income is $0.603$, whereas the implied AUC is $0.616$ for applications below the median income. 
Any disparity in terms of ROC curves is already ruled out at $\overline{\Gamma} = 1.25$ as the bounds on the AUC for applications above the median income are $[0.567, 0.653]$ and $[0.581, 0.654]$ for applications below the median income.

%%%%%%%%%%%%%%
% Conclusion %
%%%%%%%%%%%%%%
\section{Conclusion}

We developed a framework for robust design and evaluation of predictive algorithms under unobserved confounding. 
Our approach nests popular empirical strategies --- observed outcome bounds, proxy outcomes, and instrumental variables --- by bounding outcome differences between selected and unselected units (Assumption \ref{asm: bounding assumption}). 
We provide estimators that work well across bounding strategies and performance measures: conditional likelihoods for algorithm design, overall metrics like MSE and calibration, and class-specific measures like TPR and FPR. This breadth makes our framework valuable across domains where predictive algorithms face selective labels—pretrial release, consumer lending, medical diagnosis, hiring, and child welfare.

Several open questions present promising research directions. First, achieving full double robustness for positive class performance bounds remains open, though our technical results suggest possible paths forward. Second, uniform inference for breakdown frontiers would let researchers formally test how strong confounding must be to overturn conclusions \citep[][]{MastenPoirier(20)}. 
Third, we evaluate fixed algorithms; extending our results to settings where the same data is used to train and evaluate an algorithm is an important direction.
Finally, extending to multi-valued or continuous outcomes would broaden applicability, though defining class-specific performance in these settings requires care.

As algorithms increasingly drive high-stakes decisions in credit, criminal justice, and healthcare, robust evaluation under realistic confounding becomes essential. 
Our framework and these open questions offer pathways to strengthen algorithmic accountability where the stakes are highest.

%%%%%%%%%%%%%%
% References %
%%%%%%%%%%%%%%
{\singlespacing 
\bibliographystyle{chicago}
\bibliography{ref}}

%%%%%%%%%%%%%%%%%%%%%%
% Tables and Figures %
%%%%%%%%%%%%%%%%%%%%%%
\newpage \clearpage
\section*{Main Figures}

\begin{figure}[htbp]
\centering
\subfloat[Benchmark risk score vs. robust risk score.]{\includegraphics[width = 3in]{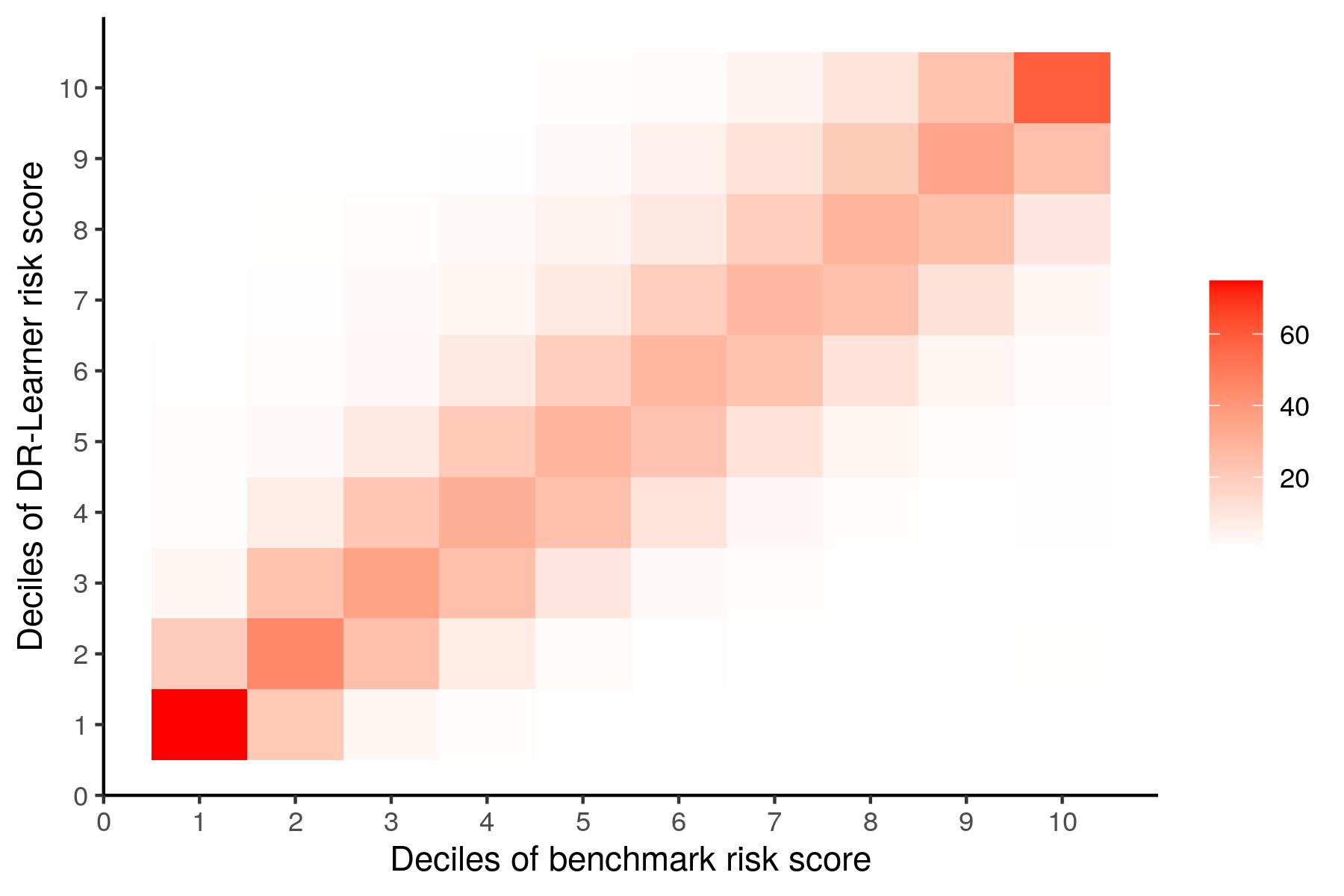}} \hspace{2.5mm}
\subfloat[Coefficients in risk score as $\overline{\Gamma}$ varies.]{\includegraphics[width=3.25in]{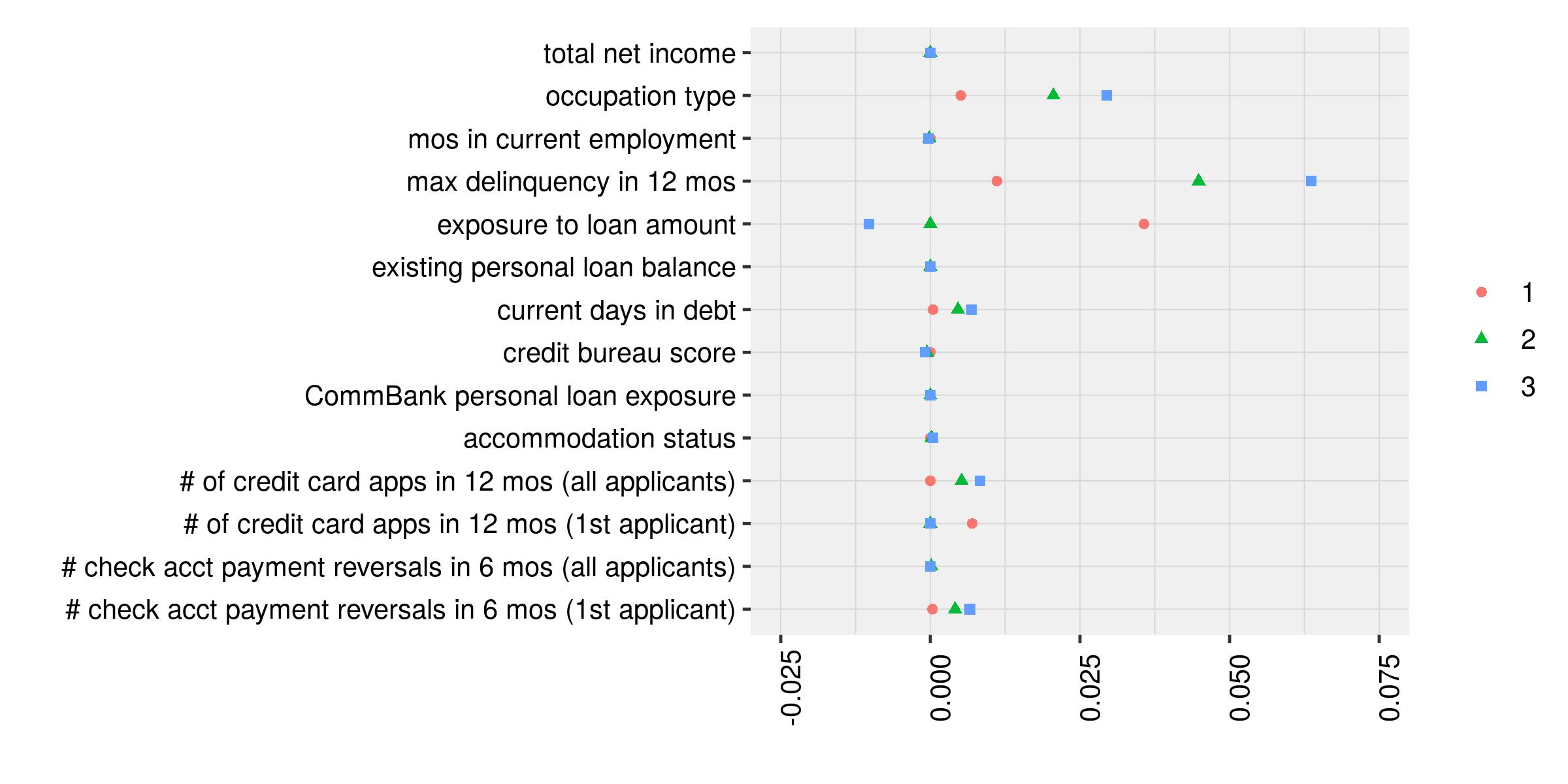}}
\caption{Estimated personal loan credit risk scores as assumptions on unobserved confounding vary.}
\floatfoot{\textit{Notes}: 
The left panel summarizes the joint distribution of a benchmark credit risk score's predictions of default risk against our robust predictions of the default risk. 
Among applications in each decile of the benchmark credit score's predicted risk distribution, the left panel plots the percentage of applications at each decile of our robust estimator's predicted risk distribution.
Our robust estimator is constructed assuming $\underline{\Gamma} = 1$, $\overline{\Gamma} = 2$, and the benchmark credit score predicts default risk among only funded applications.
The right panel summarizes how the coefficients on a subset of application-level characteristics vary with assumptions on unobserved confounding over $\overline{\Gamma} \in \{1, 2, 3\}$. 
The value $\overline{\Gamma} = 1$ corresponds to the benchmark credit score. See Section \ref{section: empirical application, main text} for further discussion.
}
\label{figure: DR-Learner, CBA application}
\end{figure}

\newpage
\begin{figure}[htbp!]
\centering
\subfloat[Mean square error]{\includegraphics[width=3in]{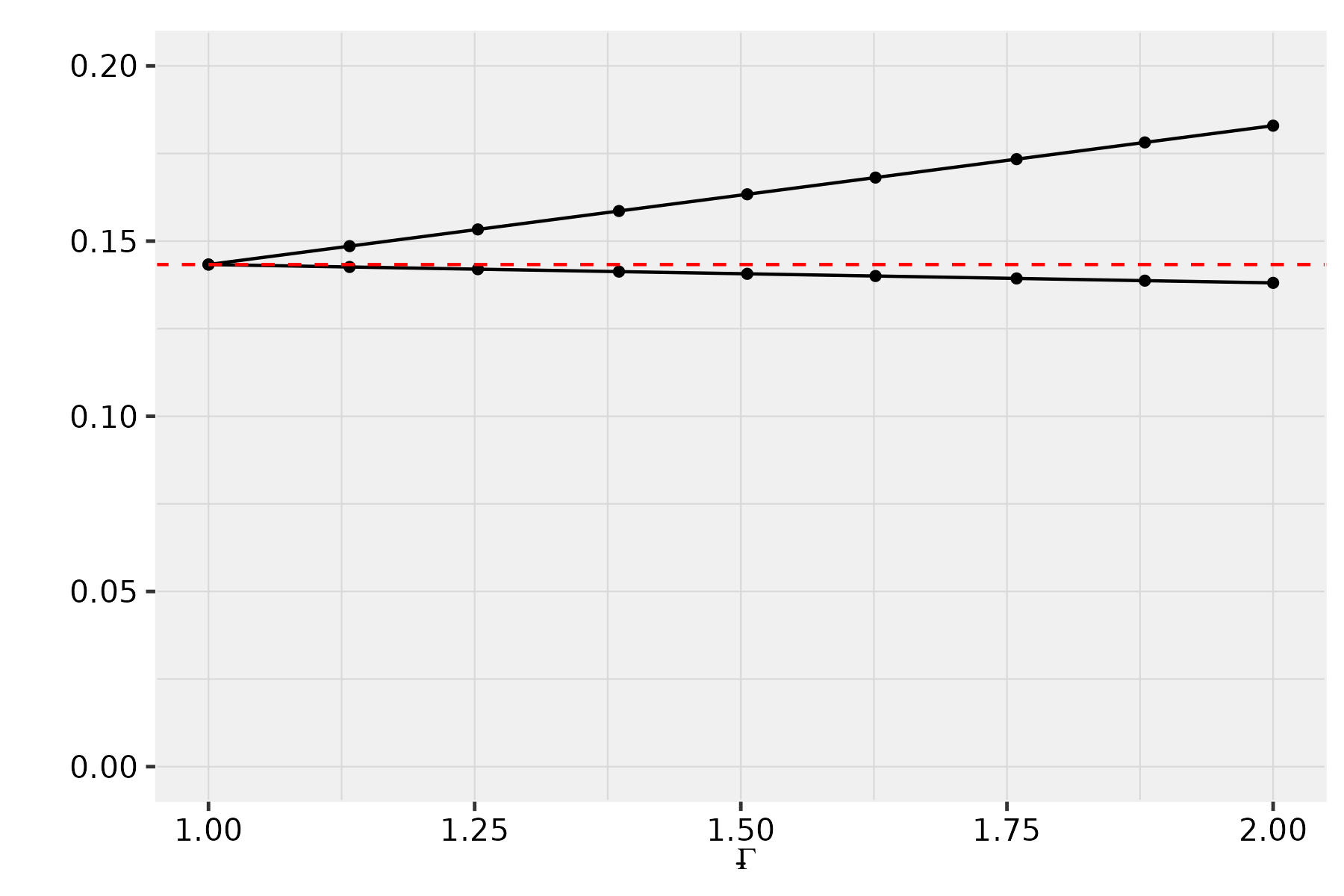}} \hspace{2.5mm}
\subfloat[ROC curve]{\includegraphics[width=3in]{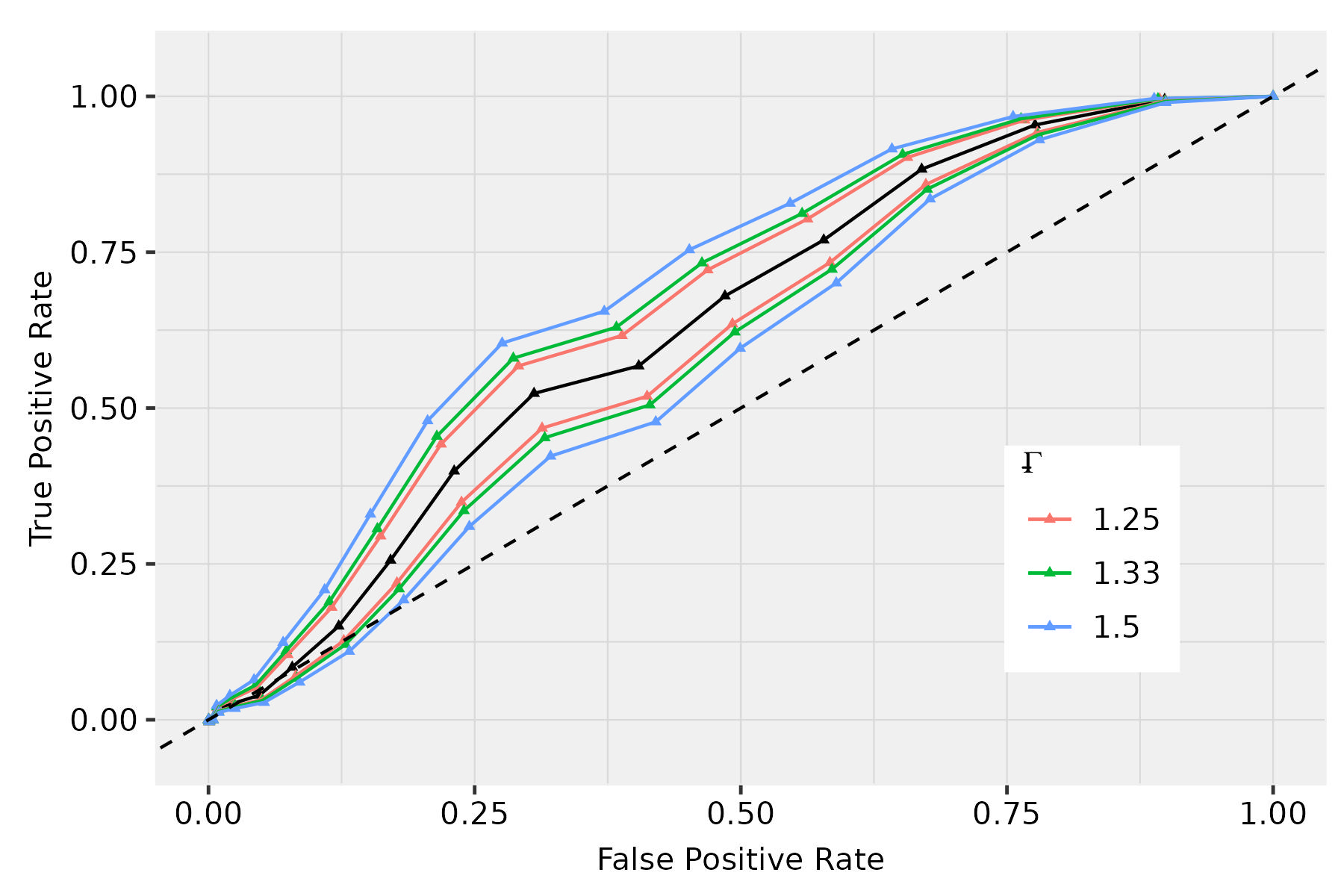}}
\caption{Bounds on mean square error and ROC curve of benchmark risk score as we vary the assumption on unobserved confounding.}
\floatfoot{\textit{Notes}: This figure summarizes how the bounds on the benchmark risk score's mean square error and ROC curve vary with our assumptions on unobserved confounding.
In Panel (A), the mean square error among only funded applications is plotted in the red dashed line. 
The bounds (black) are constructed using observed outcome bounds, varying $\overline{\Gamma} \in [1, 2]$ and setting $\underline{\Gamma} = 1$.
In Panel (B), the ROC curve among only funded applications is plotted in black.
The bounds on the ROC curve for alternative choices $\overline{\Gamma} \in \{1.25, 1.33, 1.5\}$ are plotted in different colors. See Section \ref{section: empirical application, main text} for further discussion.
}
\label{figure: full population evaluation, CBA application}
\end{figure}

\newpage
\begin{figure}[htbp]
\centering
\subfloat[Mean square error]{\includegraphics[width=3in]{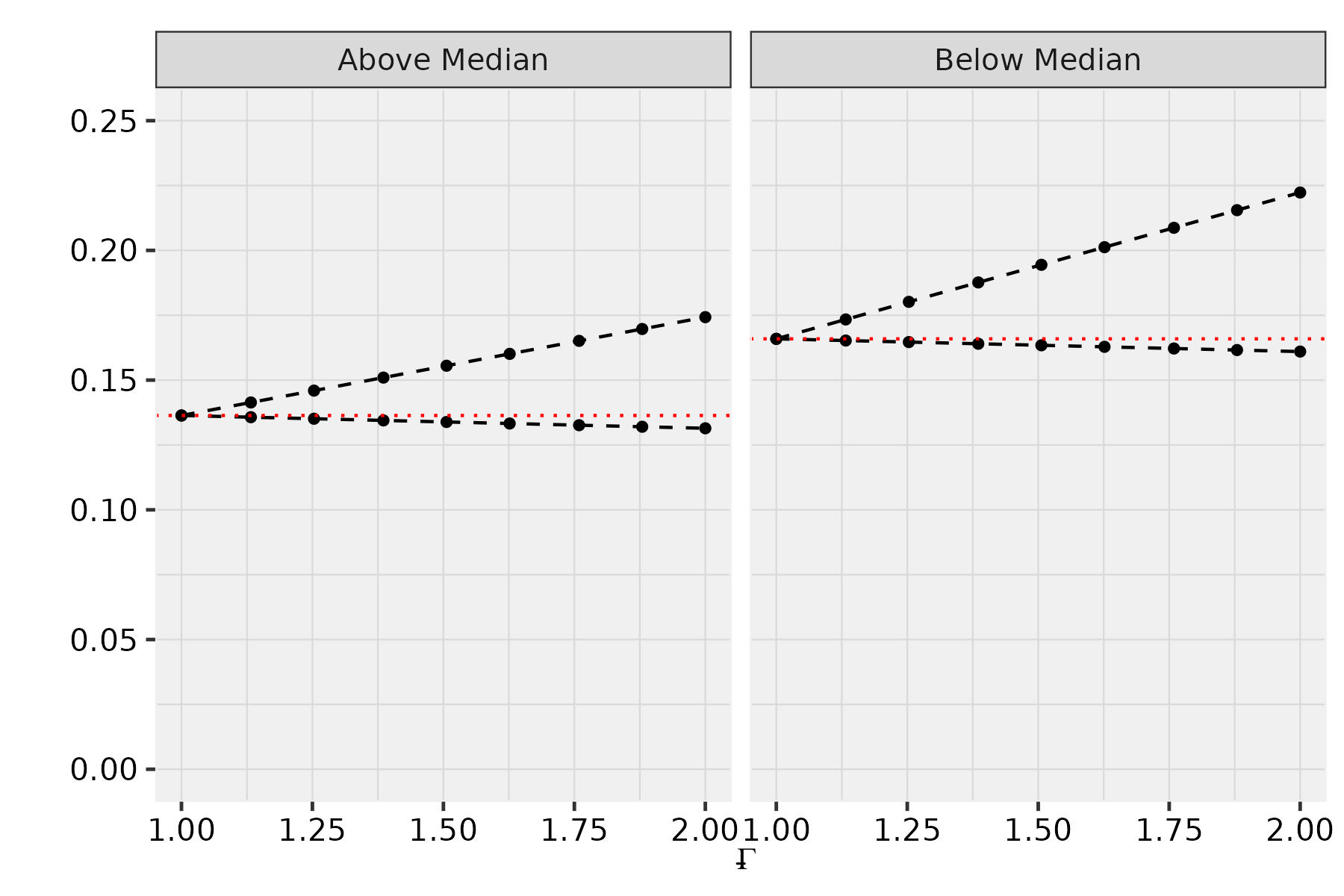}} \hspace{2.5mm}
\subfloat[ROC curve]{\includegraphics[width=3in]{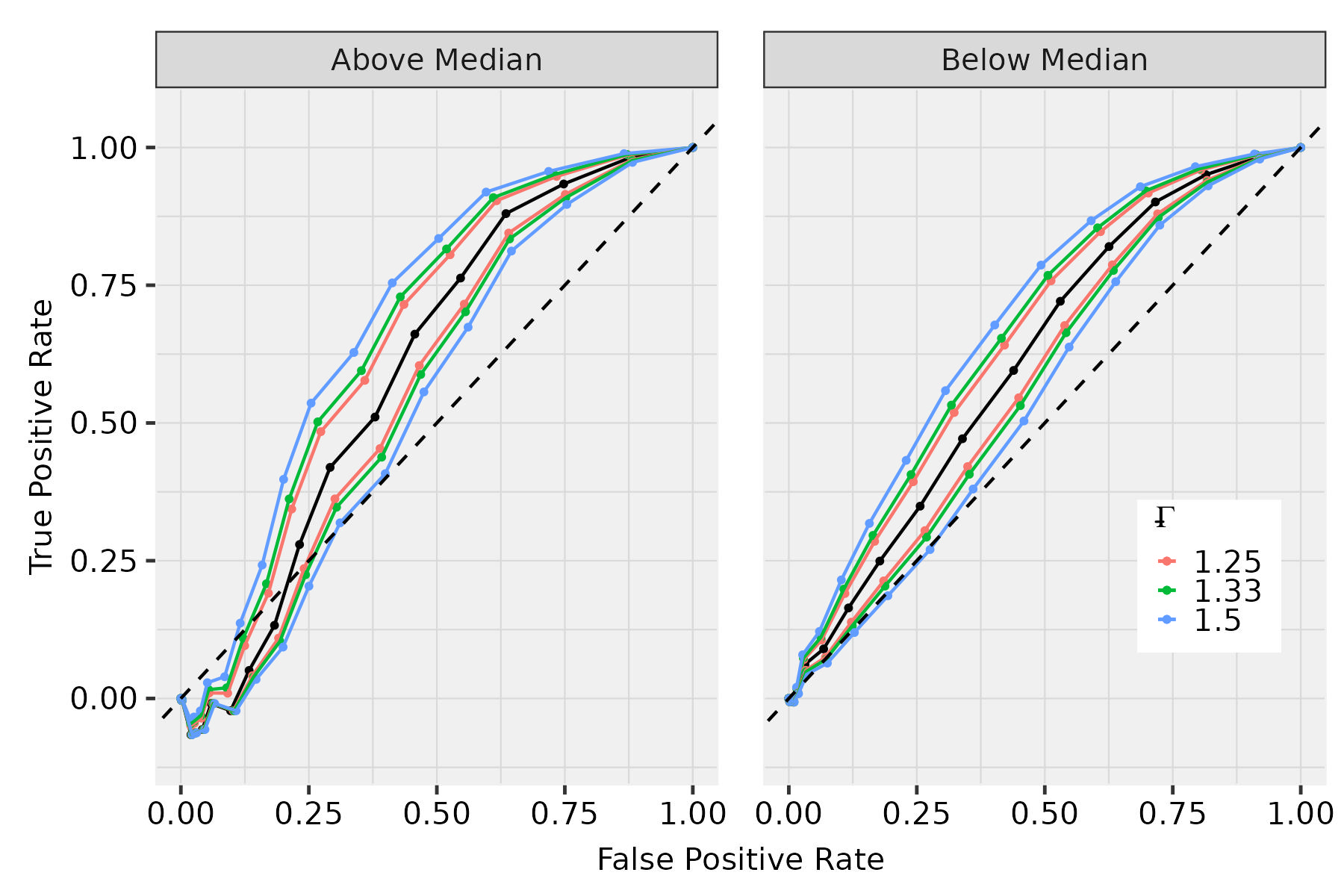}}
\caption{Bounds on mean square error and ROC curve of benchmark credit risk score across income groups as we vary the assumption on unobserved confounding.}
\floatfoot{\textit{Notes}: This figure summarizes how the bounds on the benchmark credit risk score's mean square error and ROC curve vary across income groups and as our assumptions on observed confounding vary. 
In Panel (A), the mean square among only funded applications is plotted in the red dashed line. The bounds (black) are constructed using observed outcome bounds, varying $\overline{\Gamma} \in [0, 1]$ and setting $\underline{\Gamma} = 1$.
In Panel (B), the ROC curve among only funded applications is plotted in black.
The bounds on the ROC curve for alternative choices $\overline{\Gamma} \in \{1.25, 1.33, 1.5\}$ are plotted in different colors. See Section \ref{section: empirical application, main text} for further discussion.
}
\label{figure: evaluation by income group, CBA application}
\end{figure}

%%%%%%%%%%%%%%%%%%%%%
% Online Supplement %
%%%%%%%%%%%%%%%%%%%%%
\clearpage
\newpage

\appendix
\pagenumbering{arabic} % resets `page` counter to 1
\renewcommand*{\thepage}{A-\arabic{page}}

% Table and figure numbering for appendix
\renewcommand{\thefigure}{A\arabic{figure}}
\renewcommand{\thetable}{A\arabic{table}}
\setcounter{figure}{0}
\setcounter{table}{0}

\begin{center}
\vspace{-1.8cm}{\large \textbf{Robust Design and Evaluation of Predictive Algorithms \\ under Unobserved Confounding} } \\
\large \textbf{Online Appendix} \\
Ashesh Rambachan \hspace{1em} Amanda Coston \hspace{1em} Edward H. Kennedy 
\end{center}

This online appendix contains additional results for ``Robust Design and Evaluation of Predictive Algorithms under Unobserved Confounding'' by Ashesh Rambachan, Amanda Coston, and Edward Kennedy. 
Section \ref{section: oracle inequality} contains an oracle inequality for the integrated mean square convergence rate of the pseudo-outcome regression procedures. 
Section \ref{section: main text proofs} contains the proofs of all theoretical results presented in the main text.
Section \ref{section: appendix, additional theoretical results} contains additional theoretical results discussed in the main text.
Section \ref{section: alternative bounding functions} extends our estimation results to alternative bounding functions, such as proxy outcome bounds and instrumental variable bounds.
Section \ref{section: connections to existing sensitivity analysis models} discusses connections between existing sensitivity analysis frameworks in causal inference and observed outcome bounds in our framework.
Section \ref{section: monte carlo simulations} contains Monte Carlo simulations, and Section \ref{section: additional Monte Carlo simulations and empirical application} contains additional empirical results.

%%%%%%%%%%%%%%%%%%%%%
% Oracle Inequality %
%%%%%%%%%%%%%%%%%%%%%
\section{Oracle Inequality for Pseudo-Outcome Regression}\label{section: oracle inequality}

We provide an oracle inequality on the $L_2(\p)$-error of nonparametric regression with estimated pseudo-outcomes. 
This generalizes the pointwise analysis in \cite{Kennedy(22)-towardsoptimal}. 
In Section \ref{section: DR Learner, main text} of the main text, we apply this oracle inequality to analyze our estimators for the conditional probability bounds.

We first state a $L_2(\p)$-stability condition on the second-stage nonparametric regression estimator. 
We then show that the $L_2(\p)$-stability condition is satisfied by a variety of generic linear smoothers such as linear regression, series regression, nearest neighbor matching, random forest model and several others.

\begin{assumption}\label{asm: L2 stability condition}
Suppose $\mathcal{O}_{train} = \left( V_{01}, \hdots, V_{0n} \right)$ and $\mathcal{O}_{test} = \left( V_{1}, \hdots, V_{n} \right)$ are independent train and test sets with covariates $X_i \subseteq V_{i}$.
Let (i) $\hat{f}(w) = \hat{f}(w; \mathcal{O}_{train})$ be an estimate of a function $f(w)$ using the training data; (ii) $\hat{b}(x) = \E[\hat{f}(V_i) - f(V_i) \mid X_i = x, \mathcal{O}_{train}]$ be the conditional bias of the estimator $\hat{f}$; and (iii) $\hat{\E}_n[V_i \mid X_i = x]$ be a generic nonparametric regression estimator that regresses outcomes $(V_1, \hdots, V_n)$ on covariates $(X_1, \hdots, X_n)$ in the test sample $\mathcal{O}_{test}$. 

The regression estimator $\hat{\E}_n[\cdot]$ is \textit{$L_2(\p)$-stable} with respect to distance metric $d(\cdot, \cdot)$ if 
\begin{equation}\label{eqn: stability}
\frac{\int \Big[ \widehat\E_n\{ \widehat{f}(V_i) \mid X_i=x\} - \widehat\E_n\{ {f}(V_i) \mid X_i=x\} - \widehat\E_n\{ \widehat{b}(X_i) \mid X_i = x \} \Big]^2 \ d\p(x)}{\E\left( \int \Big[ \widehat\E_n\{ f(V_i) \mid X_i = x\} - \E\{f(V_i) \mid X_i = x\} \Big]^2 d\p(x) \right)} \ \xrightarrow{p} 0
\end{equation}
whenever  
$ d(\widehat{f},f) \xrightarrow{p} 0 $.
\end{assumption}

\begin{proposition}\label{prop: linear smoothers satisfy L2 stability}
Linear smoothers of the form $\widehat\E_n\{\widehat{f}(V_i) \mid X_i=x\}=\sum_i w_i(x;X^n) \widehat{f}(V_i)$ are $L_2(\p)$-stable with respect to distance 
$$ d(\widehat{f},f)  = \| \widehat{f}-f\|_{w^2} \equiv \sum_{i=1}^n \left\{ \frac{ \| w_i(\cdot;X^n)\|^2}{\sum_j \| w_j(\cdot;X^n)\|^2} \right\} \int \left\{ \widehat{f}(v) - f(v) \right\}^2 \ d\p(v \mid X_i) , $$
whenever $1/\| \sigma \|_{w^2} = O_\p(1)$ for $\sigma(x)^2 = Var\{f(V_i) \mid X_i=x\}$. 

\begin{proof}
The proof follows an analogous argument as Theorem 1 of \cite{Kennedy(22)-towardsoptimal}.
Letting $T_n(x) = \widehat{m}(x) - \widetilde{m}(x) - \widehat\E_n\{\widehat{b}(X) \mid X=x\}$  denote the numerator of the left-hand side of \eqref{eqn: stability}, and $R_n^2= \E[\| \widetilde{m} - m \|]^2$ denote the oracle error, we will show that
$$ \| T_n \| = O_\p\left( \frac{\| \widehat{f}-f \|_{w^2}}{\|\sigma \|_{w^2}} R_n \right) $$
which yields the result when $1/\|\sigma\|_{w^2} = O_\p(1)$. 
First, note that for linear smoothers 
$$ T_n(x) = \widehat\E_n\{ \widehat{f}(V_i) - f(V_i) - \widehat{b}(X_i) \mid X_i=x\} = \sum_{i=1}^n w_i(x;X^n) \Big\{ \widehat{f}(V_i) - f(V_i) - \widehat{b}(X_i) \Big\}  $$
and this term has mean zero since
\begin{align*}
\E\Big\{ \widehat{f}(V_i) - f(V_i) - \widehat{b}(X_i) \mid \mathcal{O}_{train}, X^n \Big\} = \E\Big\{ \widehat{f}(V_i) - f(V_i) - \widehat{b}(X_i) \mid \mathcal{O}_{train}, X_i \Big\}  = 0
\end{align*}
by definition of $\widehat{b}$ and iterated expectation. Therefore,
 \begin{align}
 \E&(T_n(x)^2 \mid \mathcal{O}_{train}, X^n ) = Var\left[ \sum_{i=1}^n w_i(x;X^n)  \Big\{ \widehat{f}(V_i) - f(V_i) - \widehat{b}(X_i) \Big\} \Bigm| \mathcal{O}_{train}, X^n \right] \nonumber \\
& =  \sum_{i=1}^n w_i(x;X^n)^2 \ Var \Big\{ \widehat{f}(V_i) - f(V_i) \mid \mathcal{O}_{train}, X_i \Big\}   \label{eq:lsep1}
 \end{align}
 where the second line follows since $\widehat{f}(V_i) - f(V_i)$ are independent given the training data. Thus
 \begin{align*}
 \E\left( \| T_n \|^2 \Bigm| \mathcal{O}_{train}, X^n \right) 
 &= \int  \sum_{i=1}^n w_i(x;X^n)^2 \ Var\Big\{ \widehat{f}(V_i) - f(V_i) \mid \mathcal{O}_{train}, X_i \Big\}   \ d\p(x) \\
 &= \sum_{i=1}^n \| w_i(\cdot; X^n) \|^2 Var\Big\{ \widehat{f}(V_i) - f(V_i) \mid \mathcal{O}_{train}, X_i \Big\}   \\
 & \leq \sum_{i=1}^n \| w_i(\cdot; X^n) \|^2 \int \Big\{ \widehat{f}(v) - f(v) \Big\}^2 \ d\p(v \mid X_i)  \\
 &= \| \widehat{f}-f \|_{w^2} \sum_j \|w_j(\cdot; X^n)\|^2
 \end{align*}
 where the third line follows since $Var( \widehat{f} - f \mid \mathcal{O}_{train}, X_i ) \leq \E\{( \widehat{f}-f)^2 \mid \mathcal{O}_{train}, X_i \}$, and the fourth by definition of $\| \cdot \|_{w^2}$. 
 
Further note that $R_n^2$ equals
\begin{align}
\E&[\| \widetilde{m} - m \|]^2 =  \E\left( \int \left[ \sum_{i=1}^n w_i(x;X^n) \Big\{ f(V_i) - m(X_i) \Big\} + \sum_{i=1}^n w_i(x;X^n) m(X_i) -  m(x) \right]^2 d\p(x) \right) \nonumber \\
 &= \E\left( \int \left[ \sum_{i=1}^n w_i(x;X^n) \Big\{ f(V_i) - m(X_i) \Big\}  \right]^2  d\p(x) \right) + \E\left[ \int \left\{ \sum_{i=1}^n w_i(x;X^n) m(X_i) -  m(x) \right\}^2 d\p(x) \right] \nonumber \\
 &= \E \left\{ \int  \sum_{i=1}^n w_i(x;X^n)^2 \ \sigma(X_i)^2 \ d\p(x) \right\} + \E\left[ \int \left\{ \sum_{i=1}^n w_i(x;X^n) m(X_i) -  m(x) \right\}^2 d\p(x) \right] \nonumber  \\
 &\geq \E \sum_{i=1}^n \| w_i(\cdot; X^n) \|^2 \sigma(X_i)^2= \E\left\{  \| \sigma \|_{w^2}^2  \sum_j \| w_j(\cdot; X^n) \|^2 \right\} \label{eq:lsep2}
 \end{align}
where the second and third lines follow from iterated expectation and independence of the samples, and the fourth by definition of $\| \cdot \|_{w^2}$ (and since the integrated squared bias term from the previous line is non-negative).
Therefore,
\begin{align*}
\p \left\{   \frac{ \|\sigma\|_{w^2} \|T_n \| }{\| \widehat{f}-f \|_{w^2} R_n } \geq t  \right\} &= \E\left[ \p \left\{ \frac{ \|\sigma\|_{w^2} \|T_n \| }{\| \widehat{f}-f \|_{w^2} R_n } \geq t \Bigm| \mathcal{O}_{train}, X^n \right\} \right] \\
&\leq \left( \frac{1}{t^2 R_n^2} \right) \E\left\{  \| \sigma\|_{w^2}^2  \E\left( \frac{ \|T_n\|^2}{\| \widehat{f}-f \|_{w^2}^2  } \Bigm| \mathcal{O}_{train}, X^n  \right) \right\} \\
&\leq  \left( \frac{1}{t^2 R_n^2 } \right)  \E\left\{  \| \sigma \|_{w^2}^2 \sum_{i=1}^n \|w_i(\cdot;X^n)\|^2 \right\}  \leq \frac{1}{ t^2}
\end{align*}
where the second line follows by Markov's inequality, the third from the bound in \eqref{eq:lsep1} and iterated expectation, and the last from the bound in \eqref{eq:lsep2}. The result follows since we can always pick $t^2=1/\epsilon$ to ensure the above probability is no more than any $\epsilon$. 
\end{proof}
\end{proposition}

The $L_2(\p)$-stability condition and the consistency of $\hat{f}(\cdot)$ yields an inequality on the $L_2(\p)$-convergence of a feasible pseudo-outcome regression relative to an oracle that regresses the true unknown function $f(V_i)$ on $X_i$. 

\begin{lemma}\label{lemma: L2 oracle inequality}
Under the same setup from Assumptions \ref{asm: L2 stability condition}, define (i) $m(x) = \E[f(V_i) \mid X_i = x]$ the conditional expectation of $f(V_i)$ given $X_i$; (ii) $\hat{m}(x) := \hat{\E}_n[\hat{f}(V_i) \mid X_i = x]$ the regression of $\hat{f}(V_i)$ on $X_i$ in the test samples; (iii) $\tilde{m}(x) := \hat{\E}_n[f(V_i) \mid X_i = x]$ the oracle regression of $f(V_i)$ on $X_i$ in the test samples. 
Furthermore, let $\tilde{b}(x) := \hat{\E}_n[b(V_i) \mid X_i = x]$ be the $\hat{\E}_n$-smoothed bias and $R_n^2 = E[\| \widetilde{m} - m \|]^2$ be the oracle $L_2$-error.
If 
\begin{enumerate}
\item[i.] the regression estimator $\hat{\E}_n[\cdot]$ is $L_2(\p)$-stable with respect to distance metric $d(\cdot, \cdot)$; 
\item[ii.] $d(\hat{f}, f) \xrightarrow{p} 0$,
\end{enumerate}
then 
$$
    \| \hat{m} - \tilde{m} \| = \| \tilde{b}(\cdot) \| + o_\p(R_n).
$$
If further $\| \tilde{b} \| = o_\p\left(\sqrt{\E \| \widetilde{m} - m \|^2} \right)$, then $\widehat{m}$ is oracle efficient in the $L_2$-norm, i.e., asymptotically equivalent to the oracle estimator $\widetilde{m}$ in the sense that 
$$  
\frac{\| \widehat{m} - \widetilde{m} \| }{\sqrt{\E \|\widetilde{m} - m \|^2 }}  \ \xrightarrow{p} \ 0 
$$
and 
$$
\|\widehat{m} - m \| = \| \tilde{m} - m \| + o_\p(R_n).
$$

\begin{proof}
Note 
\begin{align*}
\| \widehat{m} - \widetilde{m} \| &\leq \| \widehat{m} - \widetilde{m} - \widetilde{b} \|  + \| \widetilde{b} \| = \| \widetilde{b} \| + o_\p\Big( \sqrt{ \E\| \widetilde{m} - m \|^2} \Big)
\end{align*}
where the first equality follows by the triangle inequality, and the second equality by the $L_2(\p)$-stability and $d(\cdot,\cdot)$-consistency of $\widehat{f}$.
\end{proof}
\end{lemma}

\noindent This generalizes Proposition 1 of \cite{Kennedy(22)-towardsoptimal}, which shows that a pointwise stability condition and consistency of $\hat{f}$ implies an oracle inequality on the pointwise convergence of a feasible pseudo-outcome regression. 
In Section \ref{section: DR Learner, main text} of the main text, we apply Lemma \ref{lemma: L2 oracle inequality} to analyze the convergence of our proposed estimators for the conditional probability bounds.

%%%%%%%%%%
% Proofs %
%%%%%%%%%%
\section{Proofs of Results in the Main Text}\label{section: main text proofs}

\subsection{Proof of Lemma \ref{lem: bounds under MOSM}}
The statements for $\cH(\mu^*(x))$ and $\cH(\perf(s;\beta))$ follow immediately since (i) both $\mu^*(x)$ and $\perf(s; \beta)$ are linear in $\delta(\cdot)$, and (ii) $\Delta$ is closed and convex.
To prove the statement for $\perf_{+}(s; \beta)$, define
$$
\perf_{+}(s; \beta, \delta) := \frac{\mathbb{E}[ \beta_{0,i} \mu_{1}(X_i) +  \beta_{0,i} \pi_0(X_i) \delta(X_i) ]}{\mathbb{E}[\mu_{1}(X_i) + \pi_0(X_i) \delta(X_i)]}.
$$
Observe that if $\widetilde{\perf}_{+}(s; \beta) \in \cH(\perf_{+}(s; \beta))$, then there exists some $\tilde{\delta} \in \Delta$ such that $\widetilde{\perf}_{+}(s; \beta) = \perf_{+}(s; \beta, \tilde \delta)$. It follows immediately that $\widetilde{\perf}_{+}(s; \beta) \in [\underline{\perf}_{+}(s; \beta), \overline{\perf}_{+}(s; \beta)]$. 
All that remains to show is that every value in the interval $[\underline{\perf}_{+}(s; \beta, \Delta), \overline{\perf}_{+}(s; \beta, \Delta)]$ is achieved by some $\delta(\cdot) \in \Delta$.

Towards this, we apply a change-of-variables. 
Let $U(\cdot) \colon \cX \rightarrow [0, 1]$ be defined as $U(x) = \frac{\delta(x) - \underline{\delta}(x)}{\overline{\delta}(x) - \underline{\delta}(x)}$. 
For any $\delta(\cdot) \in \Delta$, there exists $U(\cdot) \in [0, 1]$ such that $\perf_{+}(s; \beta, \delta) = \perf_{+}(s; \beta, U)$, where 
$$
\perf_{+}(s; \beta, U) := \frac{ \mathbb{E}[ \beta_{0,i} \mu_{1}(X_i) + \beta_{0,i} \pi_0(X_i) \underline{\delta}(X_i) + \beta_{0,i} \pi_0(X_i) (\overline{\delta}(X_i) - \underline{\delta}(X_i)) U(X_i) ]}{\mathbb{E}[\mu_{1}(X_i) + \pi_0(X_i) \underline{\delta}(X_i) + \pi_0(X_i) (\overline{\delta}(X_i) - \underline{\delta}(X_i)) U(X_i)]}. 
$$
Conversely, for any $U(\cdot) \in [0,1]$, there exists a corresponding $\delta(\cdot) \in \Delta$ such that $\perf_{+}(s; \beta, U) = \perf_{+}(s; \beta, \delta)$, where $\delta(x) = \underline{\delta}(x) + (\overline{\delta}(x) - \underline{\delta}(x)) U(x)$. 

Next, apply the Charnes-Cooper transformation with
$$
\tilde{V} = \frac{1}{\mathbb{E}[\mu_{1}(X_i) + \pi_0(X_i) \underline{\delta}(X_i) + \pi_0(X_i) (\overline{\delta}(X_i) - \underline{\delta}(X_i)) U(X_i) ]}
$$
$$
\tilde{U}(\cdot) = \frac{U(\cdot)}{\mathbb{E}[\mu_{1}(X_i) + \pi_0(X_i) \underline{\delta}(X_i) + \pi_0(X_i) (\overline{\delta}(X_i) - \underline{\delta}(X_i)) U(X_i) ]}.
$$
So, for any $U(\cdot) \in [0,1]$, there exists $\tilde{V}, \tilde{U}(\cdot)$ satisfying $\tilde{U}(\cdot) \in [0, \tilde{V}]$, $\tilde{V} \geq 0$ and $\mathbb{E}[\mu_{1}(X_i) + \pi_0(X_i) \underline{\delta}(X_i)] \tilde{V} + \mathbb{E}[\pi_0(X_i) (\overline{\delta}(X_i) - \underline{\delta}(X_i)) \tilde{U}(X_i)] = 1$ such that $\perf_{+}(s; \beta, U) = \perf_{+}(s; \beta, \tilde U, \tilde V)$, where
$$
\perf_{+}(s; \beta, \tilde{U}, \tilde{V}) = \mathbb{E}[\beta_0(X_i) \mu_{1}(X_i) + (1 - D_i) \beta_{0}(X_i) \underline{\delta}(X_i)] \tilde{V} + \mathbb{E}[\beta_0(X_i) \pi_0(X_i) (\overline{\delta}(X_i) - \underline{\delta}(X_i)) \tilde{U}(X_i)].
$$
Conversely, for any such $\tilde{U}(\cdot), \tilde{V}$, there exists $U(\cdot) \in [0,1]$ such that $\perf_{+}(s; \beta, \tilde U, \tilde V) = \perf_{+}(s; \beta, U)$.

Now consider any $\tilde{p} \in [ \underline{\perf}_{+}(s; \beta), \overline{\perf}_{+}(s; \beta) ]$, which satisfies for some $\lambda \in [0, 1]$
$$
\tilde{p} = \lambda \underline{\perf}_{+}(s; \beta) + (1 - \lambda) \overline{\perf}_{+}(s; \beta).
$$
Let $\underline{\delta}(\cdot), \overline{\delta}(\cdot)$ be the functions achieving the infimum and supremum respectively
$$
\underline{\delta}(\cdot) \in \arg \min_{\delta \in \Delta} \perf_{+}(s; \beta, \delta), \mbox{ } \overline{\delta}(\cdot) \in \arg \max_{\delta \in \Delta} \perf_{+}(s; \beta, \delta).
$$
By the change-of-variables, there exists $\tilde{\underline{V}}, \tilde{\underline{U}}(\cdot)$ and $\tilde{\overline{V}}, \tilde{\overline{U}}(\cdot)$ such that 
$$
\underline{\perf}_{+}(s; \beta) = \perf_{+}(s; \beta, \tilde{\underline{U}}(\cdot), \tilde{\underline{V}}), \mbox{ } \overline{\perf}_{+}(s; \beta) = \perf_{+}(s; \beta, \tilde{\overline{U}}(\cdot), \tilde{\overline{V}}).
$$
Therefore, $\tilde{p} = \lambda \perf_{+}(s; \beta, \tilde{\underline{U}}(\cdot), \tilde{\underline{V}}) + (1 - \lambda) \perf_{+}(s; \beta, \tilde{\overline{U}}(\cdot), \tilde{\overline{V}})$.
Since $\perf_{+}(s; \beta, \tilde U, \tilde V)$ is linear in $\tilde U, \tilde V$, we also have that 
$$
\tilde{p} = \perf_{+}(s; \beta, \lambda \tilde{\underline{U}} + (1 - \lambda) \tilde{\overline{U}}, \lambda \tilde{\underline{V}} + (1 - \lambda)\tilde{\overline{V}} ).
$$
We can therefore apply the change-of-variables in the other direction to construct the corresponding $\tilde{\delta}(\cdot) \in \Delta$, which satisfies $\tilde{p} = \perf_{+}(s; \beta, \tilde{\delta})$ by construction. $\Box$

\subsection{Proof of Lemma \ref{lem: sharp bounds on ECU under MOSM}}
Define the bounds 
\begin{align*}
& \underline{U}(d) := \E[ \left( u_{1,0,i} - (u_{1,1,i} + u_{1,0,i}) \overline{\mu}^*(X_i) \right) d(X_i) + \left( -u_{0,0,i} + (u_{0,0,i} + u_{0,1,i}) \underline{\mu}^*(X_i) \right) (1 - d(X_i))], \\
& \overline{U}(d) := \E[ \left( u_{1,0,i} - (u_{1,1,i} + u_{1,0,i}) \underline{\mu}^*(X_i) \right) d(X_i) + \left( -u_{0,0,i} + (u_{0,0,i} + u_{0,1,i}) \overline{\mu}^*(X_i) \right) (1 - d(X_i))].
\end{align*}
At each value $x \in \cX$, notice that if $d(x) = 1$, then 
$$
(-u_{1,1,i} \mu^*(x) + u_{1,0,i} (1 - \mu^*(x))) d(x) + (-u_{0,0,i} (1 - \mu^*(x)) + u_{0,1,i} \mu^*(x)) (1 - d(x)) = u_{1,0,i} - (u_{1,1,i} + u_{1,0,i}) \mu^*(x).
$$
This is minimized over $\mu^*(x) \in \cH(\mu^*(x))$ at $\mu^*(x) = \overline{\mu}^*(x)$. If $d(x) = 0$, then
$$
(-u_{1,1,i} \mu^*(x) + u_{1,0,i} (1 - \mu^*(x))) d(x) + (-u_{0,0,i} (1 - \mu^*(x)) + u_{0,1,i} \mu^*(x)) (1 - d(x)) = -u_{0,0,i} + (u_{0,0,i} + u_{0,1,i}) \mu^*(x).
$$
This is minimized over $\mu^*(x) \in \cH(\mu^*(x))$ at $\mu^*(x) = \underline{\mu}^*(x)$. The result for the lower bound immediately follows. The upper bound follows analogously. $\Box$

\subsection{Proof of Proposition \ref{prop: oracle result for learning, outcome regression bounds}}
We prove the result for our estimator of the upper bound, and the same argument applies for our estimator of the lower bound. We define $\widehat{\pi}_0(x) = 1- \widehat{\pi}_1(x)$ throughout. 
Let $\mathcal{O}_1$ denote the observations in the first fold and $\mathcal{O}_2$ denote the observations in the second fold.
We first observe that 
\begin{align*}
\| \widehat{\overline{\mu}}(\cdot; \Delta) - \overline{\mu}^*(\cdot; \Delta) \| & \leq \| \widehat{\overline{\mu}}(\cdot; \Delta) - \widehat{\overline{\mu}}_{oracle}(\cdot; \Delta) \| + \|\widehat{\overline{\mu}}_{oracle}(\cdot; \Delta) - \overline{\mu}^*(\cdot; \Delta)\| \\
& \leq \| \widehat{\overline{\mu}}(\cdot; \Delta) - \widehat{\overline{\mu}}_{oracle}(\cdot; \Delta) - \tilde{b}(\cdot) \| + \| \tilde{b}(\cdot) \| + \|\widehat{\overline{\mu}}_{oracle}(\cdot; \Delta) - \overline{\mu}^*(\cdot; \Delta)\|
\end{align*}
for $\tilde{b}(x) = \widehat{\E}_n[\hat{b}(X_i) \mid X_i = x]$ the smoothed bias and 
\begin{align*}
\hat{b}(x) = & \underbrace{\E[ \phi_{\mu,i}(\etahat) - \phi_{\mu,i}(\eta) \mid \mathcal{O}_{1}, X_i = x]}_{(a)} + (\overline{\Gamma}-1) \underbrace{\E[ \phi_{\pi \mu, i}(\etahat) - \phi_{\pi \mu,i}(\eta) \mid \mathcal{O}_1, X_i = x ]}_{(b)}.
\end{align*}
Under Assumption \ref{asm: L2 stability condition}, $\| \widehat{\overline{\mu}}(\cdot) - \widehat{\overline{\mu}}_{oracle}(\cdot) - \tilde{b}(\cdot) \| = o_\p(R_{oracle})$ by Lemma \ref{lemma: L2 oracle inequality}.
Furthermore, $\hat{b}(x)^2 \leq 2 (a)^2 + 2 (\bar{\Gamma} - 1)^2 (b)^2$, where
$$
(a)^2 = \left\{ \frac{\pi_1(x) - \pihat_1(x)}{\pihat_1(x)} (\outcomereg(x) - \outcomereghat(x)) \right\}^2 \leq \frac{1}{\epsilon^2} \left\{ (\pi_1(x) - \pihat_1(x)) (\outcomereg(x) - \outcomereghat(x)) \right\}^2,
$$
and 
$$
(b)^2 = \left\{ (\pi_0(x) - \pihat_0(x)) \outcomereghat(x) + \frac{\pi_1(x)}{\pihat_1(x)} \left( \outcomereg(x) - \outcomereghat(x) \right) \pihat_0(x) + \pihat_0(x) \outcomereghat(x) - \pi_0(x) \outcomereg(x) \right\}^2 = 
$$
$$
\left\{ (\pi_0(x) - \pihat_0(x)) \outcomereghat(x) + \frac{\pi_1(x)}{\pihat_1(x)} \left( \outcomereg(x) - \outcomereghat(x) \right) \pihat_0(x) + \pihat_0(x) ( \outcomereghat(x) - \outcomereg(x)) + \outcomereg(x) ( \pihat_0(x) - \pi_0(x)) \right\}^2 =  
$$
$$
\left\{ (\pi_0(x) - \pihat_0(x)) (\outcomereghat(x) - \outcomereg(x)) + \frac{\pihat_0(x)}{\pihat_1(x)} (\pi_1(x) - \pihat_1(x)) (\outcomereg(x) - \outcomereghat(x)) \right\}^2 =
$$
$$
\left\{ (\pi_1(x) - \pihat_1(x)) (\outcomereg(x) - \outcomereghat(x)) + \frac{\pihat_0(x)}{\pihat_1(x)} (\pi_1(x) - \pihat_1(x)) (\outcomereg(x) - \outcomereghat(x)) \right\}^2 \leq 
$$
$$
\frac{1}{\epsilon^2} \left\{ (\pi_1(x) - \pihat_1(x)) (\outcomereghat(x) - \outcomereg(x)) \right\}^2
$$
by iterated expectations and the assumption of bounded propensity score. 
Putting this together yields
$$
\| \widehat{\overline{\mu}}(\cdot) - \overline{\mu}^*(\cdot) \| \leq \|\widehat{\overline{\mu}}_{oracle}(\cdot) - \overline{\mu}^*(\cdot) \| + \sqrt{2} \epsilon^{-1} (\overline{\Gamma} - 1) \|\tilde{R}(\cdot)\| + o_\p(R_{oracle})
$$
as desired. $\Box$

\subsection{Proof of Proposition \ref{prop: regret bound for feasible plug-in decision rule}}

Recall from Lemma \ref{lem: sharp bounds on ECU under MOSM} that, for any decision rule $d(\cdot) \colon \cX \rightarrow \{0, 1\}$,
$$
\underline{U}(d) := \E[ \left( u_{1,0,i} - (u_{1,1,i} + u_{1,0,i}) \overline{\mu}^*(x) \right) d(X_i) + \left( -u_{0,0,i} + (u_{0,0,i} + u_{0,1,i}) \underline{\mu}^*(x) \right) (1 - d(X_i))] = 
$$
$$
\E[ -u_{0,0,i} + (u_{0,0,i} + u_{0,1,i}) \underline{\mu}^*(X_i) ] + \E[ \left( (u_{1,0,i} + u_{0,0,i}) - \tilde{\mu}^*(x) \right) d(X_i) ]
$$
for $\widetilde{\mu}^*(x) = (u_{1,1,i} + u_{1,0,i}) \overline{\mu}^*(x) + (u_{0,0,i} + u_{0,1,i}) \underline{\mu}^*(x)$.
Therefore, we can rewrite regret as 
$$
R(\hat{d}) = \underline{U}(d^*) - \underline{U}(\hat{d}) =
\E[ \left( c(X_i) - \tilde{\mu}^*(X_i) \right) (d^*(X_i) - \hat{d}(X_i) ],
$$
defining $c(X_i) := u_{1,0}(X_i) + u_{0,0}(X_i)$. 
It then follows that 
$$
R(\hat{d}) = \int_{x \in \cX} \left( c(x) - \tilde{\mu}^*(x) \right) \left( d^*(x) - \hat{d}(x) \right) dP(x) \leq 
\int_{x \in \cX} \left| \tilde{\mu}^*(x) - c(x) \right| 1\{ d^*(x) \neq \hat{d}(x) \} dP(x).
$$
At any fixed $X_i = x$, $\hat{d}(X_i) \neq d^*(X_i)$ implies that $| \tilde{\mu}^*(x) - \widehat{\tilde{\mu}}(x) | \geq | \tilde{\mu}^*(x) - c(x) |$.
Combining this with the previous display implies
$
R(\hat{d}) \leq \int_{x \in \cX} | \tilde{\mu}^*(x) - \widehat{\tilde{\mu}}(x) | dP(x).
$
Substituting in the definition of $\tilde{\mu}^*(x)$ and $\widehat{\tilde{\mu}}(x)$, we have
{\small
$$
| \tilde{\mu}^*(x) - \widehat{\tilde{\mu}}(x) | = 
$$
$$
| (u_{1,1}(x) + u_{1,0}(x)) \overline{\mu}^*(x) + (u_{0,0}(x) + u_{0,1}(x)) \underline{\mu}^*(x) - (u_{1,1}(x) + u_{1,0}(x)) \widehat{\overline{\mu}}(x) - (u_{0,0}(x) + u_{0,1}(x)) \widehat{\underline{\mu}}(x) | \leq
$$
$$
| \overline{\mu}^*(x) - \widehat{\overline{\mu}}(x)| + |\underline{\mu}^*(x) - \widehat{\underline{\mu}}(x)|
$$
}
by the triangle inequality and using $u_{0,0}(x), u_{0,1}(x), u_{1,0}(x), u_{1,1}(x)$ are non-negative and sum to one.
Substituting back into the bound on $R(\hat{d})$ delivers
$$
R(\hat{d}) \leq \int_{x \in \cX} | \overline{\mu}^*(x) - \widehat{\overline{\mu}}(x)| dP(x) + \int_{x \in \cX} | \underline{\mu}^*(x) - \widehat{\underline{\mu}}(x) | dP(x) =
\|  \overline{\mu}^*(x) - \widehat{\overline{\mu}}(x) \|_1 + \| \underline{\mu}^*(x) - \widehat{\underline{\mu}}(x) \|_1.
$$
Using the Cauchy-Schwarz inequality $\|  \overline{\mu}^*(x) - \widehat{\overline{\mu}}(x) \|_1^2 \leq \|  \overline{\mu}^*(x) - \widehat{\overline{\mu}}(x) \|_2^2$ and $\| \underline{\mu}^*(x) - \widehat{\underline{\mu}}(x) \|_1^2 \leq \| \underline{\mu}^*(x) - \widehat{\underline{\mu}}(x) \|_2^2$ and the inequality $(a+b)^2 \leq 2(a^2 + b^2)$,
$$
R(\hat{d})^2 \leq 2 \|  \overline{\mu}^*(x) - \widehat{\overline{\mu}}(x) \|_2^2 + 2 \| \underline{\mu}^*(x) - \widehat{\underline{\mu}}(x) \|_2^2.
$$
The result then follows by applying Proposition \ref{prop: oracle result for learning, outcome regression bounds}.
$\Box$

\subsection{Proof of Proposition \ref{prop: overall performance estimator, nonparametric outcome bounds}}

To prove Proposition \ref{prop: overall performance estimator, nonparametric outcome bounds}, we first state and prove various auxiliary lemmas.
Let $\mathcal{O}_{k}$ denote the observations in the $k$-th fold and $\mathcal{O}_{-k}$ denote the observations not in the $k$-th fold.

\begin{lemma}\label{lemma: mishler et al 2021}
Let $\beta(\cdot)$ be some function of $X_i$ such that $\|\beta(\cdot)\| \leq M$ for some $M < \infty$ and define $R_{1,n}^{k} = \|\hat{\mu}_{1,-k}(\cdot) - \outcomereg(\cdot)\| \|\pihat_{1,-k}(\cdot) - \pi_1(\cdot)\|$. 
Assume that there exists $\epsilon > 0$ s.t. $\p(\hat{\pi}_{1,-k}(X_i) \geq \epsilon) = 1$. Then, 
$$
\E[\beta(X_i) \phi_{\mu,i}(\etahat_{-k}) - \beta(X_i) \phi_{\mu,i}(\eta) \mid \mathcal{O}_{-k}]  = O_\p(R_{1,n}^{k}).
$$

\begin{proof}
We follow the proof of Lemma 3 in \cite{MishlerEtAl(21)}. Suppressing the dependence on $\mathcal{O}_{-k}$ to ease notation, observe that 
$$
\mathbb{E}[ \beta(X_i) \phi_{\mu,i}(\etahat_{-k}) - \beta(X_i) \phi_{\mu,i}(\eta) ] =
$$
$$
\mathbb{E}\left[ \beta(X_i) \left( \frac{D_i}{\hat{\pi}_1(X_i)} (Y_i - \outcomereghat(X_i)) - \frac{D}{\pi_1(X_i)} (Y_i - \outcomereg(X_i)) + (\outcomereghat(X_i) - \outcomereg(X_i)) \right) \right] \overset{(1)}{=}
$$
$$
\mathbb{E}\left[ \beta(X_i) \left( \frac{\pi_1(X_i)}{\hat{\pi}_1(X_i)} (\outcomereg(X_i) - \outcomereghat(X_i)) + (\outcomereghat(X_i) - \mu_{1}(X_i) \right) \right] = 
$$
$$
\mathbb{E}\left[ \beta(X_i) \frac{ (\outcomereghat(X_i) - \outcomereg(X_i)) (\pihat_1(X_i) - \pi_1(X_i)) }{\hat{\pi}_1(X_i)} \right] \overset{(2)}{\leq} \\
$$
$$
\epsilon^{-1} \mathbb{E}[\beta(X_i) (\outcomereghat(X_i) - \outcomereg(X_i)) (\hat{\pi}_1(X_i) - \pi_1(X_i))],
$$
where (1) follows by iterated expectations and (2) by the assumption of a bounded propensity score estimator. 
The result follows by applying the Cauchy-Schwarz inequality and using $\|\beta(\cdot )\| \leq M$ to conclude that $\| \mathbb{E}[ \beta(X_i) \phi_{\mu,i}(\etahat) - \beta(X_i) \phi_{\mu,i}(\eta)] \| = O_\p(R_{1,n}^{k})$.
\end{proof}
\end{lemma}

\begin{lemma}[Lemma 2 in \cite{KennedyEtAl(2020)-SharpInstruments}]\label{lemma: kennedy et al. 2020}
Let $\hat{\phi}(X_i)$ be a function estimated from a sample $O_i := (X_i, D_i, Y_i) \sim P(\cdot)$ i.i.d. for $i = 1, \hdots, N$ and let $\mathbb{E}_n[\cdot]$ denote the empirical average over another independent sample $O_j \sim P(\cdot)$ i.i.d. for $j = N+1, \hdots, n$. Then,
$$
\mathbb{E}_n[\hat{\phi}(O_i) - \phi(O_i)] - \mathbb{E}[\hat{\phi}(O_i) - \phi(O_i)] = O_\p\left(\frac{\|\hat{\phi}(\cdot) - \phi(\cdot)\|}{\sqrt{n}}\right).
$$
\end{lemma}

\begin{lemma}\label{lemma: extension of kennedy et al to multiplier}
Let $\beta(\cdot)$ be some function of $X_i$ such that $\|\beta(\cdot)\| \leq M$ for some $M < \infty$.
Let $\hat{\phi}(O_i)$ be a function estimated from a sample $O_i := (X_i, D_i, Y_i) \sim P(\cdot)$ i.i.d. for $i = 1, \hdots, N$ and let $\mathbb{E}_n[\cdot]$ denote the empirical average over another independent sample $O_j := (X_j, D_j, Y_j) \sim P(\cdot)$ i.i.d. for $j = N+1, \hdots, n$.
Then, 
$$
\mathbb{E}_n[\beta(X_i) \hat{\phi}(O_i) - \beta(X_i) \phi(O_i)] - \mathbb{E}[\beta(X_i) \hat{\phi}(O_i) - \beta(X_i) \phi(O_i)] = O_P\left(\frac{\|\hat{\phi}(\cdot) - \phi(\cdot)\|}{\sqrt{n}}\right).
$$
\begin{proof}
The proof follows the same argument as the proof of Lemma 2 in \cite{KennedyEtAl(2020)-SharpInstruments}. 
% Observe that, conditional on the estimation sample $\mathcal{O}^{est} = \{O_i\}_{i=1}^{N}$, $\E\{\En[\beta(X_i) (\hat\phi(O_i) - \phi(O_i))] \mid \mathcal{O}^{est} \} = \E[\beta(X_i) (\hat\phi(O_i) - \phi(O_i)) \mid \mathcal{O}^{est}] = \E[\beta(X_i) (\hat\phi(O_i) - \phi(O_i))]$. 
% Next, observe that the conditional variance is $V\left\{ (\En - \E)[\beta(X_i) (\hat\phi(O_i) - \phi(O_i))] \mid \mathcal{O}^{est} \right\} = V\left\{ \En[\beta(X_i) (\hat\phi(O_i) - \phi(O_i))] \mid \mathcal{O}^{est} \right\} = n^{-1} V(\beta(X_i) (\hat\phi(O_i) - \phi(O_i)) \mid \mathcal{O}^{est}) \leq M \| \hat\phi(\cdot) - \phi(\cdot) \| / n$.
% The result then follows by applying Chebyshev's inequality.
\end{proof}
\end{lemma}

\begin{lemma}[Convergence of plug-in influence function estimator $\phi_{\mu,i}(\etahat)$]\label{lemma: influence function bound}
Define $\|\hat{\mu}_{1,-k}(\cdot) - \outcomereg(\cdot) \| \| \hat{\pi}_{1,-k}(\cdot) - \pi_1(\cdot) \| = R_{1,n}^{k}$.
Assume (i) there exists $\delta > 0$ such that $\p(\pi_1(X_i) \geq \delta) = 1$; and (ii) there exists $\epsilon > 0$ such that $\p(\pihat_{1-k}(X_i) \geq \epsilon) = 1$.
Then, 
$$
\|\phi_{\mu,i}(\etahat_{-k}) - \phi_{\mu,i}(\eta)\| = O_\p(R_{1,n}^{k} + \| \hat{\pi}_{1,-k} - \pi_1 \| + \| \hat{\mu}_{1,-k} - \mu_1 \|).
$$
\begin{proof}
This result follows directly from the stated conditions after some algebra. Suppressing dependence on the subscript $-k$ to ease notation, observe that we can rewrite
$$
\left\| \phi_{\mu,i}(\etahat) - \phi_{\mu,i}(\eta) \right\| = 
\left\| \frac{D_i}{\hat{\pi}_1(X_i)} (Y_i - \outcomereghat(X_i)) - \frac{D_i}{\pi_1(X_i)} (Y_i - \outcomereg(X_i)) + (\outcomereg(X_i) - \outcomereghat(X_i)) \right\| \overset{(1)}{=}
$$
$$
\left\| \frac{D_i}{\pi_1(X_i)} \frac{\pi_1(X_i) - \pihat_1(X_i)}{\pihat_1(X_i)} (Y_i - \outcomereghat(X_i)) - \frac{D_i}{\pi_1(X_i)} (\outcomereghat(X_i) - \outcomereg(X_i)) + (\outcomereghat(X_i) - \outcomereg(X_i)) \right\| \overset{(2)}{\leq}
$$
$$
\left\| \frac{D_i}{\pi_1(X_i)} \frac{\pi_1(X_i) - \pihat_1(X_i)}{\pihat_1(X_i)} (Y_i - \outcomereg(X_i)) \right\| + \left\| \frac{D_i}{\pi_1(X_i)} \frac{\pi_1(X_i) - \pihat_1(X_i)}{\pihat_1(X_i)} (\outcomereg(X_i) - \outcomereghat(X_i)) \right\| +
$$
$$
\left\| \frac{D_i}{\pi_1(X_i)} (\outcomereghat(X_i) - \outcomereg(X_i)) \right\| + \left\|(\outcomereghat(X_i) - \outcomereg(X_i)) \right\| \overset{(3)}{\leq}
$$
$$
\frac{\|D_i\|}{\delta} \frac{\|\pi_1 - \pihat_1\|}{\epsilon} \|Y_i - \outcomereg(X_i) \| + \frac{\|D_i\|}{\delta} \frac{\|\pi_1 - \pihat_1\|}{\epsilon} \|\outcomereghat - \outcomereg\| + 
\frac{\|D_i\|}{\delta} \|\outcomereghat - \outcomereg\| + \left\|\outcomereghat - \outcomereg \right\|
$$
where (1) follows by adding and subtracting $\frac{D_i}{\pi_1(X_i)} (Y_i - \outcomereghat(X_i))$, (2) follows by adding and subtracting $\frac{D_i}{\pi_1(X_i)} \frac{\pi_1(X_i) - \pihat_1(X_i)}{\pihat_1(X_i)} \outcomereg(X_i)$ and applying the triangle inequality, and (3) applies the assumption of strict overlap and bounded propensity score estimator.
\end{proof}
\end{lemma}

\begin{lemma}[Convergence of plug-in influence function estimator $\phi_{\pi \mu,i}(\etahat)$]\label{lemma: convergence of pi0mu1 IF}
Define $\|\hat{\mu}_{1, -k}(\cdot) - \outcomereg(\cdot) \| \| \hat{\pi}_{1, -k}(\cdot) - \pi_1(\cdot) \| = R_{1,n}^{k}$.
Assume that (i) there exists $\delta > 0$ such that $\p(\pi_1(X_i) \geq \delta) = 1$; and (ii) there exists $\epsilon > 0$ such that $\p(\pihat_{1,-k}(X_i) \geq \epsilon) = 1$.
Then, 
$$
\| \phi_{\pi \mu,i}( \etahat_{-k}) - \phi_{\pi \mu,i}(\eta) \| = O_\p(R_{1,n}^{k} + \| \hat{\pi}_{1,-k} - \pi_1 \| + \| \hat{\mu}_{1,-k} - \mu_1 \|)
$$
\begin{proof}
This result follows directly from the stated conditions after some simple algebra. We omit the dependence on $X_i$ and $-k$. Observe that we can rewrite 
{\small
$$
\| \phi_{\pi \mu,i}( \etahat) - \phi_{\pi \mu,i}( \etahat) \| = 
$$
$$
\| ((1-D_i) - \pihat_0) \outcomereghat + \frac{D_i}{\pihat_1} (Y_i - \outcomereghat) \pihat_0 + \pihat_0 \outcomereghat - ((1 - D_i) - \pi_0) \outcomereg - \frac{D_i}{\pi_1} (Y_i - \outcomereg) \pi_0 - \pi_0 \outcomereg \| \overset{(1)}{=}
$$
$$
\| ((1-D_i)-\pihat_0) \outcomereghat + \frac{D_i}{\pi_1} \frac{\pi_1 - \pihat_1 }{ \pihat_1 } (Y_i - \outcomereghat) + \frac{D_i}{\pi_1} \pi_0 (\outcomereg - \outcomereghat) + \pihat_0 \outcomereghat - \pi_0 \outcomereg \| \overset{(2)}{\leq}
$$
$$
\| \frac{D_i}{\pi_1} \frac{\pi_1 - \pihat_1 }{ \pihat_1 } (Y_i - \outcomereg) + \frac{D_i}{\pi_1} \frac{\pi_1 - \pihat_1 }{ \pihat_1 } (\outcomereg - \outcomereghat) + \frac{D_i}{\pi_1} \pi_0 (\outcomereg - \outcomereghat) + ((1-D_i)-\pihat_0) \outcomereghat + \pihat_0 \outcomereghat - \pi_0 \outcomereg \| \overset{(3)}{=}
$$
$$
\| \frac{D_i}{\pi_1} \frac{\pi_1 - \pihat_1 }{ \pihat_1 } (Y_i - \outcomereg) + \frac{D_i}{\pi_1} \frac{\pi_1 - \pihat_1 }{ \pihat_1 } (\outcomereg - \outcomereghat) + \frac{D_i}{\pi_1} \pi_0 (\outcomereg - \outcomereghat) + ((1-D_i)-\pihat_0) (\outcomereghat - \outcomereg) + (\pi_0 - \pihat_0) \mu_1 + \pihat_0 \outcomereghat - \pi_0 \outcomereg \| \overset{(4)}{\leq}
$$
$$
\| \frac{D_i}{\pi_1} \frac{\pi_1 - \pihat_1 }{ \pihat_1 } (Y_i - \outcomereg) + \frac{D_i}{\pi_1} \frac{\pi_1 - \pihat_1 }{ \pihat_1 } (\outcomereg - \outcomereghat) + \frac{D_i}{\pi_1} \pi_0 (\outcomereg - \outcomereghat) + ((1-D_i)-\pi_0) (\outcomereghat - \outcomereg) + (\pi_0 - \pihat_0) (\outcomereghat - \outcomereg) + (\pi_0 - \pihat_0) \mu_1 \| +
$$
$$
+ \| \pihat_0 \outcomereghat - \pi_0 \outcomereg \| \overset{(5)}{=}
$$
$$
\| \frac{D_i}{\pi_1} \frac{\pi_1 - \pihat_1 }{ \pihat_1 } (Y_i - \outcomereg) + \frac{D_i}{\pi_1} \frac{\pi_1 - \pihat_1 }{ \pihat_1 } (\outcomereg - \outcomereghat) + \frac{D_i}{\pi_1} \pi_0 (\outcomereg - \outcomereghat) + ((1-D_i)-\pi_0) (\outcomereghat - \outcomereg) + (\pi_0 - \pihat_0) (\outcomereghat - \outcomereg) + (\pi_0 - \pihat_0) \mu_1 \| +
$$
$$
\|\pihat_0 (\outcomereghat - \outcomereg) - \outcomereg (\pi_0 - \pihat_0)\|
$$
}
where (1) follows by adding/subtracting $\frac{D_i}{\pi_1}(Y_i - \outcomereghat)$, (2) follows by adding/subtracting $\frac{D_i}{\pi_1} \frac{\pi_1 - \pihat_1}{\pihat_1} \mu_1$, (3) follows by adding/subtracting $\mu_1 ((1-D_i) - \pihat_0)$, (4) follows by adding/subtracting $\pi_0 (\outcomereghat - \outcomereg)$ and applying the triangle inequality once, and (5) follows by adding/subtracting $\pihat_0 \outcomereg$.
We then again apply the triangle inequality and use the assumptions of strict overlap and bounded propensity score estimator to arrive at
$$
\leq \frac{1}{\epsilon \delta} \| \pihat_1 - \pi_1 \| \|Y_i - \outcomereg\| + \frac{1}{\epsilon \delta} \|\pihat_1 - \pi_1\| \|\outcomereg - \outcomereghat\| + \frac{1-\delta}{\delta} \|\outcomereg - \outcomereghat\| + \|(1-D_i)-\pi_0\|\|\outcomereghat -\outcomereg\| +
$$
$$
\|\pihat_1 - \pi_1\| \|\outcomereghat - \outcomereg\| + \|\pihat_1 - \pi_1\| \|\outcomereg\| + (1-\epsilon)\|\outcomereghat - \outcomereg\| + \|\outcomereg\| \|\pihat_1 - \pi_1\|.
$$
The result then follows.
\end{proof}
\end{lemma}

\begin{lemma}\label{lemma: mishler et al. 2021 for pi0mu1 IF}
Let $\beta(\cdot)$ be some function of $X_i$ such that $\|\beta(\cdot)\| \leq M$ for some $M < \infty$ and define $R_{1,n}^{k}= \|\hat{\mu}_{1,-k}(\cdot) - \outcomereg(\cdot)\| \|\pihat_{1,-k}(\cdot) - \pi_1(\cdot)\|$. 
Assume that there exists $\epsilon > 0$ s.t. $\p(\hat{\pi}_{1, -k}(X_i) \geq \epsilon) = 1$. Then, 
$$
\E[ \beta(X_i) \left( \phi_{\pi \mu,i}( \etahat_{-k}) - \phi_{\pi \mu,i}( \eta) \right) \mid \mathcal{O}_{-k}] = O_\p(R_{1,n}^k)
$$

\begin{proof}
For ease of notation, we omit the dependence on $X_i$ and the subscript $-k$. The proof follows an analogous argument to Lemma \ref{lemma: mishler et al 2021}. Observe that 
$$
\E[\beta(X_i) \left( \phi_{\pi \mu,i}(\etahat) - \phi_{\pi \mu,i}(\eta) \right)] = 
$$
$$
\E[\beta(X_i) \left\{ ((1-D_i)-\pihat_0) \outcomereghat + \frac{D_i}{\pihat_1} (Y_i - \outcomereghat) \pihat_0 + \pihat_0 \outcomereghat - ((1-D_i) - \pi_0) \outcomereg - \frac{D_i}{\pi_1} (Y_i - \outcomereg) \pi_0 - \pi_0 \outcomereg \right\} ] \overset{(1)}{=}
$$
$$
\E[ \beta(X_i) \left\{ (\pi_0 - \pihat_0) \outcomereghat + \frac{\pi_1}{\pihat_1} (\outcomereg - \outcomereghat) \pihat_0 \right\} + \pihat_0 \outcomereghat - \pi_0 \outcomereg\}] \overset{(2)}{=}
$$
$$
\E[ \beta(X_i) \left\{ (\pi_0 - \pihat_0) \outcomereghat + \frac{\pi_1}{\pihat_1} (\outcomereg - \outcomereghat) \pihat_0  + \pihat_0 (\outcomereghat - \outcomereg) + \outcomereg (\pihat_0 - \pi_0) \right\} ] =
$$
$$
\E[\beta(X_i) \left\{ (\outcomereghat - \outcomereg) (\pi_0 - \pihat_0) + \frac{\pi_1 - \pihat_1}{\pihat_1} (\outcomereg - \outcomereghat) \pihat_0 \right\} ]
$$
where (1) applies iterated expectations, (2) adds/subtracts $\pihat_0 \mu_1$, and the final equality re-arranges. The result then follows by applying the assumption of bounded propensity score estimator and applying the Cauchy-Schwarz inequality.
\end{proof}
\end{lemma}

We are now ready to prove Proposition \ref{prop: overall performance estimator, nonparametric outcome bounds}.
To prove the first claim, consider our proposed estimator of the upper bound $\widehat{\overline{\perf}}(s; \beta)$. Let
\begin{align*}
& \overline{\perf}_i = \beta_{0,i} + \beta_{1,i} \phi_{\mu,i}(\eta) + \beta_{1,i} \left( 1\{\beta_{1,i} > 0\} (\overline{\Gamma} - 1) + 1\{\beta_{1,i} \leq 0\} (\underline{\Gamma} - 1) \right) \phi_{\pi \mu,i}(\eta) \\
& \widehat{\overline{\perf}}_{i} = \beta_{0,i} + \beta_{1,i} \phi_{\mu,i}(\etahat_{-K_i}) + \beta_{1,i} \left( 1\{\beta_{1,i} > 0\} (\overline{\Gamma} - 1) + 1\{\beta_{1,i} \leq 0\} (\underline{\Gamma} - 1) \right) \phi_{\pi \mu,i}(\etahat_{-K_i}).
\end{align*}
Observe $| \widehat{\overline{\perf}}(s; \beta) - \overline{\perf}(s; \beta) |$ can be written as
$$
|\En[\widehat{\overline{\perf}}_{i}] - \E[\overline{\perf}_i ] | \leq | \underbrace{\En[\overline{\perf}_i] - \E[\overline{\perf}_i]}_{(a)} | + | \underbrace{\En\left[\left( \widehat{\overline{\perf}}_{i} - \overline{\perf}_i \right)\right]}_{(b)}|.
$$
By Chebyshev's inequality, (a) is $O_\p(1/\sqrt{n})$.
Next, we can further rewrite (b) as
$$
\left| \En\left[\left( \widehat{\overline{\perf}}_{i} - \overline{\perf}_i \right)\right] \right| = \left| \sumoverfolds \En[1\{K_i = k\}] \En^k[\widehat{\overline{\perf}}_{i, -k} - \overline{\perf}_i] \right| \leq \sumoverfolds \left| \En^k[\widehat{\overline{\perf}}_{i, -k} - \overline{\perf}_i] \right|.
$$
We will show that each term in the sum is $O_\p(1/\sqrt{n} + R_{1,n}^k + R_{1,n}^k/\sqrt{n})$.
Observe that 
$$
\En^k[\widehat{\overline{\perf}}_{i, -k} - \overline{\perf}_i] | \leq
| \En^k[\widehat{\overline{\perf}}_{i, -k} - \overline{\perf}_i]  - \E[ \widehat{\overline{\perf}}_{i, -k} - \overline{\perf}_i \mid \mathcal{O}_{-k}] | + | \E[ \widehat{\overline{\perf}}_{i, -k} - \overline{\perf}_i\mid \mathcal{O}_{-k}] |,
$$
where 
$$
\widehat{\overline{\perf}}_{i, -k} - \overline{\perf}_i = \beta_{1,i} ( \phi_{\mu,i}(\etahat_{-k}) - \phi_{\mu,i}(\eta)) + \tilde{\beta}_{1,i} \left( \phi_{\pi \mu,i}(\etahat_{-k}) - \phi_{\pi \mu,i}(\eta) \right)
$$
for $\tilde{\beta}_i = \beta_{1,i} \left( 1\{\beta_{1,i} > 0\} (\overline{\Gamma} - 1) + 1\{\beta_{1,i} \leq 0\} (\underline{\Gamma} - 1) \right)$.
So $| \En^k[\widehat{\overline{\perf}}_{i, -k} - \overline{\perf}_i]  - \E[ \widehat{\overline{\perf}}_{i, -k} - \overline{\perf}_i \mid \mathcal{O}_{-k}] |$ is bounded by
$$
\underbrace{| \En^k[\beta_{1,i} \left( \phi_{\mu,i}(\etahat_{-k}) - \phi_{\mu,i}(\eta) \right)]  - \E[ \beta_{1,i} \left( \phi_{\mu,i}(\etahat_{-k}) - \phi_{\mu,i}(\eta) \right) \mid \mathcal{O}_{-k}] |}_{(c)} + 
$$
$$
\underbrace{| \En^k[\tilde{\beta}_{1,i} \left( \phi_{\pi \mu, i}(\etahat_{-k}) - \phi_{\pi \mu, i}(\eta) \right)]  - \E[ \tilde{\beta}_{1,i} \left( \phi_{\pi \mu, i}(\etahat_{-k}) - \phi_{\pi \mu, i}(\eta) \right) \mid \mathcal{O}_{-k}] |}_{(d)},
$$
where (c) is $O_\p(\|\hat{\pi}_{1,-k} - \pi_1 \|/\sqrt{n} + \|\hat{\mu}_{1,-k} + \mu_1\|/\sqrt{n} + R_{1,n}^k/\sqrt{n})$ by Lemma \ref{lemma: extension of kennedy et al to multiplier} and Lemma \ref{lemma: influence function bound}, and (d) is also $O_\p(\|\hat{\pi}_{1,-k} - \pi_1 \|/\sqrt{n} + \|\hat{\mu}_{1,-k} + \mu_1\|/\sqrt{n} + R_{1,n}^k / \sqrt{n})$ Lemma \ref{lemma: extension of kennedy et al to multiplier} and Lemma \ref{lemma: convergence of pi0mu1 IF}.
The second term  $ \E[ \widehat{\overline{\perf}}_{i, -k} - \overline{\perf}_i\mid \mathcal{O}_{k}] |$ is bounded by 
$$
\underbrace{| \E[\beta_{1,i} \left( \phi_{\mu,i}(\etahat_{-k}) - \phi_{\mu,i}(\eta) \right) \mid \mathcal{O}_{-k}] |}_{(e)} + 
| \underbrace{\E[\tilde \beta_{1,i} \left( \phi_{\pi \mu,i}(\etahat_{-k}) - \phi_{\pi \mu,i}(\eta) \right) \mid \mathcal{O}_{-k}]}_{(f)}|,
$$
where (e) is $O_\p(R_{1,n}^k)$ by Lemma \ref{lemma: mishler et al 2021} and (f) is $O_\p(R_{1,n}^k)$ by Lemma \ref{lemma: mishler et al. 2021 for pi0mu1 IF}.
Putting this together, we have shown the first claim 
$$
\left| \widehat{\overline{\perf}}(s; \beta) - \overline{\perf}(s; \beta, \Delta) \right| = 
$$
$$
O_{\p}\left(1/\sqrt{n} + \sumoverfolds R_{1,n}^k + \sumoverfolds R_{1,n}^k/\sqrt{n} + \sumoverfolds \left( \|\hat{\pi}_{1,-k} - \pi_1 \| + \|\hat{\mu}_{1,-k} + \mu_1\| \right) / \sqrt{n} \right).
$$
The result for $\widehat{\underline{\perf}}(s; \beta)$ follows the same argument. 

The second claim follows by noticing that the proof of the first claim showed that 
\begin{equation*}
\sqrt{n} \left( \begin{pmatrix} \widehat{\overline{\perf}}(s; \beta) \\ \widehat{\underline{\perf}}(s; \beta) \end{pmatrix} - \begin{pmatrix} \overline{\perf}(s; \beta) \\ \underline{\perf}(s; \beta) \end{pmatrix} \right) = \frac{1}{\sqrt{n}} \sum_{i=1}^{n} \begin{pmatrix} \overline{\perf}_i - \E[\overline{\perf}_i] \\ \underline{\perf}_i - \E[\underline{\perf}_i] \end{pmatrix} + o_\p(1)
\end{equation*}
if $R_{1,n} = o_\p(1/\sqrt{n})$. By the central limit theorem,
$$
\sqrt{n} \left( \begin{pmatrix} \widehat{\overline{\perf}}(s; \beta) \\ \widehat{\underline{\perf}}(s; \beta) \end{pmatrix} - \begin{pmatrix} \overline{\perf}(s; \beta) \\ \underline{\perf}(s; \beta) \end{pmatrix} \right) \xrightarrow{d} N\left( 0, Cov\left( \begin{pmatrix} \overline{\perf}_i \\ \underline{\perf}_i \end{pmatrix} \right) \right),
$$
from which the result follows. $\Box$

\subsection{Proof of Proposition \ref{proposition: partial double robustness for pos class estimator}}

To prove Proposition \ref{proposition: partial double robustness for pos class estimator}, we first consider the case in which $\underline{\delta}(\cdot; \eta), \overline{\delta}(\cdot; \eta)$ are known as a stepping stone. 
We prove the result for the upper bound since the argument is identical for the lower bound.
To simplify notation, we assume the data are split into two equal sized folds. 

Define
\begin{equation}\label{eqn: fold specific positive class estimator}
\widehat{\overline{\perf}}_{+}(s; \beta, \Delta_{n}) :=  \max_{\tilde{\delta} \in \Delta_{n}} \frac{ \mathbb{E}_{n}[ \beta_{0,i} \phi_{\mu,i}(\hat{\eta}) + \beta_{0,i} \tilde{\delta}_i] }{\mathbb{E}_n[ \phi_{\mu,i}(\hat{\eta}) + \tilde{\delta}_i]},
\end{equation}
where $\Delta_{n} = \left\{ \tilde{\delta} \colon (1 - D_i) \underline{\delta}(X_i; \eta) \leq \tilde{\delta}_i \leq (1 - D_i) \overline{\delta}(X_i; \eta) \mbox{ for } i = 1, \hdots, n/2 \right\}$ and $\underline{\delta}(X_i; \eta) = (\underline{\Gamma} - 1) \mu_1(X_i)$, $\overline{\delta}(X_i; \eta) = (\overline{\Gamma} - 1) \mu_1(X_i)$. 
As shorthand, define $\underline{\delta}^\prime(D_i, X_i; \eta) = (1 - D_i) \underline{\delta}(X_i; \eta)$ and $\overline{\delta}^{\prime}(D_i, X_i; \eta) = (1 - D_i) \overline{\delta}(X_i; \eta)$.

\begin{lemma}\label{lem: reduction to monotone, non-decr functions, DR estimator}
Define $\mathcal{U}$ to be the set of monotone, non-decreasing functions $u(\cdot)\colon \reals \rightarrow [0,1]$, $\Delta^{M} := \left\{ \delta^\prime(d,x) = \underline{\delta}^{\prime}(d,x; \eta) + (\overline{\delta}^{\prime}(d,x; \eta) - \underline{\delta}^{\prime}(d, x; \eta)) u(\beta_0(x)) \mbox{ for } u(\cdot) \in \mathcal{U} \right\}$, and $\Delta^{M}_{n} = \{ (\delta^{\prime}(D_1, X_1), \hdots, \delta^{\prime}(D_{n/2}, X_{n/2}) \colon \delta \in \Delta^{M} \}$. 
Then, $\overline{\perf}_{+}(s; \beta, \Delta) := \sup_{\tilde \delta \in \Delta^{M}} \perf_{+}(s; \beta, \tilde \delta)$ and 
\begin{align*}
    & \widehat{\overline{\perf}}_{+}(s; \beta, \Delta_{n}) :=  \max_{\tilde{\delta} \in \Delta_{n}^{M}} \frac{ \mathbb{E}_{n}[ \beta_{0,i} \phi_{\mu,i}(\etahat) + \beta_{0,i}  \tilde{\delta}_i ] }{\mathbb{E}_n[ \phi_{\mu,i}(\etahat) + \tilde{\delta}_i ]}.
\end{align*}

\begin{proof}
We first show this result for the fold-specific estimator $\widehat{\overline{\perf}}_{+}(s; \beta, \Delta_{n})$ by using the proof strategy of Proposition 2 in \cite{KallusZhou(21)}
By Lemma \ref{lemma: LFP to LP reduction for positive class performance},
\begin{align*}
   \widehat{\overline{\perf}}_{+}(s; \beta, \Delta_{n}) = \max_{\tilde U, \tilde V} & \hat{\alpha}^\prime \tilde{U} + \hat{c} \tilde{V} \\
   \mbox{ s.t. } & 0 \leq \tilde{U}_i \leq \tilde{V} \mbox{ for } i = 1, \hdots n/2, \mbox{ } 0 \leq \tilde{V}, \mbox{ } \hat{\gamma}^\prime \tilde{U} + \tilde{V} \hat{d} = 1.
\end{align*}
Next, define the dual program associated with this primal linear program.
Let $P_i$ be the dual variables associated with the constraints $\tilde{U}_i \leq \tilde{V}$, $Q_i$ be the dual variables associated with the constraints $\tilde{U}_i \geq 0$, and $\lambda$ be the dual variable associated with the constraint $\hat{\gamma}^\prime \tilde{U} + \tilde{V} \hat{d} = 1$.
The dual linear program is 
\begin{align*}
    \min_{\lambda, P, Q} & \lambda \\
    \mbox{ s.t. } & P_i - Q_i + \lambda \hat{\gamma}_i = \hat{\alpha}_i, \mbox{ } -\textbf{1}^\prime P + \lambda \hat{d} \geq \hat{c}, \\
    & P_i \geq 0, Q_i \geq 0 \mbox{ for } i = 1, \hdots, n/2,
\end{align*}
where $\textbf{1}$ is the vector of all ones of appropriate dimension.
By re-arranging the first constraint and substituting in the expressions for $\hat{\alpha}_i, \hat{\gamma}_i$, observe that
\begin{align*}
    & P_i - Q_i = (\beta_0 - \lambda) (\overline{\delta}^{\prime}_i - \underline{\delta}^{\prime}_i).
\end{align*}
By complementary slackness, at most only one of $P_i$ or $Q_i$ will be non-zero at the optimum, and so combined with the previous display, this implies
\begin{align*}
& P_i = \max\{\beta_{0,i} - \lambda, 0\} (\overline{\delta}^{\prime}_i - \underline{\delta}^{\prime}_i), \\
& Q_i = \max\{\lambda - \beta_{0,i}, 0\} (\overline{\delta}^{\prime}_i - \underline{\delta}^{\prime}_i).
\end{align*}
Next, notice that the constraint $-\textbf{1}^\prime P + \lambda \hat{d} \geq \hat{c}$ must be tight at the optimum.
Plugging in the previous expression for $P_i$ and the expressions for $\hat{c}, \hat{d}$, this implies that $\lambda$ satisfies
\begin{align*}
    & - \En[ \max\{\beta_{0,i} - \lambda, 0\} (\overline{\delta}^{\prime}_i - \underline{\delta}^{\prime}_i) ] = \En[ \left( \beta_{0,i} - \lambda \right) \left( \phi_{\mu,i}(\etahat) + \underline{\delta}^{\prime}_i \right) ].
\end{align*}

\noindent Finally, we consider three separate cases:
\begin{enumerate}
\item[1.] Suppose that $\lambda \geq \max_{i} \beta_{0,i}$. From the previous display, $\lambda$ must satisfy 
    $$
    0 = \En[ \left( \beta_{0,i} - \lambda \right) \left( \phi_{\mu,i}(\etahat) + \underline{\delta}^{\prime}_i \right) ] \implies \lambda = \frac{\mathbb{E}_n[\beta_{0,i} \left( \phi_{\mu,i}(\etahat) + \underline{\delta}^{\prime}_i \right)]}{\mathbb{E}_n[\phi_{\mu,i}(\etahat) + \underline{\delta}^{\prime}_i ]}.
    $$
    At this value for $\lambda$, the expressions for $P_i, Q_i$ imply that $P_i = 0$, $Q_i > 0$ for all $i$. By complementary slackness, this in turn implies that $\tilde{U}_i = 0$, or equivalently $U_i = 0$ for all $i$.

    \item[2.] Suppose that $\lambda \leq \min_{i} \beta_{0,i}$. From the previous display, $\lambda$ must satisfy 
    $$
    -\En[ (\beta_{0,i} - \lambda) (\overline{\delta}^{\prime}_i - \underline{\delta}^{\prime}_i) ] = \En[ \left( \beta_{0,i} - \lambda \right) \left( \phi_{\mu,i}(\etahat) + \underline{\delta}^{\prime}_i \right)]
    \implies \lambda = \frac{\mathbb{E}_n[\beta_{0,i} \left( \phi_{\mu,i}(\etahat) + \overline{\delta}^{\prime}_i \right)]}{\mathbb{E}_n[\phi_{\mu,i}(\etahat) + \overline{\delta}^{\prime}_i ]}.
    $$
    At this value for $\lambda$, the expressions for $P_i, Q_i$ imply that $P_i > 0$, $Q_i = 0$ for all $i$. By complementary slackness, this implies that $\tilde{U}_i = \tilde{V}$, or equivalently $U_i = 1$ for all $i$.

    \item[3.] Suppose that $\min_{i} \beta_{0,i} < \lambda < \max_{i} \beta_{0,i}$. Then, $\beta_{0,(j)} < \lambda \leq \beta_{0, (j+1)}$ for some $j$ where $\beta_{0,(1)}, \hdots, \beta_{0, (n_k)}$ are the order statistics of the sample outcomes. 
    The expressions for $P_i, Q_i$ in turn imply that $Q_i > 0$ only when $\beta_{0,i} \leq \beta_{0, (j)}$ (in which case $U_i = 0$) and $P_i > 0$ only when $\beta_{0,i} \geq \beta_{0, (j+1)}$ (in which case $U_i = 1$).
\end{enumerate}
Therefore, in all three cases, the optimal solution is such that there exists a non-decreasing function $u(\cdot) \colon \mathbb{R} \rightarrow [0,1]$ such that $U_i = u(\beta_{0,i})$ attains the upper bound.

We next prove the result for the population bound $\overline{\perf}_{+}(s; \beta, \Delta)$ via a similar argument.
We can first rewrite the bound as 
$$
\overline{\perf}_{+}(s; \beta, \Delta) := \sup_{\delta(\cdot) \in \Delta} \mathbb{E}[\mu_{1}(X_i) + (1 - D_i) \delta(X_i)]^{-1} \mathbb{E}[ \beta_{0,i} \mu_{1}(X_i) +  \beta_{0,i} (1 - D_i) \delta(X_i) ]
$$
by iterated expectations.
Then, applying the change-of-variables $\delta(X_i) = \underline{\delta}(X_i) + (\overline{\delta}(X_i) - \underline{\delta}(X_i)) U(X_i)$, we further rewrite the population bound as
$$
\overline{\perf}_{+}(s; \beta, \Delta) := \sup_{U(\cdot) \colon \cX \rightarrow [0,1]} \frac{ \E[ \beta_{0,i} \outcomereg(X_i) + \beta_{0,i} \pi_0(X_i) \underline{\delta}_i + \beta_{0,i} \pi_0(X_i) (\overline{\delta}_i - \underline{\delta}_i) U(X_i) ] }{\E[ \outcomereg(X_i) + \pi_0(X_i) \underline{\delta}_i + \pi_0(X_i) (\overline{\delta}_i - \underline{\delta}_i) U(X_i) ]}.
$$
Define $c := \mathbb{E}[\beta_{0,i} \outcomereg(X_i) + \beta_{0,i} \pi_0(X_i) \underline{\delta}_i]$, $d := \E[ \outcomereg(X_i) + \pi_0(X_i) \underline{\delta}_i ]$, and $\alpha(x) := \beta_{0}(x; s) \pi_0(x) (\overline{\delta}(x) - \underline{\delta}(x))$, $\gamma(x) := \pi_0(x) (\overline{\delta}(x) - \underline{\delta}(x))$. 
Letting $\langle f, g \rangle_{P(\cdot)}$ denote the inner product $\E[f(X_i) g(X_i)]$, we can further rewrite the population bound as
$$
\overline{\perf}_{+}(s; \beta, \Delta) := \sup_{U(\cdot) \colon \cX \rightarrow [0,1]} \frac{c + \langle \alpha, U \rangle_{P(\cdot)}}{d + \langle \gamma, U \rangle_{P(\cdot)}}.
$$
Define the change-of-variables $\tilde{U}(\cdot) = \frac{U(\cdot)}{d + \langle \gamma, U \rangle_{P(\cdot)}}$ and $\tilde{V} = \frac{1}{\langle \gamma, U \rangle_{P(\cdot)}}$. 
The previous linear-fractional optimization is equivalent to 
\begin{align*}
\sup_{\tilde{U}(\cdot), \tilde{V}} & \langle \alpha, \tilde{U} \rangle_{P(\cdot)} + c \tilde{V} \mbox{ s.t. } 0 \leq \tilde{U}(x) \leq \tilde{V} \mbox{ for all } x \in \cX, 
\mbox{ } \langle \gamma, \tilde{U} \rangle_{P(\cdot)} + \tilde{V} d = 1.
\end{align*}
Define the dual associated with this primal program. 
Let $\tilde{P}(x)$ be the dual function associated with the constraint $\tilde{U}(x) \leq \tilde{V}$, $\tilde{Q}(x)$ be the dual variables associated with the constraints $\tilde{U}(x) \geq 0$, and $\lambda$ be the dual variable associated with the constraint $\langle \gamma, \tilde{U} \rangle_{P(\cdot)} + \tilde{V} d = 1$.
The dual is
\begin{align*}
    \inf_{\lambda, \tilde{P}(\cdot), \tilde{Q}(\cdot)} & \lambda \\
    \mbox{s.t.} & \tilde{P}(x) - \tilde{Q}(x) + \lambda \gamma(x) = \alpha(x) \mbox{ for all } x \in \cX \\
    & -\langle \textbf{1}, \tilde{P} \rangle_{P(\cdot)} + \lambda d \geq c, \mbox{ } \tilde{P}(x) \geq 0, \tilde{Q}(x) \geq 0 \mbox{ for all } x \in \cX.
\end{align*}
By complementary slackness, at most only one of $\tilde P(x)$ or $\tilde Q(x)$ can be non-zero at the optimum for all $x \in \cX$. Therefore, by re-arranging the first constraint and substituting in for $\alpha(x), \gamma(x)$, observe 
\begin{align*}
    & \tilde P(x) - \tilde Q(x) = (\beta_0(x) - \lambda) \pi_0(x) (\overline{\delta}(x) - \underline{\delta}(x)),
\end{align*}
which in turn implies that
\begin{align*}
& \tilde{P}(x) = \max\{\beta_{0}(x) - \lambda, 0\} \pi_0(x) (\overline{\delta}(x) - \underline{\delta}(x)), \\
& \tilde{Q}(x) = \max\{\lambda - \beta_{0}(x), 0\} \pi_0(x) (\overline{\delta}(x) - \underline{\delta}(x)).
\end{align*}
The constraint $\langle \textbf{1}, \tilde{P} \rangle_{P(\cdot)} + \lambda d \geq c$ must be tight at the optimum. 
Plugging in the previous expression for $\tilde{P}(\cdot)$, this implies that $\lambda$ satisfies
\begin{align*}
    & -\E[\max\{\beta_{0}(X_i) - \lambda, 0\} \pi_0(X_i) (\overline{\delta}(X_i) - \underline{\delta}(X_i))] = \E[(\beta_0(X_i) - \lambda) (\outcomereg(X_i) + \pi_0(X_i) \underline{\delta}(X_i)].
\end{align*}
As in the proof for the estimator, we can consider three cases: (i) $\lambda \geq \overline{\beta}_0$, (ii) $\lambda \leq \underline{\beta}_{0}$ and (iii) $\underline{\beta}_0 < \lambda < \overline{\beta}_0$ for $\underline{\beta}_0 := \inf_{x \in \cX} \beta_0(x)$, $\overline{\beta}_0 = \sup_{x \in \cX} \beta_0(x)$. 
In each case, the optimal solution is such that there exists a non-decreasing function $u(\cdot) \colon \reals \rightarrow [0,1]$ such that $U(x) = u(\beta_0(x))$ attains the upper bound. The result follows by applying the definitions of $\underline{\delta}^{\prime}(d, x; \eta), \overline{\delta}^{\prime}(d, x; \eta)$.
\end{proof}
\end{lemma}

\begin{lemma}\label{lemma: convergence rate of positive class estimator}
Define $R_{1,n} = \|\hat{\mu}_{1}(\cdot) - \mu_{1}(\cdot)\| \|\hat{\pi}_{1}(\cdot) - \pi_{1}(\cdot)\|$.
Assume that (i) there $\delta > 0$ such that $\p(\pi_1(X_i) \geq \delta) = 1$; (ii) there exists $\epsilon > 0$ such that $\p(\hat{\pi}(X_i) \geq \epsilon) = 1$; and (iii) $\|\hat{\mu}_{1}(\cdot) - \mu_{1}(\cdot)\| = o_P(1)$ and $\|\hat{\pi}_{1}(\cdot) - \pi_{1}(\cdot)\| = o_P(1)$.
Then,
\begin{align*}
& \left\| \widehat{\overline{\perf}}_{+}(s; \beta, \Delta_n) - \overline{\perf}_{+}(s; \beta, \Delta) \right\| = O_\p\left( 1/\sqrt{n} + R_{1,n} \right).
\end{align*}

\begin{proof}
Let $\widehat{\perf}_{+}(s; \beta, \tilde{\delta}) := \En[\beta_{0,i} \phi_{\mu,i}(\etahat) + \beta_{0,i} \tilde \delta_i] / \En[\phi_{\mu,i}(\etahat) + \tilde \delta_i]$ for $\tilde \delta_i = (1 - D_i) \delta_i$. 
To prove this result, we first observe that
\begin{align*}
\left\| \widehat{\overline{\perf}}_{+}(s; \beta, \Delta_{n}) - \overline{\perf}_{+}(s; \beta, \Delta) \right\| &= \left\| \sup_{\tilde \delta \in \Delta_{n}^M} \widehat{\perf}_{+}(s; \beta, \tilde{\delta}) - \sup_{\tilde \delta \in \Delta^M} \perf_{+}(s; \beta, \tilde \delta) \right\| \\
&= \left\| \sup_{\tilde \delta \in \Delta^M} \widehat{\perf}_{+}(s; \beta, \tilde{\delta}) - \sup_{\tilde \delta \in \Delta^M} \perf{+}(s; \beta, \tilde \delta) \right\| \\
&\leq \sup_{\tilde \delta \in \Delta^M} \left\| \widehat{\perf}_{+}(s; \beta, \tilde{\delta}) - \perf_{+}(s; \beta, \tilde \delta) \right\|,
\end{align*}
where the first equality uses Lemma \ref{lem: reduction to monotone, non-decr functions, DR estimator}. 
For any $\tilde{\delta} \in \Delta^{M}$, we have that 
$$
\widehat{\perf}_{+}(s; \beta, \tilde{\delta}) - \perf_{+}(s; \beta, \tilde \delta) = 
$$
$$
\frac{\En[ \beta_{0,i} \phi_{\mu,i}(\etahat) + \beta_{0,i} \tilde{\delta}_i ] }{\En[\phi_{\mu,i}(\etahat) + \tilde{\delta}_i] } - \frac{\E[ \beta_{0,i} \phi_{\mu,i}(Y_i; \etahat) + \beta_{0,i} \tilde{\delta}_i ] }{\E[\phi_{\mu,i}(\etahat) + \tilde{\delta}_i] } = 
$$
$$
\frac{\En[(\text{\#1})]}{\En[(\text{\#2})]} - \frac{\E[(\text{\#3})]}{\E[(\text{\#4})]}  = \En[(\text{\#2})]^{-1} \left\{ \En[(\text{\#1})] - \E[(\text{\#3})] - \frac{\E[(\text{\#3})]}{\E[(\text{\#4})]} (\En[(\text{\#2})] - \E[(\text{\#4})]) \right\},
$$
where
\begin{align*}
\En[(\text{\#1})] - \E[(\text{\#3})] &= \En[\beta_{0,i} \phi_{\mu,i}(\etahat) + \beta_{0,i} \tilde \delta_i] - \E[\beta_{0,i} \phi_{\mu,i}(\eta) + \beta_{0,i} \tilde \delta_i] \\
&= \left( \En[\beta_{0,i} \phi_{\mu,i}(\etahat)] - \E[\beta_{0,i} \phi_{\mu,i}(\eta)] \right) + (\En - \E)[\beta_{0,i} \tilde \delta_i] \\
\En[(\text{\#2})] - \E[(\text{\#4})] &= \En[\phi_{\mu,i}(\etahat) + \tilde \delta_i] - \E[\phi_{\mu,i}( \eta) + \tilde \delta_i] \\
&= \left( \En[\phi_{\mu,i}(\etahat)] - \E[\phi_{\mu,i}(\eta)] \right) + (\En - \E)[\tilde \delta_i].
\end{align*}
Observe that 
\begin{align*}
& \En[(\text{\#2})] = \En[ \phi_{\mu,i}(\etahat) + \tilde \delta_i ] \geq \En[\phi_{\mu,i}(\etahat) + (1 - D_i) \underline{\delta}_i] \\
& \E[(\text{\#3})] = \E[\beta_{0,i} \phi_{\mu,i}( \eta) + \beta_{0,i} \tilde \delta_i] \leq \E[\beta_{0,i} \phi_{\mu,i}(\eta) + \beta_{0,i} (1 - D_i) \overline \delta_i] \\
& \E[(\text{\#4})] = \E[\outcomereg(X_i) + \tilde \delta_i] \geq \E[\outcomereg(X_i) + (1 - D_i) \underline{\delta}_i].
\end{align*}
Therefore, there exists $C_1 > 0$ such that $\En[(\text{\#2})] > C_1$ for all $n$ under the assumption of bounded nuisance parameter estimates. There also exists constants $C_2 < \infty$, $C_3 > 0$ such that $\E[(\text{\#3})] < C_2$ and $\E[(\text{\#4})] > C_3$. 
Putting this together, we therefore have 
$$
\left\| \widehat{\overline{\perf}}(s; \beta, \Delta_n) - \overline{\perf}_{+}(s; \beta, \Delta) \right\| \leq 
$$
$$
C_1 \left\| \underbrace{\En[\beta_{0,i} \phi_{\mu,i}(\etahat)] - \E[\beta_{0,i} \phi_{\mu,i}(\eta)]}_{(a)} \right\| + C_1 + \left\|  \underbrace{\En[\beta_{0,i} \underline{\delta}_i^\prime] - \E[\beta_{0,i} \underline{\delta}_i^\prime]}_{(b)} +  C_1 \right\| \underbrace{\sup_{U \in \cU} \left\| (\En - \E)[\beta_{0,i} (\overline{\delta}_i^\prime - \underline{\delta}_i^\prime) U(\beta_{0,i})] \right\|}_{(c)} + 
$$
$$
C_1 \frac{C_2}{C_3} \left\| \underbrace{\En[\phi_{\mu,i}(\etahat)] - \E[\phi_{\mu,i}(\eta)]}_{(d)} \right\| + C_1 \frac{C_2}{C_3} \left\| \underbrace{\En[\underline{\delta}_i] - \E[\underline{\delta}_i]}_{(e)} \right\| + C_1 \frac{C_2}{C_3} \underbrace{\sup_{U \in \cU} \left\| (\En - \E)[ (\overline{\delta}_i^\prime - \underline{\delta}_i^\prime) U(\beta_{0,i}) \right\| }_{(f)}.
$$
We analyze each term separately. 
The proof of Proposition \ref{prop: overall performance estimator, nonparametric outcome bounds} establishes that (a), (d) are $O_\p(1 / \sqrt{n} + R_{1,n})$. Moreover, (b), (e) are $O_{\p}(1/ \sqrt{n})$ by standard arguments. 
Consider (c), which we may write out as 
$$
\sup_{U \in \cU} \left| (n/2)^{-1} \sum_{i} \beta_{0, i}  (1 - D_i) (\overline{\delta}_i - \underline{\delta}_i) U(\beta_{0,i}) - \E[\beta_{0, i}  (1 - D_i) (\overline{\delta}_i - \underline{\delta}_i) U(\beta_{0,i}) ] \right|.
$$
Define $f(w, x, y, z) = w(y-x)z$ and $\mathcal{F} = \{f_{U}\}_{U \in \cU}$ to be the class of functions $f_{U} \colon  (d, \underline{\delta}, \overline{\delta}, \beta) \rightarrow (1 - d) (\overline{\delta} - \underline{\delta}) \beta u(\beta)$.
Observe $f$ is a contraction over its final argument on $[0,1]$.
We can then rewrite (c) as
$$
\sup_{f_{U} \in \mathcal{F}} \left| (n/2)^{-1} \sum_{i} f_{U}(D_i, \underline{\delta}_i, \overline{\delta}_i, \beta_{0,i}) - \E[f_{\tilde \delta}(D_i, \underline{\delta}_i, \overline{\delta}_i, \beta_{0,i}) ] \right|.
$$
Applying a standard concentration inequality (e.g., Theorem 4.10 in \cite{Wainwright_textbook}), observe that, with probability at least $1 - \delta$,
$$
\sup_{f_{U} \in \mathcal{F}} \left| (n/2)^{-1} \sum_{i} f_{U}(D_i, \underline{\delta}_i, \overline{\delta}_i, \beta_{0,i}) - \E[f_{U}(D_i, \underline{\delta}_i, \overline{\delta}_i, \beta_{0,i}) ] \right| \leq R_n(\mathcal{F}) + \sqrt{\frac{2 \log(1/\delta)}{n/2}},
$$
where $R_n(\mathcal{F})$ is the Rademacher complexity of $\mathcal{F}$.
Now we relate $R_n(\mathcal{F})$ to $R_n(\cU)$.
For any fixed $(d_1, \underline{\delta}_1, \overline{\delta}_i, \beta_{0,1}), \hdots, (d_{n/2}, \underline{\delta}_{n/2}, \overline{\delta}_{n/2}, \beta_{0, n/2})$, observe that 
\begin{align*}
\E_{\epsilon}[ \sup_{U \in \cU} \left| \sum_{i=1}^{n/2} \epsilon_i f_{U}(d_i, \underline{\delta}_i, \overline{\delta}_i, \beta_{0,i}) \right| ] &= \E_{\epsilon}[ \sup_{U \in \cU} \left| \sum_{i=1}^{n/2} \epsilon_i \beta_{0,i} ( 1 - d_i) (\overline{\delta}_i - \underline{\delta}_i) U(\beta_{0,i}) \right| ] \\
& \leq 2 (\overline{\Gamma} - \underline{\Gamma}) \E_{\epsilon}[ \sup_{U \in \cU} \left| \sum_{i=1}^{n/2} \epsilon_i U( \beta_{0,i}) \right| ]
\end{align*}
where we used that $(1 - d_i) (\overline{\delta}_i - \underline{\delta}_i) \leq (\overline{\Gamma} - \underline{\Gamma})$ and the Ledoux-Talagrand contraction inequality (e.g., Eq. (5.61) in \cite{Wainwright_textbook}). 
Dividing by $n/2$ and averaging over the observations yields $R_n(\mathcal{F}) \leq  2 (\overline{\Gamma} - \underline{\Gamma}) R_n(\cU)$.
Finally, we can bound the Rademacher complexity of $\cU$ using Dudley's entropy integral (e.g., Theorem 5.22 in \cite{Wainwright_textbook}) as 
$$
R_n(\cU) \leq \frac{C}{\sqrt{n/2}} \int_{0}^{1} \sqrt{\log( N(\xi, \cU, \| \cdot \|_{\p_n}) ) } d\xi\leq \frac{C}{\sqrt{n/2}} \int_{0}^{1} \sqrt{\log( N_{[]}(2 \xi, \cU, \| \cdot \|_{\p_n}) ) } d\xi,
$$
for some constant $C$, where $N(\xi, \cU, \| \cdot \|_{\p_n})$ is the covering number and $N_{[]}(2 \xi, \cU, \| \cdot \|_{p_n})$ is the bracketing number.
But, Theorem 2.7.5 of \cite{VDVWellner} establishes that the bracketing entropy $\log( N_{[]}(\xi, \cU, \| \cdot \|_{\p_n} )$ of the class of monotone non-decreasing functions is bounded by $(1/\xi) \log(1/\xi)$, and so $\int_{0}^{1} \sqrt{\log( N_{[]}(\xi, \cU, \| \cdot \|_{\p_n} ) )} d\xi = \sqrt{2\pi}$. 
It follows that, for any $\delta > 0$,
$$
\sup_{U \in \cU} \left\| (\En^k - \E)[\beta_{0,i} (\overline{\delta}_i^\prime - \underline{\delta}_i^\prime) U(\beta_{0,i})] \right \| \leq 
\frac{C}{\sqrt{n/2}} + \sqrt{ \frac{2\log(1/\delta)}{2n/2} }
$$
holds with probability $1 - \delta$. We therefore conclude that (c) is $O_\p( 1/\sqrt{n} )$. 
Similarly, (f) is $O_\p(1/\sqrt{n})$ by the same argument. 
\end{proof}
\end{lemma}

Now return to Proposition \ref{proposition: partial double robustness for pos class estimator}. 
We prove a more general lemma about the estimator for the upper bound on positive class performance with estimated bounding functions $\hat{\deltamax}(\cdot), \hat{\deltamin}(\cdot)$ and then show that it implies Proposition \ref{proposition: partial double robustness for pos class estimator}.
Suppose we solve the following maximization problem in each fold
$$
\widehat{\overline{\perf}}_{+}(s; \beta, \hat{\Delta}_{n}) :=  \max_{\tilde{\delta} \in \hat{\Delta}_{n}} \frac{ \mathbb{E}_{n}[ \beta_{0,i} \phi_{\mu,i}(\hat{\eta}) + \beta_{0,i} \tilde{\delta}_i] }{\En[ \phi_{\mu,i}(\hat{\eta}) + \tilde{\delta}_i]},
$$
where $\hat{\Delta}_{n} := \left\{ \delta \in \mathbb{R}^n \colon (1 - D_i) \hat{\deltamin}(X_i) \leq \delta_i \leq (1 - D_i) \hat{\deltamax}(X_i) \mbox{ for } i = 1, \hdots, n_k \right\}$.
Observe that 
$$
\|\widehat{\overline{\perf}}_{+}(s; \beta, \hat{\Delta}_{n}) - \overline{\perf}_{+}(s; \beta, \Delta)\| \leq \| \widehat{\overline{\perf}}_{+}(s; \beta, \hat{\Delta}_{n}) - \widehat{\overline{\perf}}_{+}(s; \beta, \Delta_{n}) \| + $$
$$
\|\widehat{\overline{\perf}}_{+}(s; \beta, \Delta_{n}) - \overline{\perf}_{+}(s; \beta, \Delta) \|.
$$
Lemma \ref{lemma: convergence rate of positive class estimator} analyzed the convergence rate of the second term. 
It is therefore sufficient to bound the first term.

\begin{lemma}\label{lemma: positive class estimator, estimated bounds}
Assume the same conditions as Lemma \ref{lemma: convergence rate of positive class estimator}. % and $P(\phi_{\mu,i}(\etahat) + (1 - D_i) \hat{\deltamin}_i) > c) = 1$ for some $c > 0$. 
Then,
$$
\|\widehat{\overline{\perf}}_{+}(s; \beta, \hat \Delta_{n})  - \widehat{\overline{\perf}}_{+}(s; \beta, \Delta_{n}) \| \lesssim \sqrt{\frac{1}{n/2} \sum_{i=1}^{n/2} ( \widehat{\deltamin}_i - \deltamin_i )^2 } + \sqrt{\frac{1}{n/2} \sum_{i=1}^{n/2} (\widehat{\deltamax}_i - \deltamax_i)^2 },
$$
where $a \lesssim b$ means $a \leq C b$ for some constant $C$.

\begin{proof}
Applying the change-of-variables in the Proof of Lemma \ref{section: LP reduction for positive class bounds},
\begin{align*}
    \widehat{\overline{\perf}}_{+}(s; \beta, \Delta_{n}) := \max_{0 \leq U \leq 1} & \frac{ \sum_{i=1}^{n/2} \beta_{0,i} \phi_{\mu,i}(\etahat) + \beta_{0,i} (1 - D_i) \underline{\delta}_i + \beta_{0,i} (1 - D_i) (\overline{\delta}_i - \underline{\delta}_i) U_i }{ \sum_{i=1}^{n_{k}} \phi_{\mu,i}(\etahat) + (1 - D_i) \underline{\delta}_i + (1 - D_i) (\overline{\delta}_i - \underline{\delta}_i) U_i } 
\end{align*}
\begin{align*}
    \widehat{\overline{\perf}}_{+}(s; \beta, \hat \Delta_{n}) := \max_{0 \leq U \leq 1} & \frac{ \sum_{i=1}^{n/2} \beta_{0,i} \phi_{\mu,i}(\etahat) + \beta_{0,i} (1 - D_i) \hat{\deltamin}_i + \beta_{0,i} (1 - D_i) (\hat{\deltamax}_i - \hat{\deltamin}_i) U_i }{ \sum_{i=1}^{n/2} \phi_{\mu,i}( \etahat) + (1 - D_i) \hat{\deltamin}_i + (1 - D_i) (\hat{\deltamax}_i - \hat{\deltamin}_i) U_i }.
\end{align*}
We can therefore rewrite 
$$
\| \widehat{\overline{\perf}}_{+}(s; \beta, \hat{\Delta}_{n}) - \widehat{\overline{\perf}}_{+}(s; \beta, \Delta_{n}) \| \leq
$$
{\tiny
$$
\max_{0 \leq U \leq 1} \left\| \frac{ \sum_{i=1}^{n/2} \beta_{0,i} \phi_{\mu,i}(\etahat) + \beta_{0,i} (1 - D_i) \hat{\deltamin}_i + \beta_{0,i} (1 - D_i) (\hat{\deltamax}_i - \hat{\deltamin}_i) U_i }{ \sum_{i=1}^{n/2} \phi_{\mu,i}(\etahat) + (1 - D_i) \hat{\deltamin}_i + (1 - D_i) (\hat{\deltamax}_i - \hat{\deltamin}_i) U_i } -  \frac{ \sum_{i=1}^{n/2} \beta_{0,i} \phi_{\mu,i}( \etahat) + \beta_{0,i} (1 - D_i) \underline{\delta}_i + \beta_{0,i} (1 - D_i) (\overline{\delta}_i - \underline{\delta}_i) U_i }{ \sum_{i=1}^{n/2} \phi_{\mu,i}(\etahat) + (1 - D_i) \underline{\delta}_i + (1 - D_i) (\overline{\delta}_i - \underline{\delta}_i) U_i }  \right\| = 
$$
}
$$
\max_{0 \leq U \leq 1} \| \frac{\En[(\text{\#1})]}{\En[(\text{\#2})]} - \frac{\En[(\text{\#3})]}{\En[(\text{\#4})]} \| = 
\max_{0 \leq U \leq 1} \En[(\text{\#2})]^{-1} \left\{ \underbrace{\left( \En[(\text{\#1})] - \En[(\text{\#3})] \right)}_{(a)} - \frac{\En[(\text{\#3})]}{\En[(\text{\#4})]} \underbrace{\left( \En[(\text{\#2})] - \En[(\text{\#4})] \right)}_{(b)} \right\}.
$$
Notice that we can rewrite (a), (b) as
$$
(a) = \En[ \beta_{0,i} (1 - D_i) (\hat{\deltamin}_i - \deltamin_i) (1 - U_i) + \beta_{0,i} (1 - D_i) (\hat{\deltamax}_i - \deltamax_i) U_i ]
$$
$$
(b) = \En[ (1 - D_i) (\hat{\deltamin}_i - \deltamin_i) (1 - U_i) + (1 - D_i) (\hat{\deltamax}_i - \deltamax_i) U_i].
$$
Notice that 
$$
\En[(\text{\#2})] \geq \En[\phi_{\mu,i}( \etahat) + (1 - D_i) \underline{\delta}_i] \mbox{ for all } n,
$$
$$
\En[(\text{\#3})] \leq \En[\beta_{0,i}\phi_{\mu,i}(\etahat) + \beta_{0,i} (1 - D_i) \overline{\delta}_i] \mbox{ for all } n,
$$
$$
\En[(\text{\#4})] \leq \En[\phi_{\mu,i}( \etahat) + (1 - D_i) \underline{\delta}_i] \mbox{ for all } n.
$$
So, there exists a constant $0 < C_1$ such that $\En[(\text{\#2})] > C_1$ for all $n$ under the assumption of bounded nuisance parameter estimators and the assumption on the estimated bounds and there exists constants $0 < C_2 < \infty, C_3 > 0$ such that $\E[(\text{\#3})] < C_2$, $\E[(\text{\#4})] > C_3$ under the assumption of bounded nuisance parameter estimators.
Putting this together implies that 
{\small
$$
\| \widehat{\overline{\perf}}_{+}(s; \beta, \hat{\Delta}_{n}) - \widehat{\overline{\perf}}_{+}(s; \beta, \Delta_{n}) \| \lesssim 
$$
$$
\max_{0 \leq U \leq 1} \| \En[ \beta_{0,i} (1 - D_i) (\hat{\deltamin}_i - \deltamin_i) (1 - U_i) + \beta_{0,i} (1 - D_i) (\hat{\deltamax}_i - \deltamax_i) U_i ] \| + \| \En[ (1 - D_i) (\hat{\deltamin}_i - \deltamin_i) (1 - U_i) + (1 - D_i) (\hat{\deltamax}_i - \deltamax_i) U_i] \| \leq
$$
$$
\max_{0 \leq U \leq 1} (n/2)^{-1} \sum_{i=1}^{n/2} \| \beta_{0,i} (1 - D_i) \{ (\hat{\deltamin}_i - \deltamin_i) (1 - U_i) + (\hat{\deltamax}_i - \deltamax_i) U_i \} \| + (n/2)^{-1} \sum_{i=1}^{n/2} \| (1 - D_i) \{ (\hat{\deltamin}_i - \deltamin_i) (1 - U_i) + (\hat{\deltamax}_i - \deltamax_i) U_i \} \| \leq 
$$
$$
\max_{0 \leq U \leq 1} (n/2)^{-1} \sum_{i=1}^{n/2} \|(\hat{\deltamin}_i - \deltamin_i) (1 - U_i) + (\hat{\deltamax}_i - \deltamax_i) U_i \| + (n/2)^{-1} \sum_{i=1}^{n/2} \|(\hat{\deltamin}_i - \deltamin_i) (1 - U_i) + (\hat{\deltamax}_i - \deltamax_i) U_i \} \| \lesssim 
$$
$$
\E_{n}[|\hat{\deltamin}_i - \deltamin_i|] + \E_{n}[|\hat{\deltamax}_i - \deltamax_i |].
$$
}
Then, using the inequality $\|v\|_{1} \leq \sqrt{n} \|v\|_{2}$ for $v \in \mathbb{R}^{n}$, it follows that 
$$
\| \widehat{\overline{\perf}}_{+}(s; \beta, \hat{\Delta}_{n}) - \widehat{\overline{\perf}}_{+}(s; \beta, \Delta_{n}) \| \lesssim \sqrt{\frac{1}{n/2} \sum_{i=1}^{n/2} ( \widehat{\deltamin}_i - \deltamin_i )^2 } + \sqrt{\frac{1}{n/2} \sum_{i=1}^{n/2} (\widehat{\deltamax}_i - \deltamax_i)^2 }.
$$
\end{proof}
\end{lemma}

\noindent Putting together Lemma \ref{lemma: convergence rate of positive class estimator} and Lemma \ref{lemma: positive class estimator, estimated bounds}, the result follows. $\Box$

\subsection{Proof of Corollary \ref{cor: DR learner rate for pos class}}
To prove this result, we first observe that 
$$
\sqrt{\mathbb{E}_n\left[ \left( \widehat{\delta}(X_i) - \mu_1(X_i) \right)^2 \right]} = \underbrace{\sqrt{\mathbb{E}\left[ \left( \widehat{\delta}(X_i) - \mu_1(X_i) \right)^2 \right]}}_{:= \left\| \widehat{\delta}(\cdot) - \mu_1(\cdot) \right\|}  + O_{\p}(1/\sqrt{n})
$$
by the continuous mapping theorem, since the estimated nuisance functions and estimated bounds are constructed on separate folds. 
We then apply the same argument as Proposition \ref{prop: oracle result for learning, outcome regression bounds} to arrive at
$$
\left\| \widehat{\delta}(\cdot) - \mu_1(\cdot) \right\| \leq
\| \widehat{\delta}(\cdot) - \widehat{\delta}_{oracle}(\cdot) - \tilde{b}(\cdot) \| + \| \tilde{b}(\cdot) \| + \|\widehat{\delta}_{oracle}(\cdot) - \mu_1(\cdot)\|
$$
for $\tilde{b}(x) = \widehat{\E}_n[\hat{b}(X_i) \mid X_i = x]$ the smoothed bias and $\hat{b}(x) = \E[ \phi_{\mu,i}(\etahat) - \phi_{\mu,i}(\eta) \mid \mathcal{O}_{1}, X_i = x]$.
Under Assumption \ref{asm: L2 stability condition}, $\| \widehat{\delta}(\cdot) - \widehat{\delta}_{oracle}(\cdot) - \tilde{b}(\cdot) \| = o_\p(R_{oracle})$ by Lemma \ref{lemma: L2 oracle inequality}.
Furthermore, $\hat{b}(x)^2 = (a)^2$, where
$$
\hat{b}(x)^2 = \left\{ \frac{\pi_1(x) - \pihat_1(x)}{\pihat_1(x)} (\outcomereg(x) - \outcomereghat(x)) \right\}^2 \leq \frac{1}{\epsilon^2} \left\{ (\pi_1(x) - \pihat_1(x)) (\outcomereg(x) - \outcomereghat(x)) \right\}^2,
$$
by iterated expectations and the assumption of bounded propensity score. 
Putting this together yields
$$
\| \widehat{\delta}(\cdot) - \mu_1(\cdot) \| \leq \|\widehat{\delta}_{oracle}(\cdot) - \mu_1(\cdot) \| + \epsilon^{-1} \|\tilde{R}(\cdot)\| + o_\p(R_{oracle})
$$
as desired. $\Box$

%%%%%%%%%%%%%%%%%%%%%%%%%%%%%%%%%%
% Additional Theoretical Results %
%%%%%%%%%%%%%%%%%%%%%%%%%%%%%%%%%%
\section{Additional Theoretical Results}\label{section: appendix, additional theoretical results}

\subsection{Identification and Estimation of Predictive Disparities}\label{section: identification and estimation of predictive disparities}

A central question in algorithmic fairness is whether predictive algorithms perform differently across groups defined by sensitive attributes such as race, gender, or income. 
As noted in Remark \ref{remark: fairness} of the main text, differences in predictive performance across groups formalize violations of popular fairness criteria.
Consider a binary sensitive attribute $G_i \in \{0, 1\}$ (e.g., ethnicity, gender, race, etc.), define the \textit{overall disparity} of $s(\cdot)$ as $\disp(s; \beta) := \perf_{1}(s; \beta) - \perf_{0}(s; \beta)$ for $\perf_g(s; \beta) := \e[\beta_0(X_i, s) + \beta_1(X_i; s) Y_i^* \mid G_i = g]$ the overall performance on group $G_i = g$.  
The positive class and negative class disparities $\disp_{+}(s; \beta)$ and $\disp_{-}(s; \beta)$ are defined analogously. 
In this section, we show that predictive disparities are partially identified under Assumption \ref{asm: bounding assumption}.
We propose estimators for the bounds on predictive disparities, extending our analysis of overall and positive class performance.

\subsubsection{Overall Predictive Disparities}

Define $\mathcal{H}(\disp(s; \beta))$ to be the set of overall predictive disparities consistent with Assumption \ref{asm: bounding assumption}. Further let $\beta_{0,i}^{g} := \beta_{0,i}/P(G_i = g)$, $\beta_{1,i}^{g} := \beta_{1,i}/P(G_i = g)$ for $g \in \{0,1\}$, and $\tilde{\beta}_{0,i} := G_i \beta_{0,i}^{1} - (1 - G_i) \beta_{0,i}^{0}$, $\tilde{\beta}_{1,i} := G_i \beta_{1,i}^{1} - (1 - G_i) \beta_{1,i}^{0}$.

\begin{lemma}\label{lem: overall disp bounds under MOSM}
$$
\cH(\disp(s; \beta)) = [\underline{\disp}(s; \beta), \overline{\disp}(s; \beta)],
$$
where
\begin{align*}
& \overline{\disp}(s; \beta) := \E[ \tilde{\beta}_{0,i} + \tilde{\beta}_{1,i} \outcomereg(X_i) + \tilde{\beta}_{1,i} \pi_0(X_i) ( \overline{\nu}_{i} \overline{\delta}_{i} + \underline{\nu}_{i} \underline{\delta}_i ) ] \\
& \underline{\disp}(s; \beta) := \E[ \tilde{\beta}_{0,i} + \tilde{\beta}_{1,i} \outcomereg(X_i) + \tilde{\beta}_{1,i} \pi_0(X_i) (\overline{\nu}_{i} \underline{\delta}_{i} + \underline{\nu}_{i} \overline{\delta}_i)].
\end{align*}
for $\overline{\nu}_{i} = G_i 1\{ \beta_{1,i} \geq 0\} + (1 - G_i) 1\{\beta_{1,i} \leq 0\}$ and $\underline{\nu}_{i} = G_i 1\{\beta_{1,i} < 0\} + (1 - G_i) 1\{\beta_{i,1} > 0\}$.

\begin{proof}
For $g \in \{0, 1\}$ and $\alpha_g = P(G_i = g)$, observe that 
$$
\perf_{g}(s; \beta) =
\alpha_g^{-1} \E[\beta_{0,i} 1\{G_i = g\} + \beta_{1,i} 1\{G_i = g\} \mu_{1}(X_i) + \beta_{1,i} 1\{G_i = g\} \pi_0(X_i) \delta(X_i)].
$$
% The overall disparity $\disp(s; \beta)$ can be written as 
%$$
%\alpha_1^{-1} \E[\beta_{0,i} G_i + \beta_{1,i} G_i \mu_{1}(X_i) %+ \beta_{1,i} G_i \pi_0(X_i) \delta(X_i)] - \alpha_0^{-1} \E[\beta_{0,i} (1 - G_i) + \beta_{1,i} (1 - G_i) \mu_{1}(X_i) + \beta_{1,i} (1 - G_i) \pi_0(X_i) \delta(X_i)].
%$$
We rewrite $\disp(s; \beta)$ as 
$$
\E[ \tilde{\beta}_{0,i} + \tilde{\beta}_{1,i} \outcomereg(X_i) + \tilde{\beta}_{1,i} \pi_{0}(X_i) \delta(X_i) ]
$$
using the definitions of $\tilde{\beta}_{0,i}, \tilde{\beta}_{1,i}$. 
As in the proof of Lemma \ref{lem: bounds under MOSM}, it follows that $\cH(\disp(s; \beta))$ equals the closed interval
$$
[\tilde{\beta}_{0,i} + \tilde{\beta}_{1,i} \outcomereg(X_i) + \tilde{\beta}_{1,i} \pi_{0}(X_i) \left( 1\{\tilde{\beta}_{1,i} \geq 0 \} \underline{\delta}_{i} + 1\{\tilde{\beta}_{1,i} < 0\} \overline{\delta}_i \right), 
$$
$$
\tilde{\beta}_{0,i} + \tilde{\beta}_{1,i} \outcomereg(X_i) + \tilde{\beta}_{1,i} \pi_{0}(X_i) \left( 1\{\tilde{\beta}_{1,i} \geq 0 \} \overline{\delta}_{i} + 1\{\tilde{\beta}_{1,i} < 0\} \underline{\delta}_i \right)].
$$
The result then follows by noticing that 
$$
1\{\tilde{\beta}_{1,i} \geq 0\} = 1\{(G_i - \alpha_1) \beta_{i,1} \geq 0\} = G_i 1\{ \beta_{1,i} \geq 0\} + (1 - G_i) 1\{\beta_{1,i} \leq 0\}.
$$
$$
1\{ \tilde{\beta}_{1,i} < 0 \} = 1\{(G_i - \alpha_1) \beta_{i,1} < 0\} = G_i 1\{\beta_{1,i} < 0\} + (1 - G_i) 1\{\beta_{i,1} > 0\}.
$$
\end{proof}
\end{lemma}

Since the bounds on overall disparities are linear functionals of known functions of the data and identified nuisance parameters, we construct estimators as in Section \ref{section: estimating overall performance bounds} of the main text. 
For simplicity, we develop the estimators assuming $\p(G_i = 1)$ is known and for observed outcome bounds, but they can be easily extended to the case where this is estimated and for alternative choices of bounding functions. 

\vspace{-1em}
\paragraph{Estimation Procedure for Observed Outcome Bounds:} We randomly split the data into $K$ disjoint folds.
For each fold $k$, we construct estimators of the nuisance functions $\etahat_{-k} = \left( \pihat_{1,-k}, \hat{\mu}_{1,-k} \right)$ using only the sample of observations not in the $k$-th fold.
For each observation in the $k$-th fold, we construct
\begin{align}
& \overline{\disp}_{i}(\etahat_{-k}) := \tilde{\beta}_{0,i} + \tilde{\beta}_{1,i} \phi_{\mu, i}(\etahat_{-k}) + \tilde{\beta}_{1,i} (\overline{\nu}_{i} (\overline{\Gamma} - 1) + \underline{\nu}_i (\underline{\Gamma} - 1) ) \phi_{\pi \mu, i}(\etahat_{-k}), \\
& \underline{\disp}_{i}(\etahat_{-k}) := \tilde{\beta}_{0,i} + \tilde{\beta}_{1,i} \phi_{\mu, i}(\etahat_{-k}) + \tilde{\beta}_{1,i} (\overline{\nu}_{i} (\underline{\Gamma} - 1) + \underline{\nu}_i (\overline{\Gamma} - 1)) \phi_{\pi \mu, i}(\etahat_{-k}).
\end{align}
We then estimate the bounds on overall disparities by $\widehat{\overline{\disp}}(s; \beta) := \En[\overline{\disp}_{i}(\etahat_{-k})]$ and $\widehat{\underline{\disp}}(s; \beta) := \En[\underline{\disp}_{i}(\etahat_{-k})]$.
As in Appendix \ref{section: variance estimation for overall performance bounds}, we estimate the asymptotic covariance matrix as 
\begin{align*}
n^{-1} \sum_{i=1}^{n} \begin{pmatrix} \hat{\sigma}_{i,11} & \hat{\sigma}_{i,12} \\ \hat{\sigma}_{i,12} & \hat{\sigma}_{i,22} \end{pmatrix}
\end{align*}
for $\hat{\sigma}_{i,11} = ( \overline{\disp}(O_i; \etahat_{-K(i)}) -  \widehat{\overline{\disp}}(s; \beta) )^2$, $\hat{\sigma}_{i,12} = ( \overline{\disp}(O_i; \etahat_{-K(i)}) -  \widehat{\overline{\disp}}(s; \beta) ) ( \underline{\disp}(O_i; \etahat_{-K(i)}) -  \widehat{\underline{\disp}}(s; \beta) )$, and $\hat{\sigma}_{i,22} = ( \underline{\disp}(O_i; \etahat_{-K(i)}) -  \widehat{\underline{\disp}}(s; \beta) )^2$.
Under regularity conditions and the same arguments as the proof of Proposition \ref{prop: overall performance estimator, nonparametric outcome bounds}, we can derive the rate of convergence of our proposed estimators and provide conditions under which they are jointly asymptotically normal.

\subsubsection{Positive Class Predictive Disparities}

We provide non-sharp bounds for the positive class disparity since the positive class disparity is the difference of two linear-fractional functions.
Define $\mathcal{H}(\disp_{+}(s; \beta))$ to be the set of all positive class disparities that are consistent with Assumption \ref{asm: bounding assumption}.

\begin{lemma}\label{lem: positive class disp bounds under MOSM}
\begin{align*}
& \cH(\disp_{+}(s; \beta)) \subseteq [ \underline{\disp}_{+}(s;\beta), \overline{\disp}_{+}(s; \beta)],
\end{align*}
where $\overline{\disp}_{+}(s; \beta) = \overline{\perf}_{+,1}(s; \beta) - \underline{\perf}_{+,0}(s, \beta)$, $\underline{\disp}_{+}(s; \beta) = \underline{\perf}_{+,1}(s; \beta) - \overline{\perf}_{+,0}(s, \beta)$ for, $g \in \{0,1\}$,
\begin{align*}
    & \overline{\perf}_{+,g}(s; \beta) = \sup_{\delta \in \Delta} \frac{ \E[\beta_{0,i} \mu_1(X_i) + \beta_{0,i} \pi_0(X_i) \delta(X_i) \mid G_i = g] }{ \E[ \mu_1(X_i) + \pi_0(X_i) \delta(X_i) \mid G_i = g] }, \\
    & \underline{\perf}_{+,g}(s; \beta) = \inf_{\delta \in \Delta} \frac{ \E[\beta_{0,i} \mu_1(X_i) + \beta_{0,i} \pi_0(X_i) \delta(X_i) \mid G_i = g] }{ \E[ \mu_1(X_i) + \pi_0(X_i) \delta(X_i) \mid G_i = g] }.
\end{align*}

\begin{proof}
Observe that, for $g \in \{0,1\}$ and $\delta(\cdot) \in \Delta$,
$$
\perf_{+,g}(s; \beta, \delta) = \frac{ \E[\beta_{0,i} \mu_1(X_i) + \beta_{0,i} \pi_0(X_i) \delta(X_i) \mid G_i = g] }{ \E[ \mu_1(X_i) + \pi_0(X_i) \delta(X_i) \mid G_i = g] },
$$
and so the positive class predictive disparity $\disp_{+}(s; \beta)$ can be written as $\disp_{+}(s; \beta, \delta) = \perf_{+,1}(s; \beta, \delta) - \perf_{+,0}(s; \beta, \delta)$. 
The result then follows since
$$
\sup_{\delta \in \Delta} \disp_{+}(s; \beta, \delta) \leq \sup_{\delta \in \Delta} \perf_{+,1}(s; \beta, \delta) - \inf_{\delta \in \Delta} \perf_{+,0}(s; \beta, \delta)
$$
and 
$$
\inf_{\delta \in \Delta} \disp_{+}(s; \beta, \delta) \geq \inf_{\delta \in \Delta} \perf_{+,1}(s; \beta, \delta) - \sup_{\delta \in \Delta} \perf_{+,0}(s; \beta, \delta).
$$
\end{proof}
\end{lemma}

We can therefore construct estimators for the non-sharp bounds on positive class disparities by directly applying our estimator in Section \ref{section: estimating pos class bounds} of the main text.
We separately estimate the bounds on group-specific positive class performance $\overline{\perf}_{+,g}(s; \beta)$ and $\underline{\perf}_{+,g}(s; \beta)$ and then take the appropriate difference. The same arguments as the proof of Proposition \ref{proposition: partial double robustness for pos class estimator} can be applied to show that this estimator inherits that a partial double robustness property. 

\subsection{Variance Estimation for Overall Performance Bounds}\label{section: variance estimation for overall performance bounds}

As mentioned in Section \ref{section: estimating overall performance bounds} of the main text, we develop a consistent estimator of the asymptotic covariance matrix of the estimators for the bounds on overall performance under observed outcome bounds. 
Recall from Proposition \ref{prop: overall performance estimator, nonparametric outcome bounds}, if $R_{1,n}^{k} = o_{\p}(1/\sqrt{n})$ for all folds $k$, then 
\begin{equation*}
\sqrt{n} \left( \begin{pmatrix} \widehat{\overline{\perf}}(s; \beta) \\ \widehat{\underline{\perf}}(s; \beta) \end{pmatrix} - \begin{pmatrix} \overline{\perf}(s; \beta) \\ \underline{\perf}(s; \beta) \end{pmatrix} \right) \xrightarrow{d} N\left( 0, \Sigma \right),
\end{equation*}
where $\Sigma = Cov\left( (\overline{\perf}_i, \underline{\perf}_i )^\prime \right)$ for 
\begin{align*}
& \overline{\perf}_{i}(\eta) = \beta_{0,i} + \beta_{1,i} \phi_{\mu,i}(\eta) + \beta_{1,i} \left( 1\{\beta_{1,i} > 0\} (\overline{\Gamma} - 1) + 1\{\beta_{1,i} \leq 0\} (\underline{\Gamma} - 1) \right) \phi_{\pi \mu,i}(\eta) \\
& \underline{\perf}_i(\eta) = \beta_{0,i} + \beta_{1,i} \phi_{\mu,i}(\eta) + \beta_{1,i} \left( 1\{\beta_{1,i} > 0\} (\underline{\Gamma} - 1) + 1\{\beta_{1,i} \leq 0\} (\overline{\Gamma} - 1) \right) \phi_{\pi \mu,i}(\eta)
\end{align*}
and  $\E[\overline{\perf}_i] = \overline{\perf}(s; \beta)$, $\E[\underline{\perf}_i] = \underline{\perf}(s; \beta)$. Consider the estimator 
$$
\widehat{\Sigma} = \frac{1}{n} \sum_{i=1}^{n} \begin{pmatrix} \widehat{\overline{\perf}}_i(\etahat_{-K_i}) - \widehat{\overline{\perf}}(s; \beta) \\ \widehat{\underline{\perf}}_i(\etahat_{-K_i}) - \widehat{\underline{\perf}}(s; \beta) \end{pmatrix} \begin{pmatrix} \widehat{\overline{\perf}}_{i}(\etahat_{-K_i}) - \widehat{\overline{\perf}}(s; \beta) \\ \widehat{\underline{\perf}}_{i}(\etahat_{-K_i}) - \widehat{\underline{\perf}}(s; \beta) \end{pmatrix}^\prime.
$$
To show $\widehat{\Sigma} \xrightarrow{p} \Sigma$, it suffices to show convergence in probability for each entry. We prove this directly by extending Lemma 1 in \cite{DornGuoKallus(21)}.

\begin{lemma}\label{lemma: convergence of pairs of functions}
Let $\phi_1, \phi_2$ be any two square integrable functions.
Let $\hat{\phi}_{1,n} = \left( \hat{\phi}_{1}(O_i), \hdots, \hat{\phi}_{1}(O_n) \right), \hat{\phi}_{2,n} = \left( \hat{\phi}_{2}(O_i), \hdots, \hat{\phi}_{2}(O_n) \right)$ be random vectors satisfying 
\begin{align*}
& \|\hat{\phi}_{1,n} - \phi_{1,n}\|_{L_2(\p_n)} := \sqrt{ n^{-1} \sum_{i=1}^{n} (\hat{\phi}_{1}(O_i) - \phi_1(O_i))^2} = o_\p(1), \\
& \|\hat{\phi}_{2,n} - \phi_{2,n}\|_{L_2(\p_n)} := \sqrt{ n^{-1} \sum_{i=1}^{n} ( \hat{\phi}_{2}(O_i) - \phi_2(O_i) )^2 } = o_\p(1),
\end{align*}
where $\phi_{1,n} = \left( \phi_{1}(O_i), \hdots, \phi_{1}(O_n) \right)$ and $\phi_{2,n} = \left( \phi_{2}(O_i), \hdots, \phi_{2}(O_n) \right)$.
Define $\p_n$ to be the empirical distribution.
Then, the second moments of $\p_n$ converge in probability to the respective second moments of $\left( \phi_1(O_i), \phi_2(O_i) \right) \sim P$

\begin{proof}
Let $\hat{\phi}_{i,1} = \hat{\phi}_{i}(O_i)$ and define $\phi_{i,1}, \hat{\phi}_{i,1}, \phi_{i,2}$ analogously. 
To prove this result, we first show that $\En[\hat{\phi}_{1,i}^2] = \E[\phi_{1,i}^2] + o_\p(1)$ since the same argument applies for $\phi_{2,i}$.
Observe that 
$$
n^{-1} \sum_{i=1}^{n} \hat{\phi}_{1,i} - \E[\phi_{i,1}^2] = n^{-1} \sumoveri (\hat{\phi}_{i,1}^2 - \phi_{i,1}^{2}) + (\En - \E)[\phi_{i,1}^2],
$$
where $(\En - \E)[\phi_{i,1}^2] = o_\p(1)$. Furthermore, we can rewrite the first term as
$$
n^{-1} \sumoveri (\hat{\phi}_{i,1}^2 - \phi_{i,1}^{2}) = n^{-1} \sumoveri \left( \hat{\phi}_{i,1} - \phi_{i,1} \right) \left( \hat{\phi}_{i,1} + \phi_{i,1} \right) = 
$$
$$
n^{-1} \sumoveri \left( \hat{\phi}_{i,1} - \phi_{i,1} \right) \left( \hat{\phi}_{i,1} - \phi_{i,1} + 2\phi_{i,1} \right) \leq \| \hat{\phi}_{i,n} - \phi_{i,1} \| \left( \| \hat{\phi}_{i,1} - \phi_{i,1} \| + 2\| \phi_{i,1}\| \right) = o_\p(1),
$$
where the last inequality applies the Cauchy-Schwarz inequality and triangle inequality.
We next show that $\En[\hat{\phi}_{i,1} \hat{\phi}_{i,2}] = \E[\phi_{i,1} \phi_{i,2}] + o_\p(1)$.
Observe that 
$$
n^{-1} \sumoveri \hat{\phi}_{i,1} \hat{\phi}_{i,2} - \E[\phi_{i,1} \phi_{i,2}] = n^{-1} \sumoveri \left( \hat{\phi}_{i,1} \hat{\phi}_{i,2} - \phi_{i,1} \phi_{i,2} \right) + (\En - \E)[ \phi_{i,1} \phi_{i,2} ],
$$
where $(\En - \E)[ \phi_{i,1} \phi_{i,2} ] = o_\p(1)$. 
We can further rewrite the first term as 
$$
n^{-1} \sumoveri \left( \hat{\phi}_{i,1} \hat{\phi}_{i,2} - \phi_{i,1} \phi_{i,2} \right) = n^{-1} \sumoveri \left( \hat{\phi}_{i,1} (\hat{\phi}_{i,2} - \phi_{i,2}) + \phi_{i,2} (\hat{\phi}_{i,1} - \phi_{i,1}) \right) = 
$$
$$
n^{-1} \sumoveri \phi_{i,1} (\hat{\phi}_{i,2} - \phi_{i,2}) + n^{-1} \sumoveri (\hat{\phi}_{i,1} - \phi_{i,1}) (\hat{\phi}_{i,2} - \phi_{i,2}) + n^{-1} \sumoveri \phi_{i,2} (\hat{\phi}_{i,1} - \phi_{i,1}) \leq
$$
$$
\|\phi_{1, n}\| \|\hat{\phi}_{2,n} - \phi_{2,n}\| + \| \hat{\phi}_{1,n} - \phi_{1,n} \| \| \hat{\phi}_{2,n} - \phi_{2,n} \| + \| \phi_{2,n} \| \|\hat{\phi}_{1,n} - \phi_{1,n} \| = o_\p(1),
$$
where the last inequality applies Cauchy-Schwarz inequality. 
\end{proof}
\end{lemma}

We show that the conditions of Lemma \ref{lemma: convergence of pairs of functions} are satisfied for $\widehat{\overline{\perf}}_{i}(\etahat_{-K_i})$ and $\widehat{\underline{\perf}}_{i}(\etahat_{-K_i})$. 
The convergence of $\hat{\Sigma}$ follows by the continuous mapping theorem since we already established the convergence of the first moments in Proposition \ref{prop: overall performance estimator, nonparametric outcome bounds}. 

\begin{lemma}\label{lem: empirical l2 convergence of overall performance}
Under the same assumptions as Proposition \ref{prop: overall performance estimator, nonparametric outcome bounds}, for each fold $k$,
\begin{align*}
& \| \widehat{\overline{\perf}}_i(\etahat_{-k}) - \overline{\perf}_i(\eta) \|_{L_2(\p_{n}^k)} = o_\p(1) \\
& \| \widehat{\underline{\perf}}_i(\etahat_{-k}) - \underline{\perf}_i(\eta) \|_{L_2(\p_{n}^k)} = o_\p(1) 
\end{align*}
conditionally on $\mathcal{O}_{-k}$.

\begin{proof}
    We prove the result for $\widehat{\overline{\perf}}_i(\etahat_{-k})$ since the analogous argument applies for $\widehat{\underline{\perf}}_i(\etahat_{-k})$.
    Following the proof of Lemma \ref{lemma: influence function bound} and Lemma \ref{lemma: convergence of pi0mu1 IF}, we observe that 
    $$
    \| \widehat{\overline{\perf}}_i(\etahat_{-k}) - \overline{\perf}_i(\eta) \|_{L_2(\p_{n}^k)} \leq (a) + (b)
    $$
    for
    $$
    (a) = M ( \frac{\|D_i\|_{L_2(\p_{n}^k)}}{\delta} \frac{\|\pi_1 - \pihat_1\|_{L_2(\p_{n}^k)}}{\epsilon} \|Y_i - \outcomereg(X_i) \|_{L_2(\p_{n}^k)} + $$ 
    $$
    \frac{\|D_i\|_{L_2(\p_{n}^k)}}{\delta} \frac{\|\pi_1 - \pihat_1\|_{L_2(\p_{n}^k)}}{\epsilon} \|\outcomereghat - \outcomereg\|_{L_2(\p_{n}^k)} + 
    \frac{\|D_i\|_{L_2(\p_{n}^k)}}{\delta} \|\outcomereghat - \outcomereg\|_{L_2(\p_{n}^k)} + \left\|\outcomereghat - \outcomereg \right\|_{L_2(\p_{n}^k)} )
    $$
    and 
    $$
    (b) = \tilde{M} ( \frac{1}{\epsilon \delta} \| \pihat_1 - \pi_1 \|_{L_2(\p_{n}^k)} \|Y_i - \outcomereg\|_{L_2(\p_{n}^k)} + \frac{1}{\epsilon \delta} \|\pihat_1 - \pi_1\|_{L_2(\p_{n}^k)} \|\outcomereg - \outcomereghat\|_{L_2(\p_{n}^k)} + \frac{1-\delta}{\delta} \|\outcomereg - \outcomereghat\|_{L_2(\p_{n}^k)} + 
    $$
    $$
    \|(1-D_i)-\pi_0\|_{L_2(\p_{n}^k)} \|\outcomereghat -\outcomereg\|_{L_2(\p_{n}^k)} + \|\pihat_1 - \pi_1\|_{L_2(\p_{n}^k)} \|\outcomereghat - \outcomereg\|_{L_2(\p_{n}^k)} + \|\pihat_1 - \pi_1\|_{L_2(\p_{n}^k)} \|\outcomereg\|_{L_2(\p_{n}^k)} + 
    $$
    $$
    (1-\epsilon) \|\outcomereghat - \outcomereg\|_{L_2(\p_{n}^k)} + \|\outcomereg\|_{L_2(\p_{n}^k)} \|\pihat_1 - \pi_1\|_{L_2(\p_{n}^k)})
    $$
    and constants $M, \tilde{M}$. However, by Markov's Inequality, 
    $$
    (a) = \mathcal{O}_{\p}( \|\pi_1 - \pihat_1\|_{L_2(\p)} +  \|\pi_1 - \pihat_1\|_{L_2(\p)} \|\outcomereghat - \outcomereg\|_{L_2(\p)} + \|\outcomereghat - \outcomereg\|_{L_2(\p)})
    $$
    and 
    $$
    (b) = \mathcal{O}_{\p}( \|\pi_1 - \pihat_1\|_{L_2(\p)} +  \|\pi_1 - \pihat_1\|_{L_2(\p)} \|\outcomereghat - \outcomereg\|_{L_2(\p)} + \|\outcomereghat - \outcomereg\|_{L_2(\p)}).
    $$
    The result is then immediate.
\end{proof}
\end{lemma}

\noindent In our Monte Carlo simulations and empirical application, we use this formula for calculating standard errors. 
In the proof of Proposition \ref{prop: overall performance estimator, nonparametric outcome bounds}, we show that our estimator is asymptotically equivalent to the centered average of the true bounds, so bootstrap-based inference would also be valid.

\subsection{Linear Programming Reduction for Bounds on Positive Class Performance}\label{section: LP reduction for positive class bounds}

As mentioned in Section \ref{section: estimating pos class bounds} of the main text, the optimization program defining $\widehat{\overline{\perf}}_{+}(s; \beta)$ can be characterized as a linear program by applying the Charnes-Cooper transformation \citep[][]{CharnesCooper(62)}. 
Lemma \ref{lemma: LFP to LP reduction for positive class performance} states this reduction for arbitrary bounding functions $\underline{\delta}_i := \underline{\delta}(X_i; \eta)$, $\overline{\delta}_i := \overline{\delta}(X_i; \eta)$ and for any fold of observations.

\begin{lemma}\label{lemma: LFP to LP reduction for positive class performance}
Let $n^k$ denote the number of observations in any fold $k$ and $\E_{n}^{k}[\cdot]$ the sample average over all observations in the $k$-th fold.
Define $\widehat{c}^{k} = \e_n^k[ \beta_{0,i} \phi_{\mu,i}(\etahat) + \beta_{0,i} (1 - D_i) \widehat{\underline{\delta}}(X_i)]$, $\hat{d}^{k} = \En^k[(\phi_{\mu,i}(\etahat) + (1 - D_i) \widehat{\underline{\delta}}(X_i)]$, $\widehat{\alpha}_i = n_k^{-1} \beta_{0,i} (1 - D_i) (\widehat{\overline{\delta}}(X_i) - \widehat{\underline{\delta}}(X_i))$, $\widehat{\gamma}_i = n_k^{-1} (1 - D_i) (\widehat{\overline{\delta}}(X_i) - \widehat{\underline{\delta}}(X_i))$, and $\widehat{\alpha} = (\hat{\alpha}_1, \hdots, \hat{\alpha}_n)$, $\widehat{\gamma} = (\hat{\gamma}_1, \hdots, \hat{\gamma}_n)$. 
Then, 
\begin{align*}
\widehat{\overline{\perf}}_{+}^{k}(s; \beta) = \max_{\tilde U \in \mathbb{R}^{n^k}, \tilde V \in \mathbb{R}} & \hat{\alpha}^\prime \tilde{U} + \hat{c}^k \tilde{V} \\
\mbox{ s.t. } & 0 \leq \tilde{U}_i \leq \tilde{V} \mbox{ for } i = 1, \hdots n_k, \\
& 0 \leq \tilde{V}, \mbox{ } \hat{\gamma}^\prime \tilde{U} + \tilde{V} \hat{d}^k = 1.
\end{align*}
$\widehat{\underline{\perf}}_{+}^{k}(s; \beta, \Delta_n)$ is optimal value of the corresponding minimization problem.

\begin{proof}
We first use the change-of-variables $\delta(X_i) = \widehat{\underline{\delta}}(X_i) + (\widehat{\overline{\delta}}(X_i) - \widehat{\underline{\delta}}(X_i)) U_i$ for $U_i \in [0,1]$ to rewrite $\widehat{\overline{\perf}}_{+}^{k}(s; \beta)$ as
\begin{align*}
    \widehat{\overline{\perf}}_{+}^{k}(s; \beta) := \max_{U} & \frac{ \En^k[ \beta_{0,i} \phi_{\mu,i}(\etahat) + \beta_{0,i} (1 - D_i) \widehat{\underline{\delta}}(X_i) + \beta_{0,i} (1 - D_i) (\widehat{\overline{\delta}}(X_i) - \widehat{\underline{\delta}}(X_i)) U_i ] }{ \En^k[ \phi_{\mu,i}(\etahat) + (1 - D_i) \widehat{\underline{\delta}}(X_i) + (1 - D_i) (\widehat{\overline{\delta}}(X_i) - \widehat{\underline{\delta}}(X_i)) U_i ] }  & \\
    \mbox{ s.t. } & 0 \leq U_i \leq 1 \mbox{ for } i = 1, \hdots, n_k,
\end{align*}
where $U = \left( U_1, \hdots, U_n \right)^\prime$. 

Next, define $\hat{c}^k = \En^k[ \beta_{0,i} \phi_{\mu,i}(\etahat) + \beta_{0,i} (1 - D_i) \widehat{\underline{\delta}}(X_i) ]$, $\hat{d} = \En^k[ \phi_{\mu,i}(\etahat) + (1 - D_i) \widehat{\underline{\delta}}(X_i) ]$, $\hat \alpha_i := n_k^{-1} \beta_{0,i} (1 - D_i) (\overline{\delta}(X_i) - \widehat{\underline{\delta}}(X_i))$, $\hat \gamma_i := n_k^{-1} (1 - D_i) (\widehat{\overline{\delta}}(X_i) - \widehat{\underline{\delta}}(X_i))$. We can further rewrite the estimator as 
$$
\widehat{\overline{\perf}}_{+}^k(s; \beta) = \max_{U} \frac{\hat{\alpha}^\prime U + \hat{c}^k}{\hat{\gamma}^\prime U + \hat{d}^k} \mbox{ s.t. } 0 \leq U_i \leq 1 \mbox{ for } i = 1, \hdots, n_k,
$$
where $\hat \alpha = \left( \hat \alpha_1, \hdots, \hat \alpha_n \right)^\prime$, $\hat \gamma = \left( \hat \gamma_1, \hdots, \hat \gamma_n \right)^\prime$. 
Finally, applying the Charnes-Cooper transformation with $\tilde{U} = \frac{U}{\hat{\gamma}^\prime U + \hat{d}^k}$, $\tilde{V} = \frac{1}{\hat{\gamma}^\prime U + \hat{d}^k}$, this linear-fractional program is equivalent to the linear program
\begin{align*}
   \widehat{\overline{\perf}}_{+}^{k}(s; \beta) = \max_{\tilde U, \tilde V} \, & \hat{\alpha}^\prime \tilde{U} + \hat{c}^k \tilde{V} \\
   \mbox{ s.t. } & 0 \leq \tilde{U}_i \leq \tilde{V} \mbox{ for } i = 1, \hdots n_k, \\
   & 0 \leq \tilde{V}, \mbox{ } \hat{\gamma}^\prime \tilde{U} + \tilde{V} \hat{d}^k = 1.
\end{align*}
\end{proof}
\end{lemma}

%%%%%%%%%%%%%%%%%%%%%%%%%%%%%%%%%%%%%%%
% Estimation under Alternative Bounds %
%%%%%%%%%%%%%%%%%%%%%%%%%%%%%%%%%%%%%%%
\section{Estimation under Alternative Bounding Functions}\label{section: alternative bounding functions}

In the main text, we developed estimation procedures for observed outcome bounds (Section \ref{section: observed outcome bounds}). 
In this appendix, we extend our estimation framework to two additional classes of bounding functions: proxy outcome bounds (Section \ref{section: proxy variable bounds}) and instrumental variable bounds (Section \ref{section: instrumental variable bounds}). 
Both approaches leverage additional data --- proxy outcomes or quasi-experimental variation --- to sharpen inference about unobserved confounding.
While this complicates notation, we nonetheless can extend our estimators for the bounds on the conditional likelihood and predictive performance measures.

\subsection{Estimation under Proxy Outcome Bounds}\label{section: proxy outcome bounds}

Suppose the proxy outcome $\tilde{Y}_i \in \{0, 1\}$ is always observed (for both selected and unselected units) and statistically related to the true outcome $Y_i^*$. 
We assume that the proxy outcome is equally informative about the true outcome for selected and unselected units: that is, for all for all $x \in \cX$,
\begin{equation}\label{equation: proxy outcome assumption}
    \p(Y_i^* = \tilde{Y}_i \mid D_i = 0, X_i = x) = \gamma_1(x).
\end{equation}

\noindent Equation \ref{equation: proxy outcome assumption} states that the probability the true and proxy outcomes agree is the same regardless of selection status, conditional on covariates.  
In consumer lending, this says that default on credit cards is equally predictive of default on personal loans for both funded and unfunded applicants with the same credit profile.

Our next result characterizes the sharp bounds on the confounding function $\delta(x)$ under Equation \ref{equation: proxy outcome assumption}.

\begin{lemma}\label{lemma: proxy outcome bounds, general case}
Under Equation \eqref{equation: proxy outcome assumption}, the confounding function satisfies, for all $x \in \cX$, 
$$
    \underline{\delta}(x; \eta) \leq \delta(x) \leq \overline{\delta}(x; \eta)
$$
for $\underline{\delta}(x; \eta) = | 1 - \gamma_1(x) - \tilde{\mu}_0(x) | - \mu_1(x)$ and $\overline{\delta}(x; \eta) = 1 - | \gamma_1(x) - \tilde{\mu}_0(x)| - \mu_1(x)$.

\begin{proof}
We consider a particular $x \in \cX$ and drop the dependence on $x$ in our notation. Let $p(y^*, \tilde{y}) = \p(Y_i^* = y^*, \tilde{Y}_i = \tilde{y} \mid D_i = 0)$ for all $(y^*, \tilde{y}) \in \{0, 1\}^2$. 
Observe that $\p(Y_i^* = 1 \mid D_i = 0) = p(1,1) + p(1,0)$ and $\p(\tilde{Y}_i^* = 1 \mid D_i = 0) = p(1,1) + p(0,1)$. 
The upper bound on $\p(Y_i^* = 1 \mid D_i = 0)$ can be expressed as the optimal value of the following linear program
\begin{align*}
    \max_{p(0,0), p(0,1), p(1,0), p(1,1)} & \, p(1,1) + p(1,0) \\
    \mbox{ s.t. } & p(1,1) + p(0,0) = \gamma_1, \\
    & p(1, 1) + p(0,1) = \tilde{\mu}_0, \\
    & p(0,0) + p(0, 1) + p(1, 0) + p(1,1) = 1, \\
    & p(0,0), p(0,1), p(1,0), p(1,1) \geq 0.
\end{align*}
The constraints $p(1, 1) + p(0,1) = \tilde{\mu}_0$ and $p(1,1) + p(0,0) = \gamma_1$ imply $p(0,1) = \tilde{\mu}_0 - p(1,1)$ and $p(0,0) = \gamma_1 - p(1,1)$  respectively. 
The non-negativity constraints then imply $0 \leq p(1,1) \leq \min\{ \gamma_1, \tilde{\mu}_0 \}$.
Substituting in, the constraint $p(0,0) + p(0, 1) + p(1, 0) + p(1,1) = 1$ then implies 
$$
    p(1,0) = 1 - \gamma_1 - \tilde{\mu}_0 + p(1,1),
$$
and the non-negativity constraint implies $\max\{0, \gamma_1 + \tilde{\mu}_0 - 1 \} \leq p(1,1)$. We can therefore rewrite the linear program as
\begin{align*}
    \max_{p(1,1)} & \, 2p(1,1) + (1 - \gamma_1 - \tilde{\mu}_0) \\
    \mbox{ s.t. } & \max\{0, \gamma_1 + \tilde{\mu}_0 - 1 \} \leq p(1,1) \leq \min\{ \gamma_1, \tilde{\mu}_0 \}.
\end{align*}
The upper bound on $\p(Y_i^* = 1 \mid D_i = 0)$ is therefore 
$$
2 \min\{ \gamma_1, \tilde{\mu}_0 \} + (1 - \gamma_1 - \tilde{\mu}_0) = \min \{ 1 + \gamma_1 - \tilde{\mu}_0, 1 - \gamma_1 + \tilde{\mu}_1 \} = 1 - | \gamma_1 - \tilde{\mu}_0|.
$$
Analogously, the lower bound is 
$$
2 \max\{0, \gamma_1 + \tilde{\mu}_1 - 1\} + (1 - \gamma_1 - \tilde{\mu}_0) = \max\{ 1 - \gamma_1 - \tilde{\mu}_0, \tilde{\mu}_0 + \gamma_1 - 1 \} = | 1 - \gamma_1 - \tilde{\mu}_0|.
$$
The result is then immediate.
\end{proof}
\end{lemma}

In many applications, the proxy satisfies additional restrictions that simplify the bounding functions. 
The bounding functions stated in Equation \ref{eqn: proxy assumption} of the main text immediately follow from Lemma \ref{lemma: proxy outcome bounds, general case} if further $\tilde{\mu}_0(x) \leq \gamma_1(x)$ and $\tilde{\mu}_0(x) + \gamma_1(x) \leq 1$ for all $x \in \cX$. 
These restrictions are satisfied by the proxy outcomes used in \cite{BlattnerNelson(21)} (see their Table 3) and \cite{MullainathanObermeyer(21)} (see their Table 2) on average over the covariates. 

For the remainder of this section, we extend our estimators for the conditional probability, positive class performance, and overall performance to the simplified bounding functions based on a proxy outcome $\tilde{Y}_i$ satisfying $\tilde{\mu}_0(x) \leq \gamma_1(x)$ and $\tilde{\mu}_0(x) + \gamma_1(x) \leq 1$ for all $x \in \cX$. The simplified bounding functions are:
\begin{align*}
    & \underline{\delta}(x; \eta) = 1 - \gamma_1(x) - \tilde{\mu}_0(x) - \mu_1(x), \\ 
    & \overline{\delta}(x; \eta) = 1 - \gamma_1(x) + \tilde{\mu}_0(x) - \mu_1(x).
\end{align*}
We observe that  
\begin{align*}
& \phi_{\tilde{\mu},i}(\eta) = \tilde{\mu}_0(X_i) + \frac{1 - D_i}{1 - \pi_1(X_i)} (\tilde{Y}_i -  \tilde{\mu}_0(X_i)), \\
& \phi_{\gamma,i}(\eta) = \gamma_1(X_i) - \frac{D_i}{\pi_1(X_i)} ( Y_i \tilde{Y}_i + (1 - Y_i) (1 - \tilde{Y}_i) - \gamma_1(X_i)) \\
& \phi_{\pi \tilde{\mu}, i}(\eta) = \pi_0(X_i) \tilde{\mu}_0(X_i) + ((1 - D_i) - \pi_0(X_i)) \tilde{\mu}_0(X_i) +  \frac{1 - D_i}{1 - \pi_1(X_i)} (\tilde{Y}_i -  \tilde{\mu}_0(X_i)) \pi_0(X_i), \\
& \phi_{\pi \gamma, i}(\eta) = \pi_0(X_i) \gamma_1(X_i) + ((1 - D_i) - \pi_0(X_i)) \gamma_1(X_i) + \frac{D_i}{\pi_1(X_i)} ( Y_i \tilde{Y}_i + (1 - Y_i) (1 - \tilde{Y}_i) - \gamma_1(X_i)) \pi_0(X_i)
\end{align*}
are influence functions for $\e[\tilde{\mu}_0(X_i)], \e[\gamma_1(X_i)]$, $\e[\pi_i(X_i) \tilde{\mu}_0(X_i)]$, and $\e[\pi_0(X_i) \gamma_1(X_i)]$ respectively by standard calculations \citep[e.g.,][]{Kennedy(22)-IFreview, Hines_2022}.

To extend our estimators to the general proxy outcome bounds in Lemma \ref{lemma: proxy outcome bounds, general case}, we use smooth approximations to the absolute value functions. 
The lower bounding function can also be written as $\max\{1 - \gamma_1(x) - \tilde{\mu}_0(x), \tilde{\mu}_0(x) + \gamma_1(x) - 1 \} - \mu_1(x)$ and the upper bounding function as $\min\{ 1- \gamma_1(x) + \tilde{\mu}_0(x), 1 + \gamma_1(x) - \tilde{\mu}_0(x)\}$. 
We apply the log-sum-exponential function to approximate these pointwise maximum and minimum operations as we discuss below in Appendix \ref{section: log-sum-exp bounds for IV}.
Alternatively, for the overall performance estimator, we could invoke a margin condition as discussed in \citet[][]{LevisEtAl(23)-CovariateAssistedIV, semenova2023adaptive}.

\subsubsection{Bounds on the Conditional Likelihood}\label{section: proxy outcome, pseudo outcome regression}

We extend our nonparametric regression estimator for the proxy outcome bounds on the conditional probability. 
The construction is identical to Section \ref{section: DR Learner, main text}, except we modify the pseudo-outcomes that are constructed in the second-stage. 
We again illustrate the procedure using two folds for simplicity. 

\vspace{-1em}
\paragraph{Estimation Procedure for Proxy Outcome Bounds:} 
Split the data into two folds.
Using the first fold, we construct estimates $\etahat = \left( \hat{\pi}_1(\cdot),  \hat{\mu}_1(\cdot), \widehat{\tilde{\mu}}_0(\cdot), \widehat{\gamma}_1(\cdot) \right)$. 
On the second fold, we construct the pseudo-outcome $\phi_{\mu, i}(\etahat) + (1 - D_i) - \phi_{\pi \gamma, i}(\etahat) - \phi_{\pi \tilde{\mu}, i}(\etahat) - \phi_{\pi \mu, i}(\etahat)$ for the lower bound and the pseudo-outcome $\phi_{\mu, i}(\etahat) + (1 - D_i) - \phi_{\pi \gamma, i}(\etahat) + \phi_{\pi \tilde{\mu}, i}(\etahat) - \phi_{\pi \mu, i}(\etahat)$ for the upper bound. 
We then regress the estimated pseudo-outcomes on the covariates $X_i$ using a researcher-specified non-parametric regression procedure satisfying the $L_2(\p)$-stability condition (Assumption \ref{asm: L2 stability condition}). 

\vspace{-1em}
\paragraph{Bound on Integrated Mean Square Error Convergence:}
For modified definitions of the smoothed doubly robust residuals and smoothed bias, we again derive a bound on the integrated mean square error of our feasible estimators relative to that of an infeasible oracle nonparametric regression that has access to the true nuisance functions using the same arguments as Proposition \ref{prop: oracle result for learning, outcome regression bounds}. 
We state the result for completeness, but skip the proof for brevity.

\begin{proposition}\label{prop: proxy outcome, target regression}
Let $\widehat{\E}_n[\cdot \mid X_i = x]$ denote the second-stage pseudo-outcome regression estimator.
Suppose $\widehat{\E}_n[\cdot \mid X_i = x]$ satisfies the $L_2(\p)$-stability condition (Assumption \ref{asm: L2 stability condition}), and $\p(\epsilon \leq \hat{\pi}_{1}(X_i) \leq 1 - \epsilon) = 1$ for some $\epsilon > 0$. 
Define $\tilde{R}_1(x) = \widehat{\E}_n[ (\pi_1(X_i) - \pihat_1(X_i)) (\outcomereg(X_i) - \outcomereghat(X_i)) \mid X_i = x ]$, $\tilde{R}_2(x) = \widehat{\E}_n[ (\pi_1(X_i) - \pihat_1(X_i)) (\hat{\tilde{\mu}}_0(X_i) - \tilde{\mu}_0(X_i)) \mid X_i = x ]$, $\tilde{R}_3(x) = \widehat{\E}_n[ (\pi_1(X_i) - \pihat_1(X_i)) (\hat{\gamma}_1(X_i) - \gamma_1(X_i)) \mid X_i = x ]$, and $R_{oracle}^2 = \E[ \| \widehat{\overline{\mu}}_{oracle}(\cdot) - \overline{\mu}^*(\cdot) \|^2]$. 
Then, 
\begin{align*}
& \| \widehat{\overline{\mu}}(\cdot) - \overline{\mu}^*(\cdot) \| \leq \|\widehat{\overline{\mu}}_{oracle}(\cdot) - \overline{\mu}^*(\cdot) \| + \epsilon^{-1} \left( \| \tilde{R}_1(\cdot) \| + \| \tilde{R}_2(\cdot) \| +  \| \tilde{R}_3(\cdot) \| \right) + o_\p(R_{oracle})
\end{align*}
The analogous result holds for $\widehat{\underline{\mu}}(x)$.
\end{proposition}

\subsubsection{Bounds on Overall Performance}

Under proxy outcome bounds, the upper bound on overall performance can be written as 
$$
\overline{\perf}(s; \beta) = \e[\beta_{0,i} + \beta_{1,i} \mu_1(X_i) + \beta_{1,i} (1 - D_i) - \beta_{1,i} ( \pi_0(X_i) \gamma_1(X_i) + \pi_0(X_i) \mu_1(X_i) + \pi_0(X_i) \tilde{\mu}_0(X_i) ) ].
$$
The lower bound can be written analogously. 
Like in Section \ref{section: estimating overall performance bounds} of the main text, both bounds are linear functionals of known functions of the data and identified nuisance parameters. 
This linearity enables us to construct debiased estimators using standard arguments.

\vspace{-1em}
\paragraph{Estimation Procedure for Proxy Outcome Bounds:} 
We sketch the construction of our estimators based on $K$-fold cross-fitting. 
We randomly split the data into $K$ disjoint folds.
For each fold $k$, we estimate the nuisance functions $\etahat_{-k}$ using all observations not in the $k$-th fold and construct
$$
\overline{\perf}_i(\etahat_{-k}) = \beta_{0,i} + \beta_{1,i} \phi_{\mu,i}(\etahat_{-k}) + \beta_{1,i} (1 - D_i) - \beta_{1,i} ( \phi_{\pi \gamma, i}(\etahat_{-k}) + \phi_{\pi \mu, i}(\etahat_{-k}) + \phi_{\pi \tilde \mu, i}(\etahat_{-k}) )
$$
for each observation $i$ in the $k$-th fold. We take the average across all observations and return $\widehat{\overline{\perf}}(s; \beta) = \e_n[ \widehat{\overline{\perf}}(s; \beta)(\etahat_{-K_i}) ]$.
The estimator for the lower bound is defined analogously. 

\vspace{-1em}
\paragraph{Consistency and Asymptotic Normality:} Using the same arguments as in the proof of Proposition \ref{prop: overall performance estimator, nonparametric outcome bounds}, we can again show that these estimators converge to the proxy outcome bounds on overall performance and are jointly asymptotically normal. 
We next state the result, but skip the proof for brevity.

\begin{proposition}\label{prop: proxy outcome, overall performance}
Define $R_{1,n}^{k} = \| \muhat_{1,-k}(\cdot) - \mu_1(\cdot) \|  \| \pihat_{1,-k}(\cdot) - \pi_1(\cdot) \|$, $R_{2,n}^{k} = \| \tilde{\muhat}_{0,-k}(\cdot) - \tilde{\mu}_{0}(\cdot) \| \| \pihat_{1,-k}(\cdot) - \pi_1(\cdot) \|$, and $R_{3,n}^{k} = \| \hat{\gamma}_{1, -k}(\cdot) - \gamma_1(\cdot) \| \| \pihat_{1,-k}(\cdot) - \pi_1(\cdot) \|$ for each fold $k = 1, \hdots, K$.
Assume (i) there exists some $M < \infty$ such that $\| \beta_1(\cdot) \| \leq M$; (ii) $\p(\pi_1(X_i) \geq \delta) = 1$ for some $\delta > 0$, (iii) there exists $\epsilon > 0$ such that $\p(\pihat_{1,-k}(X_i) \geq \epsilon) = 1$ for each fold $k = 1, \hdots, K$, and (iv) $\| \pihat_{1,-k}(\cdot) - \pi_1(\cdot) \| = o_\p(1), \| \muhat_{1,-k}(\cdot) - \mu_1(\cdot) \| = o_\p(1), \| \tilde{\muhat}_{0,-k}(\cdot) - \tilde{\mu}_{0}(\cdot) \| = o_\p(1)$, and $\| \hat{\gamma}_{1, -k}(\cdot) - \gamma_1(\cdot) \| = o_\p(1)$ for each fold $k = 1, \hdots, K$. 
Then, 
$$
    \left| \widehat{\overline{\perf}}(s; \beta) - \overline{\perf}(s; \beta) \right| = O_\p\left( 1/\sqrt{n} + \sum_{k=1}^{K} (R_{1,n}^k + R_{2,n}^k + R_{3,n}^k) \right)
$$
and the analogous result holds for $\widehat{\underline{\perf}}(s; \beta)$. 
If further $R_{1,n}^k, R_{2,n}^k, R_{3,n}^k = o_\p(1/\sqrt{n})$ for all folds $k = 1, \hdots, K$, then
\begin{equation*}
    \sqrt{n} \left( \begin{pmatrix} \widehat{\overline{\perf}}(s; \beta) \\ \widehat{\underline{\perf}}(s; \beta)\end{pmatrix} - \begin{pmatrix} \overline{\perf}(s; \beta) \\ \underline{\perf}(s; \beta) \end{pmatrix} \right) \xrightarrow{d} N\left( 0, \Sigma \right)
\end{equation*}
for covariance matrix $\Sigma = Cov\left( (\overline{\perf}_i(\eta), \underline{\perf}_i(\eta))^\prime \right)$. 
\end{proposition}

\subsubsection{Bounds on Positive Class Performance}

We analogously estimate the proxy outcome bounds on positive class performance by solving sample linear fractional programs as in Section \ref{section: estimating pos class bounds} of the main text.
Building on Proposition \ref{proposition: partial double robustness for pos class estimator}, all that needs to be modified is the choice estimator for the bounding functions. 
As an illustration, we sketch out how the sample linear fractional program can be combined with our pseudo-regression procedure for proxy outcome bounds discussed earlier in Appendix \ref{section: proxy outcome, pseudo outcome regression}.

\vspace{-1em}
\paragraph{Estimation Procedure for Proxy Outcome Bounds:}
We split the data into three folds. 
On the first fold, we estimate the nuisance functions $\widehat{\mu}_1(\cdot), \widehat{\pi}_1(\cdot), \widehat{\gamma}_1(\cdot)$, and $\widehat{\widetilde{\mu}}_0(\cdot)$. 
On the second fold, we then construct the pseudo-outcomes $\phi_{\mu, i}(\etahat) + (1 - D_i) - \phi_{\pi \gamma, i}(\etahat) - \phi_{\pi \tilde{\mu}, i}(\etahat) - \phi_{\pi \mu, i}(\etahat)$ for the lower bound and the pseudo-outcome $\phi_{\mu, i}(\etahat) + (1 - D_i) - \phi_{\pi \gamma, i}(\etahat) + \phi_{\pi \tilde{\mu}, i}(\etahat) - \phi_{\pi \mu, i}(\etahat)$ for the upper bound. 
We regress the estimated pseudo-outcomes on the covariates $X_i$ using a researcher-specified non-parametric regression procedure, yielding estimators $\widehat{\underline{\delta}}(X_i)$, $\widehat{\overline{\delta}}(X_i)$ respectively.
Finally, on the third fold, we estimate the upper bound on positive class performance by solving
\begin{equation*}
    \widehat{\overline{\perf}}_{+}(s; \beta) = \max_{\tilde{\delta} \in \widehat{\Delta}_{n}} \, \frac{ \e_{n}[ \beta_{0,i} \phi_{\mu,i}(\etahat) + \beta_{0, i} \tilde{\delta}_i ]}{\e_{n}[\phi_{\mu,i}(\etahat) + \tilde{\delta}_i ]},
\end{equation*}
where now $\widehat{\Delta}_{n} = \{ (1 - D_i) \widehat{\underline{\delta}}(X_i) \leq \tilde{\delta}_i \leq (1 - D_i) \widehat{\overline{\delta}}(X_i) \}$.
These estimators are again equivalent to solving linear programs by the Charnes-Cooper transformation.
Proposition \ref{proposition: partial double robustness for pos class estimator} can be applied, and this estimator has the same partial double robustness property.
The error of this estimator again depends on the root mean square error of the estimated bounding functions. If the nonparametric regression procedure further satisfies the $L_2(\p)$-stability condition, then we can further control these errors using the results in Appendix \ref{section: proxy outcome, pseudo outcome regression}. 

\subsection{Estimation under Instrumental Variable Bounds}\label{section: log-sum-exp bounds for IV}

Suppose we observe an instrumental variable $Z_i \in \mathcal{Z}$ with finite support that generates quasi-random variation in selection decisions but does not directly affect outcomes as discussed in Section \ref{section: instrumental variable bounds}. 
The instrument provides identifying power under two assumptions: (i) $Z_i$ exogenously shifts the selection decision but not the true outcome, $Y_i^* \indep Z_i \mid X_i$; and (ii) $Z_i$ is relevant, meaning there exists $z, z^\prime \in \mathcal{Z}$ with $\p(D_i = 1 \mid X_i = x, Z_i = z) \neq \p(D_i = 1 \mid X_i = x, Z_i = z^\prime)$.
Under these conditions, classic results by \citet[][]{Manski(94)-SampleSelectionBounds} imply bounds on the confounding function. 
As notation, let $p = |\cZ|$ denote the number of unique instrument values, $\lambda_{z}(x) := \mathbb{E}[Y_i D_i \mid X_i = x, Z_i = z]$ and $\kappa_{z}(x) := \p(D_i = 0 \mid X_i = x, Z_i = z)$.

Recall that for any value of the instrument $z \in \cZ$, the confounding function satisfies, for all $x \in \cX$, 
$$
\underline{\delta}_{z}(x; \eta) \leq \delta(x) \leq \overline{\delta}_z(x; \eta),
$$
where $\underline{\delta}_{z}(x) := (\lambda_z(x) - \mu_1(x))/\pi_0(x)$ and $\overline{\delta}_{z}(x) := (\kappa_{z}(x) + \lambda_{z}(x) - \mu_1(x))/\pi_0(x)$.
We will first extend our estimators for instrumental variable bounds at a particular instrument value $z \in \cZ$. 
We will then discuss how to extend our estimators to a smooth approximation of the intersection bounds, for all $x \in \cX$, 
\begin{equation*}
    \max_{z \in \cZ} \underline{\delta}_{z}(x; \eta) \leq \delta(x) \leq \min_{z \in \cZ} \overline{\delta}_z(x; \eta).
\end{equation*}

To extend our estimators, it will be convenient to rewrite the instrumental variable bounds as, for each $z \in \cZ$, 
\begin{equation*}
    \lambda_z(x) - \mu_1(x) \leq \pi_0(x) \delta(x) \leq \kappa_{z}(x) + \lambda_z(x) - \mu_1(x).
\end{equation*}
We also observe that 
\begin{align*}
& \phi_{\lambda, z, i}(\eta) := \frac{1\{Z_i = z\}}{\p(Z_i = z \mid X_i)} \left( Y_i D_i - \lambda_{z}(X_i) \right) + \lambda_{z}(X_i), \\
& \phi_{\kappa, z, i}(\eta) := \frac{1\{ Z_i = z \}}{\p(Z_i = z \mid X_i)} \left( (1 - D_i) - \kappa_{z}(X_i) \right) + \kappa_{z}(X_i)
\end{align*}
are the influence functions for $\e[\lambda_{z}(X_i)]$ and $\e[\kappa_{z}(X_i)]$ respectively.

\subsubsection{Bounds on the Conditional Likelihood}\label{section: IV bounds, pseudo outcome regression}

We briefly extend our nonparametric regression estimator for for the instrumental variable bounds at a particular instrument value $z \in \cZ$. 
The construction is identical to Section \ref{section: DR Learner, main text}, except we modify the pseudo-outcomes that are constructed in the second-stage. 
We again illustrate the procedure using two folds for simplicity. 

\vspace{-1em}
\paragraph{Estimation Procedure for Single-Instrument Bounds:}
Split the data into two folds.
Using the first fold, we construct nuisance function estimates $\etahat = (\hat{\lambda}_{z}(\cdot), \hat{\kappa}_{z}(\cdot), \hat{\p}(Z_i = z \mid X_i = \cdot))$.
On the second fold, we construct the pseudo-outcome $\phi_{\kappa,z,i}(\etahat), \phi_{\lambda, z, i}(\etahat)$ for the upper bound and $\phi_{\lambda, z, i}(\etahat)$ for the lower bound. 
We regress the estimated pseudo-outcomes on the covariates $X_i$ using a researcher-specified nonparametric regression procedure satisfying the $L_2(\p)$-stability condition (Assumption \ref{asm: L2 stability condition}). 

\vspace{-1em}
\paragraph{Bound on Integrated Mean Square Error Convergence:}
For modified definitions of the smoothed doubly robust residuals and smoothed bias, we can again derive a bound on the integrated mean square error of our feasible estimators relative to that of an infeasible oracle nonparametric regression that has access to the true nuisance functions using the same arguments as Proposition \ref{prop: oracle result for learning, outcome regression bounds}. 
We state the result for completeness, but skip the proof for brevity.

\begin{proposition}\label{prop: oracle result for learning, instrumental variable bounds, fixed z}
Suppose $\widehat{\E}_n[\cdot \mid X_i = x]$ satisfies the $L_2(\p)$-stability condition (Assumption \ref{asm: L2 stability condition}) and $\p(\epsilon \leq \widehat{\p}(Z_i = z \mid X_i = x)) = 1$ for some $\epsilon > 0$.
Define $\tilde{R}_2(x) = \widehat{\E}_n[ (\p(Z_i = z \mid X_i = x) - \widehat{\p}(Z_i = z \mid X_i = x)) (\hat{\lambda}_z(x) - \lambda_z(x)) \mid X_i = x]$, $\tilde{R}_{3}(x) = \widehat{\E}_n[ (\p(Z_i = z \mid X_i = x) - \widehat{\p}(Z_i = z \mid X_i = x)) (\hat{\kappa}_z(x) - \kappa_{z}(x)) \mid X_i = x ]$, and $R^2_{oracle}(z) = \E[ \|\widehat{\overline{\mu}}(\cdot) - \overline{\mu}^*(\cdot) \|^2 ]$.
Then,
\begin{align*}
& \| \widehat{\overline{\mu}}(\cdot) - \overline{\mu}(\cdot) \| \leq \| \widehat{\overline{\mu}}_{oracle}(\cdot) - \overline{\mu}(\cdot) \| + \epsilon^{-1} \left( \| \tilde{R}_2(\cdot)\| + \| \tilde{R}_3(\cdot) \| \right) + o_\p(R_{oracle}(z)), \\
& \| \widehat{\underline{\mu}}(\cdot) - \underline{\mu}(\cdot) \| \leq \| \widehat{\underline{\mu}}_{oracle}(\cdot) - \underline{\mu}(\cdot) \| + \epsilon^{-1} \| \tilde{R}_2(\cdot) \| + o_\p(R_{oracle}(z)).
\end{align*}
\end{proposition}

\subsubsection{Bounds on Overall Performance}

Under instrumental variable bounds at a single instrument value $z \in \cZ$, the upper bound on overall performance can be written as 
$$
\overline{\perf}(s; \beta) = \e[ \beta_{0,i} + \beta_{i,1} \lambda_z(X_i) + \beta_{i,1} 1\{ \beta_{i,1} > 0 \} \kappa_{z}(X_i) ].
$$
Like in Section \ref{section: estimating overall performance bounds}, the bounds are again linear functionals of known functions of the data and identified nuisance parameters.
We construct debiased estimators using standard arguments. 

\vspace{-1em}
\paragraph{Estimation Procedure for Single-Instrument Bounds:}
We randomly split the data into $K$ disjoint folds. 
For each fold $k$, we estimate the nuisance functions $\etahat_{-k}$ using all observations not in the $k$-th fold and construct
$$
\overline{\perf}_i(\etahat_{-k}) = \beta_{0,i} + \beta_{i,1} \phi_{\lambda, z, i}(\etahat_{-k}) + \beta_{i,1} 1\{ \beta_{i,1} > 0 \} \phi_{\kappa, z, i}(\etahat_{-k})
$$
for each observation $i$ in the $k$-th fold. 
We take the average across all observations and return $\widehat{\overline{\perf}}(s; \beta) = \e_n[ \widehat{\overline{\perf}}(s; \beta)(\etahat_{-K_i}) ]$.
The estimator for the lower bound is defined analogously. 

\vspace{-1em}
\paragraph{Consistency and Asymptotic Normality:} Using the same arguments as in the proof of Proposition \ref{prop: overall performance estimator, nonparametric outcome bounds}, we can again show that these estimators converge quickly to the proxy outcome bounds on overall performance and are jointly asymptotically normal. 
We next state the result, but skip the proof for brevity.

\begin{proposition}\label{prop: instrumental variable bounds, fixed z, overall performance}
Define $R_{2,n}^{k} = \| \hat{\lambda}_{z,-k}(\cdot) - \lambda_{z}(\cdot) \| \| \hat{\p}(Z_i = z \mid X_i = \cdot) - \p(Z_i = z \mid X_i = \cdot) \|$ and $R_{3,n}^{k} = \| \hat{\kappa}_{z, -k}(\cdot) - \kappa_{z}(\cdot) \| \| \hat{\p}(Z_i = z \mid X_i = \cdot) - \p(Z_i = z \mid X_i = \cdot) \|$.
Assume (i) there exists some $M < \infty$ such that $\| \beta_1(\cdot) \| \leq M$; (ii) $\p\{ \p(Z_i = z \mid X_i) \geq \delta \} = 1$; (iii) there exists $\epsilon > 0$ such that $\p\{ \hat{\p}_{-k}(Z_i = z \mid X_i) \geq \epsilon \} = 1$ for all folds $k$; and (iv) $\| \hat{\lambda}_z(\cdot) - \lambda_z(\cdot) \| = o_\p(1)$, $ \|\hat{\kappa}_z(\cdot) - \kappa_{z}(\cdot)\| = o_\p(1)$, and $\|\hat{\p}_{-k}(Z_i = z \mid X_i) - \p(Z_i = z \mid X_i)\| = o_\p(1)$ for all folds $k$.
Then,
\begin{align*}
    & \left| \widehat{\overline{\perf}}(s; \beta) - \overline{\perf}(s; \beta) \right| = O_\p\left( 1 / \sqrt{n} + \sumoverfolds (R_{2,n}^{k} + R_{3,n}^{k}) \right),
\end{align*}
and the analogous result holds for $\widehat{\underline{\perf}}(s; \beta)$.
If further $R_{2,n}^{k} = o_\p(1/\sqrt{n})$ and $R_{3,n}^{k} = o_\p(1/\sqrt{n})$ for all folds $k$, then
$$
\sqrt{n} \left( \begin{pmatrix} \widehat{\overline{\perf}}(s; \beta) \\ \widehat{\underline{\perf}}(s; \beta) \end{pmatrix} - \begin{pmatrix} \overline{\perf}(s; \beta) \\ \underline{\perf}(s; \beta) \end{pmatrix}\right) \xrightarrow{d} N\left( 0, \Sigma \right)
$$
for covariance matrix $\Sigma = Cov\left( (\overline{\perf}_i, \underline{\perf}_i)^\prime \right)$.
\end{proposition}

\subsubsection{Bounds on Positive Class Performance}

We analogously estimate the instrumental variable bounds on positive class performance by solving sample linear fractional programs as in Section \ref{section: estimating pos class bounds} of the main text.
All that needs to be modified is the choice estimator for the bounding functions. 
As an illustration, we sketch out how the sample linear fractional program can be combined with our pseudo-regression procedure discussed earlier in Appendix \ref{section: IV bounds, pseudo outcome regression}.

\vspace{-1em}
\paragraph{Estimation Procedure under Single-Instrument Bounds:} 
We split the data into three folds. 
Using the first fold, we construct nuisance function estimators 
$\pihat_{1}(\cdot), \hat{\mu}_{1}(\cdot), \hat{\lambda}_{z}(\cdot), \hat{\kappa}_{z}(\cdot), \hat{\p}(Z_i = z \mid X_i = \cdot)$. 
Using the second fold, we construct the pseudo-outcome $\phi_{\kappa,z,i}(\etahat), \phi_{\lambda, z, i}(\etahat)$ for the upper bound and $\phi_{\lambda, z, i}(\etahat)$ for the lower bound. 
We regress the estimated pseudo-outcomes on the covariates $X_i$ using a researcher-specified nonparametric regression procedure, yielding estimators $\widehat{\overline{\delta}}(X_i), \widehat{\underline{\delta}}(X_i)$ respectively. 
Finally, on the third fold, we construct the fold-specific estimate of the upper bound on positive class performance by solving 
\begin{equation*}
\widehat{\overline{\perf}}_{+}(s; \beta) = \max_{\tilde{\delta} \in \widehat{\Delta}_{n}} \, \frac{ \e_{n}[ \beta_{0,i} \phi_{\mu,i}(\etahat) + \beta_{0, i} \tilde{\delta}_i ]}{\e_{n}[\phi_{\mu,i}(\etahat_{-k}) + \tilde{\delta}_i ]},
\end{equation*}
where now $\widehat{\Delta}_{n} = \{ (1 - D_i) \widehat{\underline{\delta}}(X_i) \leq \tilde{\delta}_i \leq (1 - D_i) \widehat{\overline{\delta}}(X_i) \}$.
Proposition \ref{proposition: partial double robustness for pos class estimator} can be applied, and this estimator has the same partial double robustness property.
The error of this estimator again depends on the root mean square error of the estimated bounding functions. 
If the nonparametric regression procedure further satisfies the $L_2(\p)$-stability condition, then we can further control these errors using the results in Appendix \ref{section: IV bounds, pseudo outcome regression}. 

\subsubsection{Smooth Approximations to Intersection Bounds}\label{sec:smooth_approx}

We now discuss how our estimators can be extended to the intersection bounds 
\begin{equation*}
    \max_{z \in \cZ} \underline{\delta}_{z}(x; \eta) \leq \delta(x) \leq \min_{z \in \cZ} \overline{\delta}_z(x; \eta)
\end{equation*}
for all $x \in \cX$, where $\underline{\delta}_{z}(x) := (\lambda(x, z) - \mu_1(x))/\pi_0(x)$ and $\overline{\delta}_{z}(x) := (\pi_0(x,z) + \lambda(x, z) - \mu_1(x))/\pi_0(x)$. 
Rather than working directly with these intersection bounds, we instead consider bounding functions based on smooth approximations to the pointwise minimum and pointwise maximum functions, following \cite{LevisEtAl(23)-CovariateAssistedIV}'s analysis of covariate-assisted Balke-Pearl bounds on the average treatment effects. 
We focus on the log-sum-exponential function but other choices are possible such as the Boltzmann operator.
Exploring different choices of smooth approximations is an interesting question for future work.

For a researcher-specified $\alpha > 0$, the \textit{log-sum-exponential function} $g_{\alpha}(\cdot) \colon \mathbb{R}^{p} \rightarrow \mathbb{R}$ is
\begin{equation*}
    g_{\alpha}(v) = \frac{1}{\alpha} \log\left( \sum_{j=1}^{p} \exp(\alpha v_j) \right).
\end{equation*}
The log-sum-exponential function approximates the pointwise maximum function and satisfies the inequality
$$
\max \, \{ v_1, \hdots, v_p \} \leq g_{\alpha}(v) \leq \max \, \{ v_1, \hdots, v_p \} + \frac{\log(p)}{\alpha}.
$$
The choice of parameter $\alpha > 0$ therefore determines the quality of the approximation $g_{\alpha}(\cdot)$ provides to the pointwise maximum function. 
Furthermore, $\nabla g_{\alpha}(v) = \frac{v}{\ell^{T} v}$ for $\ell$ the $p$-dimensional vector of ones and $\nabla^2 g_{\alpha}(v) = \frac{\alpha}{(\ell^T v)^2} ((\ell^T v) diag(v) - v v^T)$.
Analogously, we can approximate the pointwise minimum with the function $g_{-\alpha}(v)$, which satisfies
$$
\min \, \{v_1, \hdots, v_p\} - \frac{\log(p)}{\alpha} \leq g_{-\alpha}(v) \leq \min \, \{v_1, \hdots, v_p \}.
$$

We will consider bounding functions that apply the log-sum-exponential function to the $\underline{\delta}_{z}(x; \eta)$ and $\overline{\delta}_{z}(x; \eta)$ respectively. We define the pointwise bounding functions on $\pi_0(x) \delta(x)$ as 
$$
\underline{\delta}(x; \eta) = g_{\alpha}\left( \lambda_{z_1}(x) - \mu_1(x), \hdots, \lambda_{z_p}(x) - \mu_1(x) \right) \mbox{ and } 
$$
$$
\overline{\delta}(x; \eta) = g_{-\alpha}\left( \kappa_{z_1}(x) + \lambda_{z_1}(x) - \mu_1(x), \hdots, \kappa_{z_p}(x) + \lambda_{z_p}(x) - \mu_1(x) \right)
$$
for some choice $\alpha > 0$ such that there exists constants $C_1, C_2$ such that $\| \nabla g_{\alpha}(v)\|_{\infty}, \|\nabla g_{-\alpha}(v)\|_{\infty} \leq C_1$and $\| \nabla^2 g_{\alpha}(v)\|_{op} \leq C_2$ uniformly over $v$.
To then extend our estimators, we notice that 
\begin{align*}
& \phi_{\lambda_z,i}(\eta) := \frac{1\{Z_i = z\}}{\p(Z_i = z \mid X_i)} ( Y_i D_i - \lambda_z(X_i) ) + \lambda_z(X_i), \\
& \phi_{\kappa_z,i}(\eta) := \frac{1\{Z_i = z\}}{\p(Z_i = z \mid X_i)} ((1 - D_i) - \kappa_z(X_i)) + \kappa_{z}(X_i)
\end{align*}
are the influence functions for $\e[\lambda_z(X_i)]$ and $\e[\kappa_z(X_i)]$ respectively. Furthermore,
\begin{align*}
& \phi_{\underline{\delta},i}(\eta) :=  g_{\alpha}(\underline{\delta}_{z_1}(X_i; \eta), \hdots, \underline{\delta}_{z_p}(X_i; \eta)) + \sum_{j=1}^{p} \frac{\partial g_{\alpha}(\underline{\delta}_{z_1}(X_i; \eta), \hdots, \underline{\delta}_{z_p}(X_i; \eta))}{\partial \underline{\delta}_{z_j}(X_i; \eta)} \left( \phi_{\lambda_{z_j}, i}(X_i) - \phi_{\mu,i}(X_i) \right) \\
& \phi_{\overline{\delta}, i}(\eta) := g_{\alpha}(\overline{\delta}_{z_1}(X_i; \eta), \hdots, \overline{\delta}_{z_p}(X_i; \eta)) + \sum_{j=1}^{p} \frac{\partial g_{\alpha}(\overline{\delta}_{z_1}(X_i; \eta), \hdots, \overline{\delta}_{z_p}(X_i; \eta))}{\partial \overline{\delta}_{z_j}(X_i; \eta)} \left( \phi_{\kappa_{z_j},i}(\eta) + \phi_{\lambda_{z_j}, i}(X_i) - \phi_{\mu,i}(X_i) \right)
\end{align*}
are the influence functions for $\e[\underline{\delta}(X_i; \eta)]$, $\e[\overline{\delta}(X_i; \eta)]$ respectively, by Theorem 3 in \cite{LevisEtAl(23)-CovariateAssistedIV} and standard calculations involving influence functions. 
With this in hand, we then extend our estimators for the bounds on overall performance, positive class performance, and the conditional likelihood by substituting in the appropriate influence functions for the upper bounding and lower bounding functions. 

\vspace{-1em}
\paragraph{Alternative Approach under Margin Conditions.}
An alternative is to invoke a margin condition as in \citet{LevisEtAl(23)-CovariateAssistedIV} and \citet{semenova2023adaptive}. The margin condition assumes that, for each $x \in \mathcal{X}$, the maximal and minimal values in the intersection bounds are sufficiently well-separated from the remaining instrument values. 
Under such margin conditions, \citet{LevisEtAl(23)-CovariateAssistedIV} and \citet{semenova2023adaptive} show that one can construct estimators directly targeting the averages of intersection bounds. 
For our overall performance bounds, these results apply directly. 
Extending this approach to estimate the conditional likelihood bounds and the positive class performance bounds is non-trivial.  Developing these extensions is an interesting direction for future work.

%%%%%%%%%%%%%%%%%%%%%%%%%%%%%%%
% Sensitivity Analysis Models %
%%%%%%%%%%%%%%%%%%%%%%%%%%%%%%%
\section{Connections to Sensitivity Analysis Models}\label{section: connections to existing sensitivity analysis models}

A substantial literature in causal inference develops sensitivity analysis frameworks for assessing robustness to unobserved confounding. 
While these frameworks are often presented in treatment effect settings, they can be translated to our selective labels context. 
This appendix establishes formal connections between our framework---specifically Assumption \ref{asm: bounding assumption} with observed outcome bounds discussed in Section \ref{section: observed outcome bounds} of the main text---and several influential sensitivity analysis models.

This is valuable for three reasons. 
First, pedagogically, researchers familiar with sensitivity analysis in causal inference may find it easier to interpret observed outcome bounds by relating them to established frameworks. 
Second, these connections show that our estimation results extend to settings where researchers invoke these alternative sensitivity models. 
Rather than developing separate estimation procedures for each model, researchers could apply our results by translating their sensitivity assumptions into the corresponding values of $\underline{\Gamma}, \overline{\Gamma}$ (although the resulting bounds may not be sharp). 
Finally, most sensitivity analysis models in causal inference place restrictions on selection propensities---how unobservables affect the probability of treatment/selection. 
In contrast, Assumption \ref{asm: bounding assumption} with observed outcome bounds restricts outcome differences---how outcomes differ between selected and unselected units conditional on observables. 
While conceptually distinct, Bayes' rule establishes precise mathematical connections between these two approaches.

\subsection{Marginal Sensitivity Model}
The \textit{marginal sensitivity model} (MSM) has received substantial recent attention in the causal inference literature \citep[][]{Tan(06), ZhaoSmallBhattacharya(19), KallusMaoZhou(18)-IntervalEstimation, DornGuo(21), DornGuoKallus(21), KallusZhou(21), JinRenCandes(21)}. 
The MSM specifies that the selection mechanism satisfies a multiplicative bound on odds ratios: that is, for some $\Lambda \geq 1$, 
\begin{equation}\label{eqn: marginal sensitivity model}
  \Lambda^{-1} \leq \frac{\p(D_i = 1 \mid X_i, Y_i^*)}{\p(D_i = 0 \mid X_i, Y_i^*)} \frac{\p(D_i = 0 \mid X_i)}{\p(D_i = 1 \mid X_i)} \leq \Lambda \mbox{ holds with probability } 1.
\end{equation}
The MSM bounds the ratio of conditional-to-marginal odds of selection. When $\Lambda = 1$, this reduces to $\p(D_i = 1 \mid X_i, Y^*_i) = \p(D_i = 1 \mid X_i)$ i.e., unconfoundedness. 
Larger values of $\Lambda$ allow greater departure from unconfoundedness. 
In other words, the parameter $\Lambda$ directly quantifies the strength of confounding on the odds ratio scale.

We can relate the MSM to Assumption \ref{asm: bounding assumption} under observed outcome bounds via Bayes' rule.

\begin{proposition}\label{proposition: MSM implies MOSM}
\hfill 
\begin{enumerate}
\item[i.] Suppose $(X_i, D_i, Y_i^*) \sim \p(\cdot)$ satisfies the MSM (\ref{eqn: marginal sensitivity model}) for some $\Lambda \geq 1$. Then, $\p(\cdot)$ satisfies Assumption \ref{asm: bounding assumption} with $\underline{\delta}(x; \eta) = (\Lambda^{-1} - 1) \mu_{1}(x)$ and $\overline{\delta}(x; \eta) = (\Lambda - 1) \mu_{1}(x)$.

\item[ii.] Suppose $(X_i, D_i, Y_i^*) \sim \p(\cdot)$ satisfies Assumption \ref{asm: bounding assumption} with observed outcome bounds for some $\underline{\Gamma}, \overline{\Gamma} > 0$. Then, $\p(\cdot)$ satisfies
$$
\overline{\Gamma}^{-1} \leq \frac{ \p(D_i = 1 \mid Y_i^* = 1, X_i) \p(D_i = 0 \mid X_i) }{ \p(D_i = 0 \mid Y_i^* = 1, X_i) \p(D_i = 1 \mid X_i) } \leq \underline{\Gamma}^{-1}.
$$
\end{enumerate}
\begin{proof}
For brevity, we omit the conditioning on $X_i$. 
Consider the first claim. 
By Bayes' rule, $\frac{ \p(D_i = 1 \mid Y_i^*) \p(D_i = 0) }{ \p(D_i = 0 \mid Y_i^*) \p(D_i = 1) } = \frac{ \p(Y_i^* \mid D_i = 1) }{ \p(Y_i^* \mid D_i = 0) }$. 
The MSM therefore implies bounds $\Lambda^{-1} \leq \frac{ \p(Y_i^* \mid D_i = 1) }{ \p(Y_i^* \mid D_i = 0) } \leq \Lambda$, which can be equivalently written as $\Lambda^{-1} \p(Y_i^* \mid D_i = 1) \leq \p(Y_i^* \mid D_i = 0) \leq \Lambda \p(Y_i^* \mid D_i = 1)$. Adding and subtracting $\p(Y_i^* = 1 \mid D_i = 1)$ then delivers the first claim.

Consider the second claim. Observed outcome bounds implies $\overline{\Gamma}^{-1} \leq \frac{\p(Y_i^* = 1 \mid D_i = 1)}{\p(Y_i^* = 1 \mid D_i = 0)} \leq \underline{\Gamma}^{-1}$.
But, by Bayes' rule, $\frac{\p(Y_i^* = 1 \mid D_i = 1)}{\p(Y_i^* = 1 \mid D_i = 0)} = \frac{ \p(D_i = 1 \mid Y_i^* = 1) \p(D_i = 0) }{ \p(D_i = 0 \mid Y_i^* = 1) \p(D_i = 1) }$, and so $\overline{\Gamma}^{-1} \leq \frac{ \p(D_i = 1 \mid Y_i^* = 1) \p(D_i = 0) }{ \p(D_i = 0 \mid Y_i^* = 1) \p(D_i = 1) } \leq \underline{\Gamma}^{-1}$
holds.
\end{proof}
\end{proposition}

\noindent Proposition \ref{proposition: MSM implies MOSM} shows that any analysis conducted under the MSM with parameter $\Lambda$ can be conducted using observed outcome bounds with $\underline{\Gamma} = \Lambda^{-1}$ and $\overline{\Gamma} = \Lambda$.
In the opposite direction, observed outcome bounds imply a specific MSM-style restriction on selection propensities, but only for the positive class $Y_i^* = 1$. 

\subsection{Rosenbaum's $\Gamma$-Sensitivity Model}
Rosenbaum's \textit{$\Gamma$-sensitivity  model} \citep[e.g.,][]{Rosenbaum(87), Rosenbaum(02)} is perhaps the most widely used framework for sensitivity analysis in observational studies. 
It bounds how much unobservables can differentially affect selection propensities across different outcome values: specifically, for some $\Gamma \geq 1$, $(X_i, D_i, Y_i^*) \sim \p(\cdot)$ satisfies 
\begin{equation}\label{eqn: rosenbaum sensitivity analysis}
   \Gamma^{-1} \leq \frac{\p(D_i = 1 \mid X_i, Y_i^* = y^*}{\p(D_i = 0 \mid X_i, Y_i^* = y^*)} \frac{\p(D_i = 0 \mid X_i, Y_i^* = \tilde{y}^*)}{\p(D_i = 1 \mid X_i, Y_i^* = \tilde{y}^*)} \leq \Gamma
\end{equation}
for all $y^*, \tilde{y}^* \in \{0, 1\}$ and with probability one. Notice that $\Gamma = 1$ again nests unconfoundedess.

\cite{AronowLee(2013)} and \cite{MiratrixWagerZubizaretta(18)} use a version of Rosenbaum's $\Gamma$-sensitivity model to construct bounds on a finite-population from a random sample with unknown selection probabilities.
\cite{yadlowsky2018bounds} applies Rosenbaum's $\Gamma$-sensitivity analysis model to derive bounds on the conditional average treatment effect and the average treatment effect. 
We refer the reader to Section 7 of \cite{ZhaoSmallBhattacharya(19)} for an in-depth comparison of the marginal sensitivity model and Rosenbaum's $\Gamma$-sensitivity model. 

We can relate Rosenbaum's $\Gamma$-sensitivity model to Assumption \ref{asm: bounding assumption} under observed outcome bounds.

\begin{proposition}\label{prop: rosenbaum sensitivity analysis implies MOSM}
Suppose $(X_i, D_i, Y_i^*) \sim \p(\cdot)$ satisfies Rosenbaum's sensitivity analysis model (\ref{eqn: rosenbaum sensitivity analysis}) for some $\Gamma > 1$. Then $P(\cdot)$ satisfies Assumption \ref{asm: bounding assumption} with $\underline{\delta}(x; \eta) = (\Gamma^{-1} - 1) \outcomereg(x)$ and $\overline{\delta}(x; \eta) = (\Gamma - 1) \outcomereg(x)$.
\begin{proof}
For brevity, we omit the conditioning on $X_i$ throughout the proof.
As a first step, we again apply Bayes' rule and observe that $
\frac{ \p(Y_i^* \mid D_i = 1) }{ \p(Y_i^* \mid D_i = 0) } = \frac{ \p(D_i = 1 \mid Y_i^*) \p(D_i = 0) }{ \p(D_i = 0 \mid Y_i^*) \p(D_i = 1) }$.
Then, further notice that $\frac{\p(D_i = 0)}{\p(D_i = 1)} = \frac{ \sum_{y \in \{0, 1\}} P(D_i = 0 \mid Y_i^* = y) P(Y_i^* = y) }{\sum_{y \in \{0, 1\}} P(D_i = 1 \mid Y_i^* = y) P(Y_i^* = y)}.$
Letting $y^* = \arg \max_{y^*\in \{0, 1\}} \frac{P(D_i = 0 \mid Y_i(0) = y_0, Y_i(1) = y_1)}{P(D_i = 1 \mid Y_i(0) = y_0, Y_i(1) = y_1)}$, the quasi-linearity of the ratio function implies that 
$$
\frac{ \sum_{y \in \{0, 1\}} P(D_i = 0 \mid Y_i^* = y) P(Y_i^* = y) }{\sum_{y \in \{0, 1\}} P(D_i = 1 \mid Y_i^* = y) P(Y_i^* = y)} \leq \frac{ P(D_i = 0 \mid Y_i^* = y^*) }{ P(D_i = 1 \mid Y_i^* = y^*) }
$$
This implies, for any $y \in \{0, 1\}$,
$$
\frac{ \p(Y_i^* = y \mid D_i = 1) }{ \p(Y_i^* = y \mid D_i = 0) } \leq
 \frac{ \p(D_i = 1 \mid Y_i^* = y) }{ \p(D_i = 0 \mid Y_i^* = y) } \frac{ P(D_i = 0 \mid Y_i^* = y^*) }{ P(D_i = 1 \mid Y_i^* = y^*) } \leq \Gamma,
$$
where the last inequality is implied by Rosenbaum's sensitivity analysis model (\ref{eqn: rosenbaum sensitivity analysis}). From this, we follow the same argument as the proof of Proposition \ref{proposition: MSM implies MOSM} to show that $\overline{\delta}(x; \eta) = (\Gamma - 1) \mu_1(x)$. The proof for the lower bound follows an analogous argument.
\end{proof}
\end{proposition}

\noindent Rosenbaum's $\Gamma$-sensitivity model with parameter implies observed outcome bounds with $\underline{\Gamma} = \Gamma^{-1}$ and $\overline{\Gamma} = \Gamma$.
However, the reverse direction does not hold --- observed outcome bounds are weaker than Rosenbaum's model.

\subsection{Partial $c$-Dependence} 
A recent strand of the econometrics literature proposes the \textit{partial c-dependence model} for conducting sensitivity analyses on unobserved confounding in causal inference \citep[e.g.,][]{MastenPoirier(18), MastenPoirier(20)}. 
Cast into our setting, partial $c$-dependence specifies that, for some $c > 0$, 
\begin{equation}\label{equation: partial c-dependence}
    \sup_{y \in \{0, 1\}} \, | \p(D_i = 1 \mid Y_i^* = y, X_i = x) - \p(D = 1 \mid X_i = x) | \leq c
\end{equation}
for all $x \in \cX$. 
Partial $c$-dependence bounds how much observing the outcome changes the selection problem. 
When $c = 0$, we have unconfoundedness. 
The parameter $c$ represents the maximum percentage point difference in selection probabilities between different outcome values. 
The next result shows that Assumption \ref{asm: bounding assumption} and partial c-dependence model can be related to one another through Bayes' rule.

\begin{proposition}\label{prop: c-dependence}
\hfill
\begin{enumerate}
\item[(i)] Suppose $(X_i, D_i, Y_i^*) \sim \p(\cdot)$ satisfies $\left| \p(Y_i^* = 1 \mid D_i = 0, X_i = x) - \p(Y_i^* = 1 \mid D_i = 1, X_i = x) \right| \leq C$ for all $x \in \mathcal{X}$. Then, $ \sup_{y \in \{0, 1\}} \, | \p(D_i = 1 \mid Y_i^* = y, X_i = x) - \p(D = 1 \mid X_i = x) | \leq  c(x)$
for all $x \in \cX$, where $c(x) = C \ \max\{ \frac{V(D_i \mid X_i = x)}{\p(Y_i^* = 1 \mid X_i = x)}, \frac{V(D_i \mid X_i = x)}{\p(Y_i^* = 0 \mid X_i = x)} \}$.

\item[(ii)] Suppose $(X_i, D_i, Y_i^*) \sim \p(\cdot)$ satisfies $\sup_{y \in \{0, 1\}} \, | \p(D_i = 1 \mid Y_i^* = y, X_i = x) - \p(D = 1 \mid X_i = x) | \leq c$ for all $x \in \mathcal{X}$. 
Then, $\left| \p(Y_i^* = 1 \mid D_i = 0, X_i = x) - \p(Y_i^* = 1 \mid D_i = 1, X_i = x) \right| \leq c \ \min \{ \frac{V(D_i \mid X_i = x)}{\p(Y_i^* = 1 \mid X_i = x)}, \frac{V(D_i \mid X_i = x)}{\p(Y_i^* = 0 \mid X_i = x)} \} $ for all $x \in \mathcal{X}$.
\end{enumerate}
 
\begin{proof}
For brevity, we omit the conditioning on $X_i$ throughout the proof.
To show the first result, we first rewrite $| \p(Y_i^* = 1 \mid D_i = 0) - \p(Y_i^* = 1 \mid D_i = 1) | \leq C$
as $| \p(D_i = 0 \mid Y_i^* = 1) \p(D_i = 1) - \p(D_i = 1\mid Y_i^* = 1) \p(D_i = 0) | \leq C \frac{Var(D_i)}{\p(Y_i^* = 1)}$.
We can further rewrite the left hand side to arrive at $| \p(D_i = 1 \mid Y_i^* = 1, X_i = x) - \p(D_i = 1 \mid X_i = x) | \leq C \frac{Var(D_i)}{\p(Y_i^* = 1)}$.
Analogously, we can rewrite the bound as $
| \p(Y_i^* = 0 \mid D_i = 1, X_i = x) - \p(Y_i^* = 0 \mid D_i = 0, X_i = x) | \leq C$,
which in turn implies $| \p(D_i = 1 \mid Y_i^* = 0, X_i = x) - \p(D_i = 1 \mid X_i = x) \leq C \frac{Var(D_i)}{\p(Y_i^* = 0)}$ by the same argument.
The second result follows by the same argument.
\end{proof}
\end{proposition}

\noindent As mentioned, partial $c$-dependence has an interpretation in terms of percentage point changes in selection probabilities, while observed outcome bounds has an interpretation in terms of changes in outcome probabilities. 
In applications like pretrial release or consumer lending, the appropriate value of $c$ may be challenging to calibrate --- it requires articulating assumptions about how judge and lender decision rates differ across defendants and applicants who would versus would not experience adverse outcomes.
Observed outcome bounds ask researchers to directly articulate assumptions about outcome rates directly (e.g., "rejected applicants are at most twice as likely to default"). 
This may align more naturally with available domain knowledge or proxy data. 

Of course, researchers should choose the parameterization that best aligns with their substantive knowledge and available evidence.
Moreover, 
Proposition \ref{prop: c-dependence} provides explicit formulas that allow researchers to translate between the two frameworks, enabling those who find it easier to specify partial $c$-dependence to map to implied observed outcome bounds (and vice versa).

\subsection{Tukey's Factorization and Outcome Models}
A alternative approach to sensitivity analysis specifies outcome models via Tukey's factorization that directly parameterize the relationship between $\p(Y^*_i \mid D_i = 0, X_i)$ and $\p(Y^*_i \mid D_i = 1, X_i)$ or $\p(Y^*_i \mid X_i)$ \citep[e.g.,][]{ RotnitzkyEtAl(01), BirminghamEtAl(03), brumback2004sensitivity}.
For example, \cite{RobinsEtAl(00), FranksEtAl(19), Scharfstein(21)} assume $\p(Y_i^* \mid D_i = 0, X_i) = \p(Y_i^* \mid D_i = 1, X_i) \frac{ \exp(\gamma_t s_t(Y_i^*)) }{ C(\gamma_t; X_i)}$, where $\gamma_t$ is a chosen parameter and $s_t(\cdot)$ is a chosen ``tilting function.''
For particular fixed choices of $\gamma_t$, $s_t(\cdot)$, such a model is sufficient to point identify various quantities of interest such as the conditional probability and the predictive performance estimands we consider.

The key difference relative to our framework is how uncertainty is communicated. 
Researchers in this literature report how conclusions vary for alternative choices of $\gamma_t$ or $s_t(\cdot)$.
By contrast, we suggest that researchers should articulate bounding functions that encode their beliefs about unobserved confounding, and then report the fully identified set consistent with those bounds. 
These bounding functions can formalize common heuristics without requiring a fully specified outcome model that can be difficult to justify.

An alternative approach places bounds on the mean difference in potential outcomes under treatment and control \cite{luedtke2015statistics, diaz2013sensitivity, diaz2018sensitivity}. This is analogous to the ``approximate mean independence'' assumption in \citet[][]{Manski(03)}.
Our identification framework extends this approach by placing bounds on the covariate-conditional difference in means.

%%%%%%%%%%%%%%%%%%%%%%%%%%%
% Monte Carlo Simulations %
%%%%%%%%%%%%%%%%%%%%%%%%%%%
\section{Monte Carlo Simulations}\label{section: monte carlo simulations}

In this section, we illustrate that the finite sample behavior of our proposed estimators. 
We simulate training and evaluation datasets satisfying Assumption \ref{asm: bounding assumption} under observed outcome bounds.
We draw $X_i = \left( X_{i,1}, \hdots, X_{i,d} \right)^\prime \sim N(0, I_d)$ and $D_i$ conditional on $X_i$ according to $\p(D_i = 1 \mid X_i = x) = \sigma\left( \frac{1}{2 \sqrt{d_{\pi}}} \sum_{d = 1}^{d_\pi} X_{i,d} \right)$ for some $d_{\pi} \in \{1, \hdots, d\}$ and $\sigma(a) = \frac{\exp(a)}{1 + \exp(a)}$. 
We finally draw $Y_i^*$ conditional on $(D_i, X_i)$ according to 
\begin{align*}
& \p(Y_i^* = 1 \mid D_i = 1, X_i = x) = \sigma\left( \frac{1}{2 \sqrt{d_{\mu}}} \sum_{d = 1}^{d_{\mu}} X_{i,d} \right), \mbox{ and } \p(Y_i^* = 1 \mid D_i = 0, X_i = x) = \Gamma_{true} \sigma\left( \frac{1}{2 \sqrt{d_{\mu}}} \sum_{d = 1}^{d_{\mu}} X_{i,d} \right)
\end{align*}
for some $d_{\mu} \in \{1, \hdots, d\}$ and $0 < \Gamma_{true} < 1$ 
We set $d = 50$, $d_{\pi} = 20$, $d_{\mu} = 25$, and $\Gamma_{true} = 0.75$.

\subsection{Behavior of Estimated Conditional Probability Bounds}

We compare three estimators for the upper bound on the conditional likelihood $\overline{\mu}^*(\cdot)$ with $\underline{\Gamma} = 2/3$, $\overline{\Gamma} = 3/2$: first, our proposed estimators; second, the infeasible oracle that uses the true observed conditional probability and propensity score; and finally, a plug-in learner that does not use sample splitting nor pseudo-outcomes. 
Our estimator and the oracle procedure are constructed using a single split of the evaluation data.
The first-stage nuisance functions $\eta = (\pi_1(\cdot), \mu_1(\cdot))$ are estimated using cross-validated Lasso logistic regressions, and the second-stage nonparametric regression uses cross-validated Lasso.
Proposition \ref{prop: oracle result for learning, outcome regression bounds} established that the integrated mean-square error of our estimators converge at fast rates to the integrated mean-square error of the infeasible oracle.
By contrast, the plug-in estimator will inherit errors from the estimation of the nuisance functions.

\begin{figure}[hb!]
\centering
\includegraphics[width=0.5\textwidth]{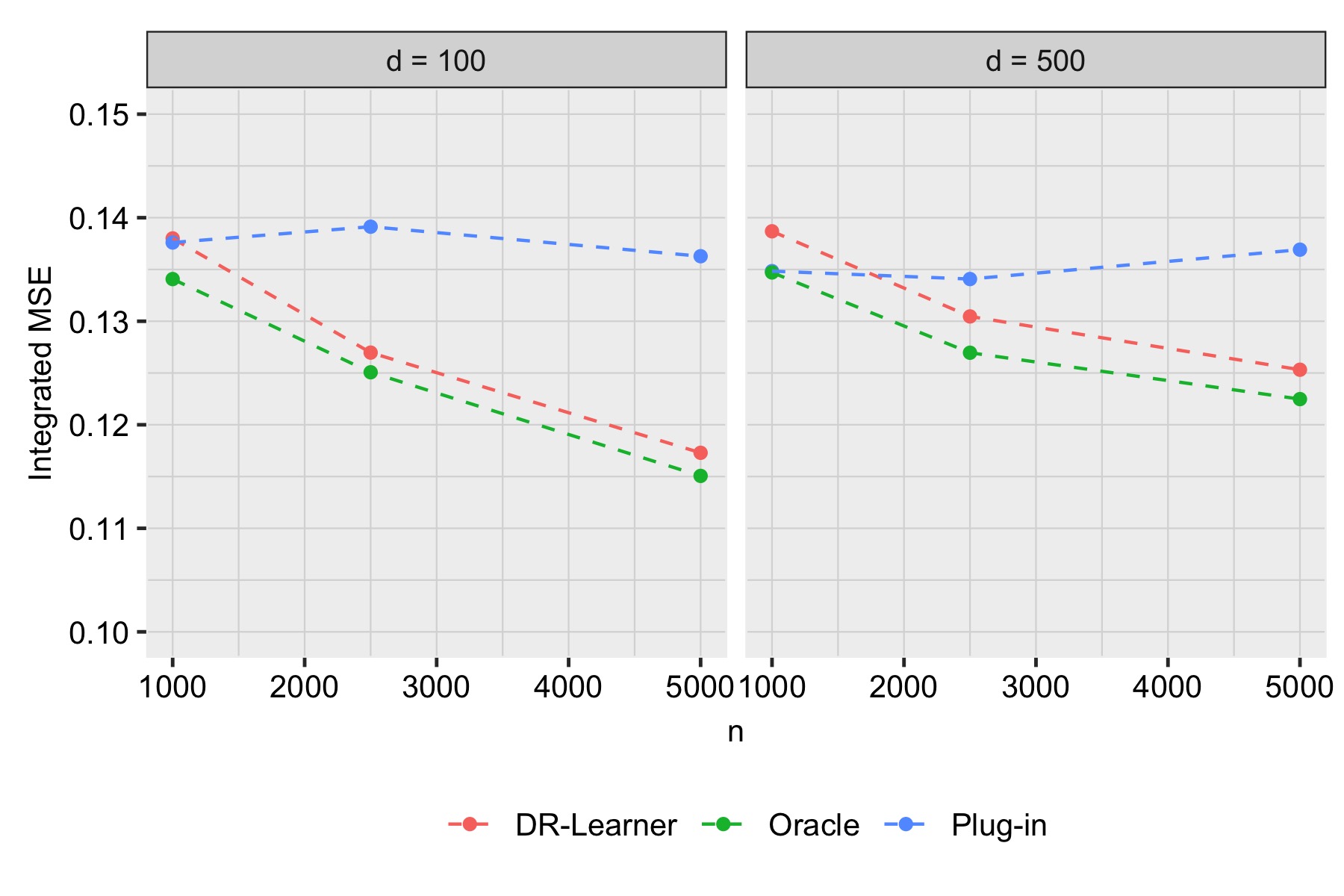}
\caption{Average integrated mean square of our estimator, the oracle learner, and the plug-in learner for the upper bound on the conditional probability $\overline{\mu}^*(\cdot)$.}
\floatfoot{\textit{Notes}: 
This figure plots the average integrated mean square error of our estimator, the oracle learner, and the plug-in learner for the upper bound on the conditional probability $\overline{\mu}^*(\cdot)$ across Monte Carlo simulations.
We report results for $n \in \{1000, 2500, 5000\}$, $d \in \{100, 500\}$, and observed outcome bounds with $\underline{\Gamma} = 2/3$, $\overline{\Gamma} = 3/2$. 
The results are computed over 1,000 simulations. See Appendix \ref{section: monte carlo simulations} for further discussion.
}
\label{fig: monte carlo simulations, DR Learner performance}
\end{figure}
 
Across 1000 simulated datasets of varying size $n \in \{1000, 2500, 5000\}$ and dimension $d \in \{100, 500\}$, Figure \ref{fig: monte carlo simulations, DR Learner performance} reports the average integrated mean square error of our estimator $\widehat{\overline{\mu}}(\cdot)$, the oracle learner $\widehat{\overline{\mu}}_{oracle}(\cdot)$, and the plug-in learner for the true bound $\overline{\mu}^*(\cdot)$.
As $n$ grows large, the average integrated mean square error of our estimator converges towards zero alongside the integrated mean square error of the oracle learner as expected.
While it is competitive for smaller sample sizes ($n = 1000$), the plug-in learner performs relatively poorly as $n$ grows larger. 
At $n = 5000$ and $d = 500$, our estimator's integrated mean square error is only 1.9\% larger than that of the oracle learner, whereas the plug-in learner's integrated mean square error is 18.4\% larger.
By leveraging both sample-splitting and pseudo-outcomes based on influence functions, our estimator substantively improves upon simple plug-in estimation approaches.

\subsection{Behavior of Estimated Predictive Performance Bounds}

To evaluate our estimators for the bounds on positive class performance and overall performance, we simulate a training dataset $(X_i, D_i, Y_i)$ for $i = 1, \hdots, n_{train}$ and estimate a predictive algorithm $s(\cdot)$ that predicts $Y_i = 1$ only on the selected data $D_i = 1$.
We evaluate the constructed predictive algorithm's positive class performance $\perf_{+}(s; \beta)$ and overall performance $\perf(s; \beta)$ using evaluation datatsets $(X_i, D_i, Y_i)$ for $i = 1, \hdots, n$ simulated from the same data generating process and observed outcome bounds for alternative choices of $\underline{\Gamma}, \overline{\Gamma}$.

\begin{table}[htpb!]
\begin{tabular}{c | c c c c }
$n$ & Estimate & SD & Bias \\
\hline \hline 
500 & 0.545 & 0.016 & 0.001 \\
1000 & 0.544 & 0.010 & 0.000 \\
2500 & 0.544 & 0.007 & 0.000
\end{tabular}
\caption{Bias properties for estimator for the upper bound on the true positive rate.}
\floatfoot{\textit{Notes}: 
This table summarizes the average bias of $\widehat{\overline{\perf}}_{+}(s;\beta)$ for the upper bounds on the true positive rate and its standard deviation across simulations.
We report these results for $n \in \{500, 1000, 2500\}$ and observed outcome bounds with $\underline{\Gamma} = 2/3$, $\overline{\Gamma} = 3/2$.
The results are computed over 1,000 simulations. See Appendix \ref{section: monte carlo simulations} for further discussion.
}
\label{table: nonparametric bounds, tpr, bias table}
\end{table}

We explore how well our proposed estimator recover the bound on the true positive rate $\overline{\perf}_{+}(s; \beta)$ across $1000$ simulated datasets of varying size $n \in \{500, 1000, 2500\}$ and a fixed choice $\underline{\Gamma} = 2/3$ and $\overline{\Gamma} = 3/2$.
We construct our estimator using three splits of the evaluation data.
We estimate the first-stage nuisance functions using random forests in the first fold, and the estimated bounding functions using cross-validated Lasso in the second fold. 
Online Appendix Table \ref{table: nonparametric bounds, tpr, bias table} reports the average bias, and Online Appendix Figure \ref{figure: nonparametric outcome bounds, tpr, distributions} visualizes the density of our estimates across simulations. 
Their bias quickly converges to zero as the size of the evaluation data grows large.
We further explore how the performance of proposed estimators varies as we vary the researcher's assumptions on the strength of unobserved confounding.
Online Appendix Figure \ref{figure: nonparametric outcome bounds, sensitivity analysis} reports the simulation distribution of our proposed estimators for both the upper and lower bounds as we vary $\overline{\Gamma} = \widetilde{\Gamma}$ for $\widetilde{\Gamma} \in \{1, \hdots, 2.5\}$ and set $\underline{\Gamma} = 1/\widetilde{\Gamma}$.

\begin{figure}[!t]
\centering
\includegraphics[width=3in]{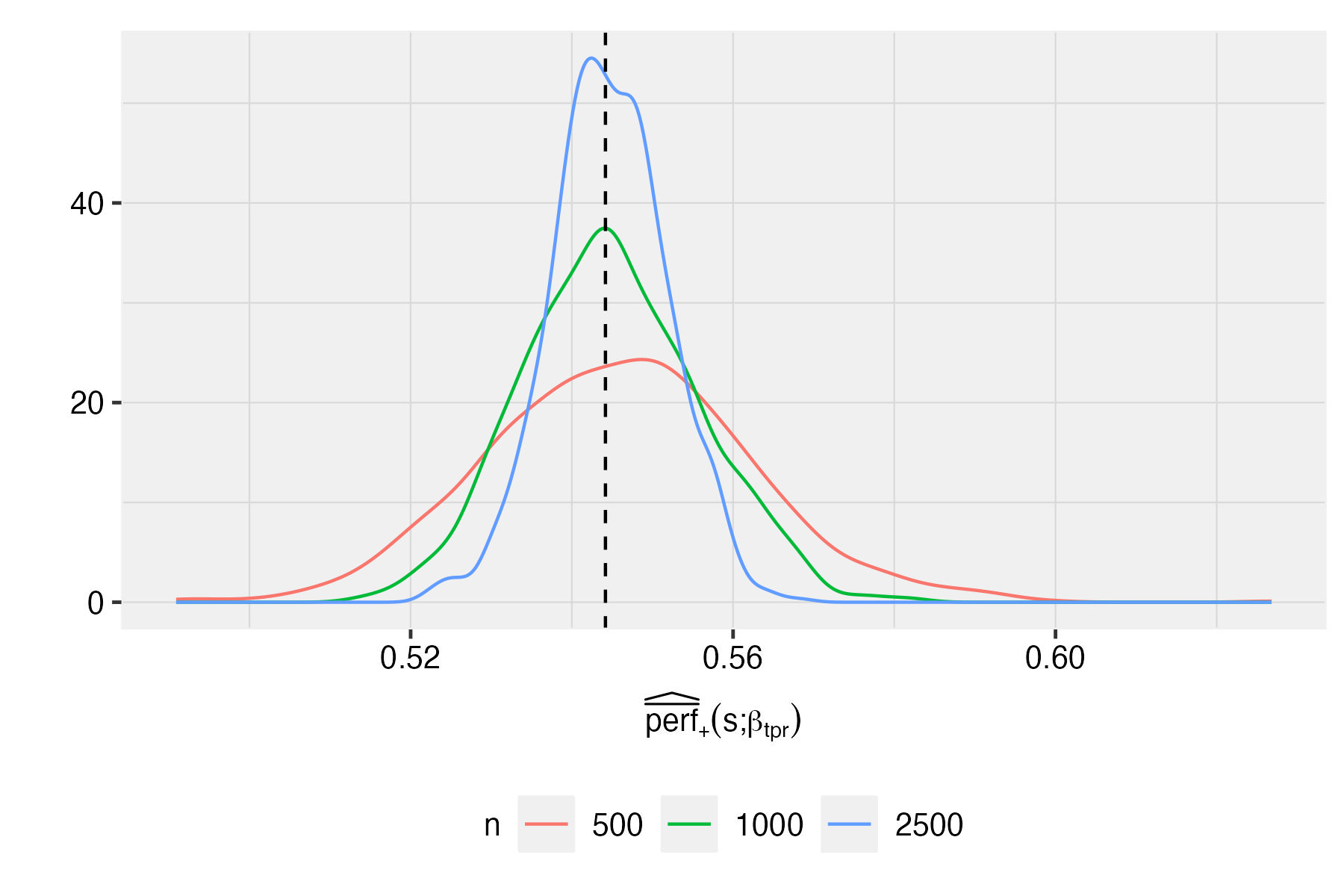}
\caption{Distribution of estimator for the upper bound on the true positive rate across Monte Carlo simulations with observed outcome bounds}
\floatfoot{\textit{Notes}: This figure plots the distribution of the positive class performance estimator for the upper bound on the true positive rate.
We report these results for $n \in \{500, 1000, 2500\}$ (color) and observed outcome bounds with $\underline{\Gamma} = 2/3$, $\overline{\Gamma} = 3/2$.
The vertical dashed lines show the true upper bound on the true positive rate $\overline{\perf}_{+}(s; \beta)$.
The results are computed over 1,000 simulations. See Appendix \ref{section: monte carlo simulations} for further discussion. 
}
\label{figure: nonparametric outcome bounds, tpr, distributions}
\end{figure}

\begin{figure}[h!]
\centering
\includegraphics[width = 3in]{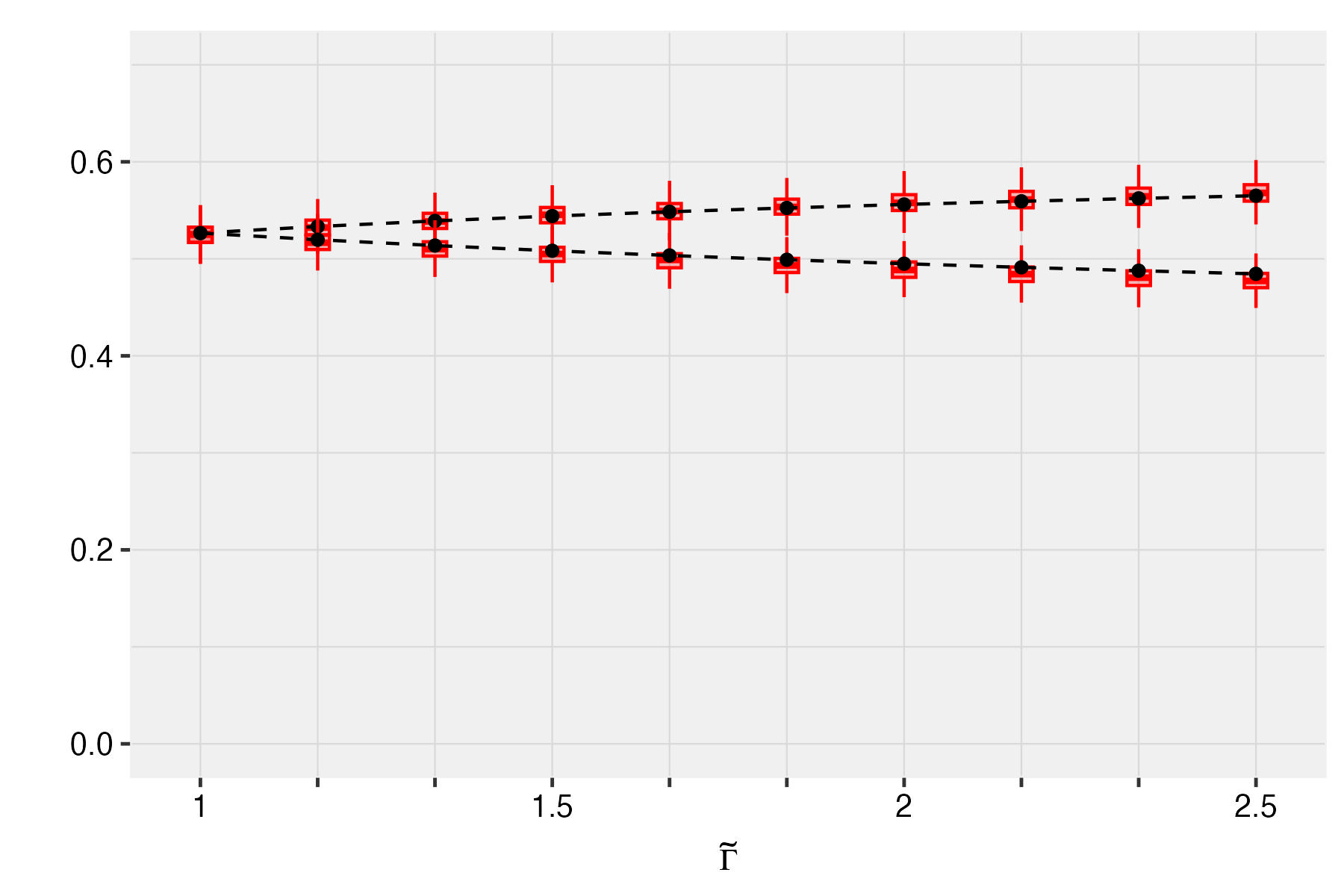}
\caption{Estimated bounds on the true positive rate as $\underline{\Gamma}, \overline{\Gamma}$ varies.}
\floatfoot{\textit{Notes}: 
This figure illustrates box-plots (red) for the distribution of estimators of the bounds on the true positive rate as $\underline{\Gamma} = 1 / \tilde{\Gamma}$, $\overline{\Gamma} = \tilde{\Gamma}$ varies.
The dashed black lines show the true upper and lower bounds for each value of $\tilde{\Gamma}$. 
Results are reported for $n = 1000$ and computed over $1000$ simulations. 
}
\label{figure: nonparametric outcome bounds, sensitivity analysis}
\end{figure}

Our proposed estimators for the overall performance bounds follow a standard debiased construction.
We briefly illustrate that our proposed estimators for the overall performance bounds also converge quickly and confidence intervals based on the asymptotic normal approximation have good coverage properties.
For $\underline{\Gamma} = 2/3$, $\overline{\Gamma} = 3/2$ and focusing on the lower bound on mean square error and the upper bound on accuracy, Online Appendix Table \ref{table: nonparametric outcome, overall performance, bias and coverage} reports the average bias of our estimators and the estimated coverage rate of 95\% nominal confidence intervals.
Our proposed estimators are approximately unbiased for the true bounds. 
Their estimated standard errors slightly underestimate the true standard errors when the size of the evaluation dataset is small, but are quite accurate for $n \geq 1000$.
Consequently, confidence intervals based on the asymptotic normal approximation have approximately 95\% coverage. 
Our proposed estimators are approximately normally distributed in finite samples and concentrate around the true bounds (Online Appendix Figure \ref{figure: nonparametric outcome bounds, overall performance, distributions}).

\begin{table}[htbp!]
\begin{center}
\subfloat[Lower bound on mean square error]{
\begin{tabular}{c | c c c c }
$n$ & Bias & SD & $\widehat{\sigma}$ & Coverage \\
\hline \hline 
500 & -0.001 & 0.011 & 0.011 & 0.950 \\
1000 & -0.001 & 0.008 & 0.008 & 0.939 \\
2500 & -0.001 & 0.004 & 0.008 & 0.939
\end{tabular} 
} \hspace{0.5em}
\subfloat[Upper bound on accuracy]{
\begin{tabular}{c | c c c c}
$n$ & Bias & SD & $\widehat{\sigma}$ & Coverage \\
\hline \hline 
500 & 0.001 & 0.011 & 0.011 & 0.950 \\
1000 & 0.001 & 0.008 & 0.008 & 0.934 \\
2500 & 0.001 & 0.005 & 0.005 & 0.942 
\end{tabular}
}
\end{center}
\caption{Bias and coverage properties with observed outcome bounds.}
\floatfoot{\textit{Notes}: 
This table summarizes the average bias of $\widehat{\overline{\perf}}(s; \beta)$ for the upper bound on MSE and $\widehat{\underline{\perf}}(s; \beta)$ for the lower bound on accuracy, the standard deviations of our estimators (SD), their average estimated standard errors ($\widehat{\sigma}$), and the coverage rate of nominal 95\% confidence intervals.
We report these results for $n \in \{500, 1000, 2500\}$ and observed outcome bounds with $\underline{\Gamma} = 2/3$, $\overline{\Gamma} = 3/2$.
The results are computed over 1,000 simulations. See Appendix \ref{section: monte carlo simulations} for further discussion.
}
\label{table: nonparametric outcome, overall performance, bias and coverage}
\end{table}

\begin{figure}[!t]
\centering
\subfloat[Upper bound on mean square error]{\includegraphics[width = 3in]{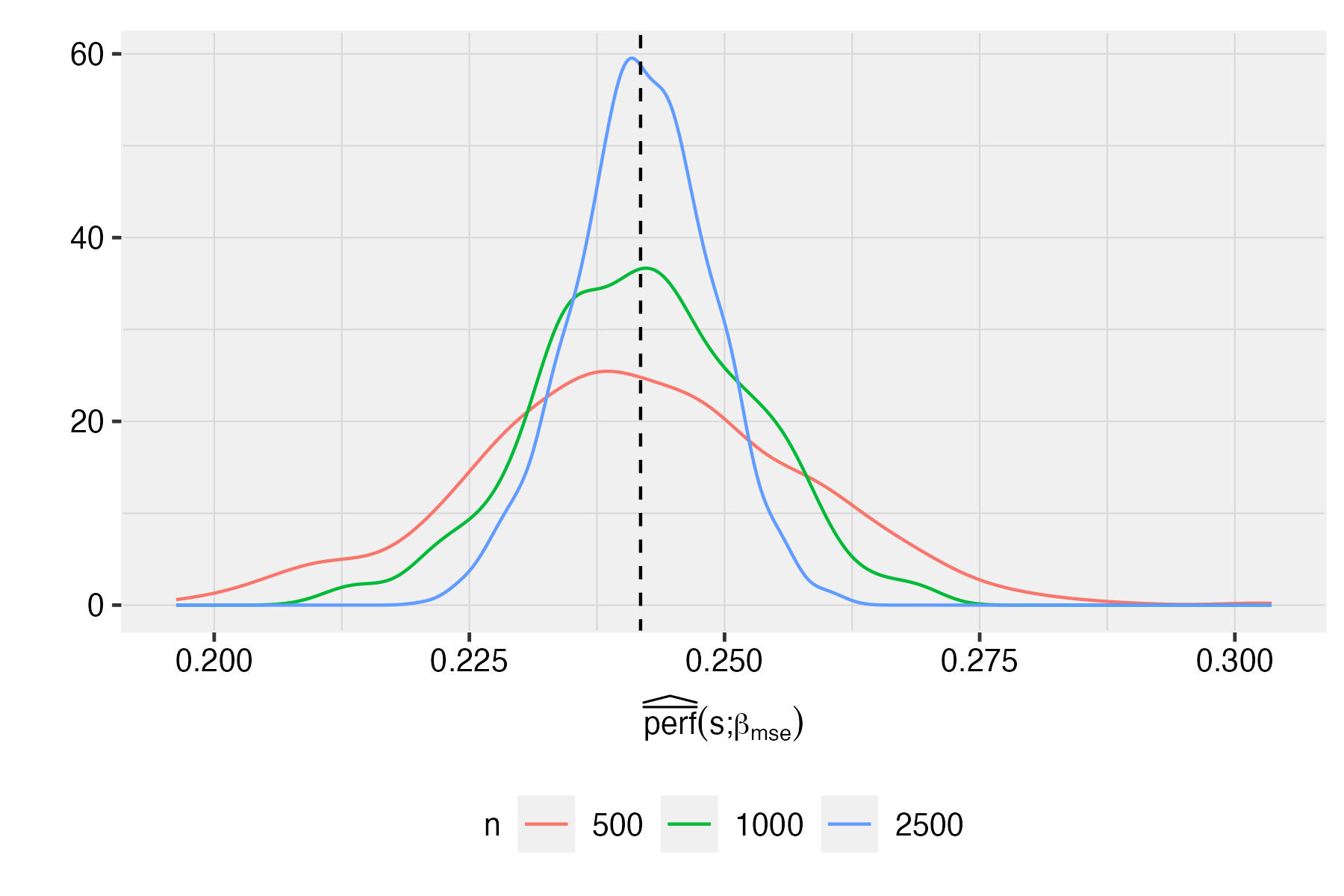}} \hspace{5mm}
\subfloat[Lower bound on accuracy]{\includegraphics[width=3in]{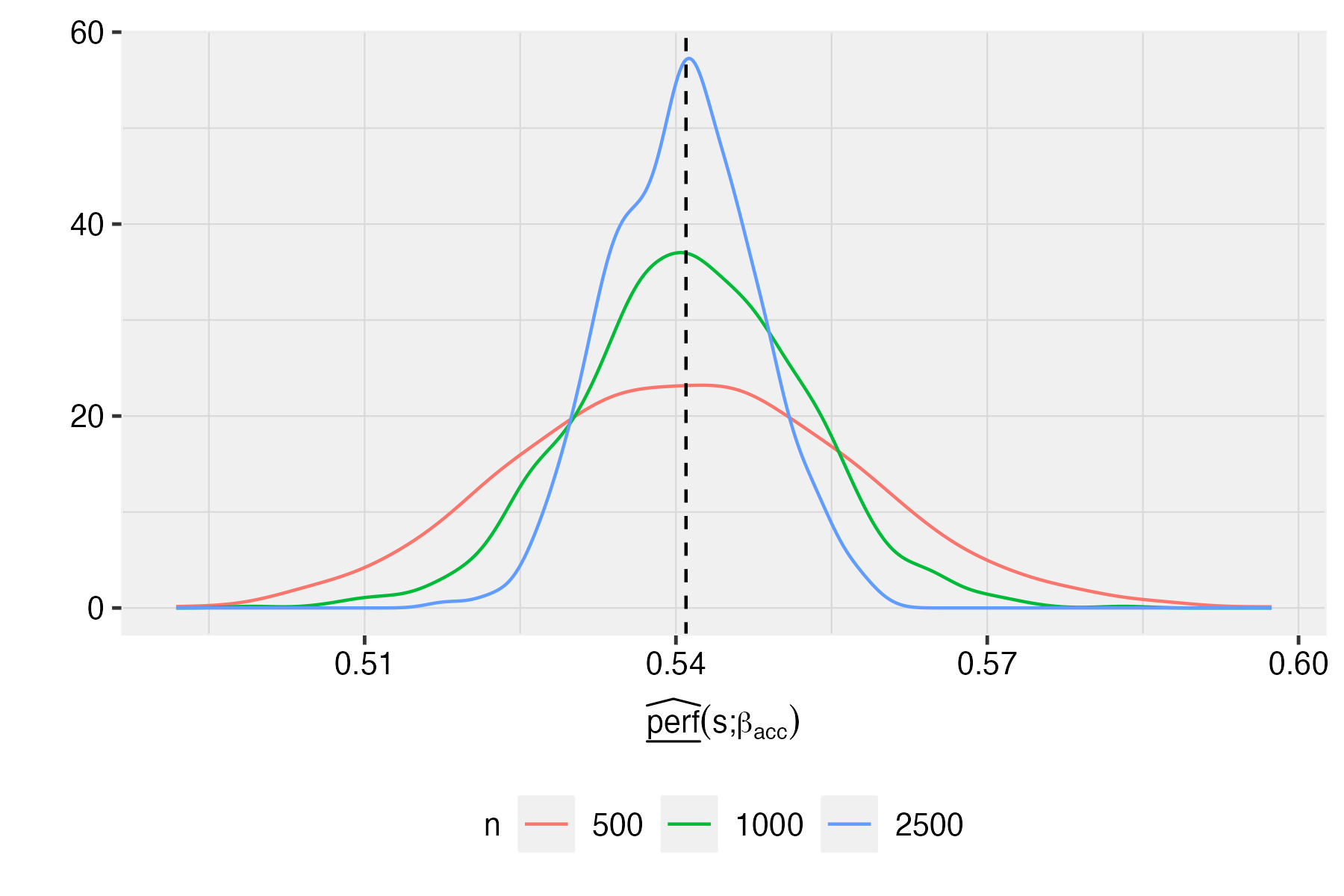}}
\caption{Distribution of overall performance estimators across Monte Carlo simulations with observed outcome bounds.}
\floatfoot{\textit{Notes}: This figure plots the distribution of the overall performance estimator for the upper bound on the mean square error $\overline{\perf}(s; \beta)$ (A) and the lower bound on the accuracy $\underline{\perf}(s; \beta)$ (B).
We report these results for $n \in \{500, 1000, 2500\}$ (color).
The vertical dashed lines show the true upper bound on mean square error $\overline{\perf}(s; \beta)$ and the true lower bound on accuracy $\underline{\perf}(s; \beta)$.
 The overall performance estimators assume that $\underline{\Gamma} = 2/3$, $\overline{\Gamma} = 3/2$.
The results are computed over 1,000 simulations. 
}
\label{figure: nonparametric outcome bounds, overall performance, distributions}
\end{figure}

%%%%%%%%%%%%%%%%%%%%%%%%%%%%%%%%%%%%%%%%%%%%
% Empirical Application: Additional Tables %
%%%%%%%%%%%%%%%%%%%%%%%%%%%%%%%%%%%%%%%%%%%%
\section{Additional Tables for Empirical Application}\label{section: additional Monte Carlo simulations and empirical application}

In this section, we report an additional table for the empirical application that is referenced in Section \ref{section: empirical application, main text} of the main text.  
Table \ref{table: DR-Learner, CBA detailed variable description} provides detailed descriptions of the variable names in right panel of Figure \ref{figure: DR-Learner, CBA application} in Section \ref{section: DR-Learner, CBA application} of the main text.
Personal loans applications can have multiple listed applicants, so some variables refer to just the first listed applicant.

\begin{table}[htbp]
\begin{adjustbox}{angle=90}
\begin{tabular}{l | l}
\textbf{Variable name} & \textbf{Detailed description} \\
\hline \hline
Total net income & 
\begin{tabular}{l} Total net income for all applicants \\ on the personal loan application \end{tabular} \\ \hline
Occupation type & Industry code of 1st applicant's occupation. \\ \hline
Mos in current employment & \begin{tabular}{c} Number of months 1st applicant \\ has held current job. \end{tabular} \\ \hline
Max delinquency in 12 mos & \begin{tabular}{c} Maximum delinquency over last 12 months \\ (home loan, personal loan or credit card). \end{tabular} \\ \hline
Exposure to loan amount & Exposure to requested personal loan amount. \\ \hline
Existing personal loan balance & Existing personal loan balance of applicants. \\ \hline
Current days in debt & Current number of days in debt of all applicants. \\ \hline
Credit bureau score & External credit score. \\ \hline
Accommodation status & \begin{tabular}{c} Type of accommodation applicant currently \\  occupies (e.g., owned, rented, etc). \end{tabular} \\ \hline
\# of credit card apps in 12 mos (all applicants) & \begin{tabular}{c} Number of credit card applications submitted \\ by all applications in last 12 months. \end{tabular} \\ \hline
\# of credit card apps in 12 mos (1st applicant) & \begin{tabular}{c} Number of credit card applications submitted \\ by 1st applicant in last 12 months. \end{tabular} \\ \hline
\# of check acct payment reversals in 6 mos (all applicants) & \begin{tabular}{c} Number of checking account payment \\ reversals by all applicants in last 6 months. \end{tabular} \\ \hline
\# of check acct payment reversals in 6 mos (1st applicants) & \begin{tabular}{c} Number of checking account payment \\ reversals by first applicant in last 6 months. \end{tabular}
\end{tabular}
\end{adjustbox}
\caption{Detailed description of variable names in right panel of Table \ref{figure: DR-Learner, CBA application}.}
\label{table: DR-Learner, CBA detailed variable description}
\end{table}

\end{document}